\documentclass[epj,nopacs]{svjour}

\usepackage{latexsym}
\usepackage{graphics}
\usepackage{amssymb}
\usepackage{amsmath}
\usepackage[english]{babel}
\usepackage{hyperref}
\usepackage{color}
\usepackage{graphicx}
\usepackage{subfigure}
\usepackage{longtable}
\usepackage{bm}
\usepackage{mwe}
\usepackage{float}
\usepackage{textcomp}
\usepackage{mathtools}
\usepackage{dcolumn}
\usepackage{epsf}
\usepackage[T1]{fontenc}
\usepackage{cuted}
\usepackage{microtype}
\usepackage{cite}
\usepackage{balance}

\begin{document}

\title{Tilt in quadratic gravity II}
\author{Waleska P. F. de Medeiros\inst{1}\thanks{\emph{Present address:} waleskademedeiros@gmail.com}
\and Daniel M{\"u}ller\inst{2}\thanks{\emph{Present address:} dmuller@unb.br}
\and Oliver F. Piattella \inst{3}\thanks{\emph{Present address:} of.piattella@uninsubria.it}
 \and Matheus J. Lazo\inst{1}\thanks{\emph{Present address:} matheusjlazo@gmail.com}
\and Dinalva A. Sales\inst{1}\thanks{\emph{Present address:} dinalvaires@gmail.com}
}                     

\offprints{Waleska P. F. de Medeiros}\mail{waleskademedeiros@gmail.com}          
\institute{Instituto de Matem{\'a}tica, Estat{\'i}stica e F{\'i}sica, Universidade Federal do Rio Grande, Caixa Postal 474, Rio Grande, Rio Grande do Sul 96201-900, Brazil \and Instituto de F{\'i}sica, Universidade de Bras{\'i}lia, Caixa Postal 04455, 70919-970 Bras{\'i}lia, Brazil \and Department of Science and High Technology, University of Insubria, Via Valleggio 11, 22100, Como, Italy}

\date{Received: date / Revised version: date}

\abstract{We investigate a tilted fluid component on a Bianchi V geometry in the theories of General Relativity (GR) and Quadratic Gravity (QG). The main objective of this work is the study of how the properties of matter can modify the future evolution of the attractors and their consequences on the regions of initial conditions of the solutions. As is well known, QG contains the Ruzmaikina-Ruzmaikin (RR) solution. This solution describes the slow-roll regime of Starobinsky's inflationary model, which is currently the best one due to the excellent agreement with Cosmic Microwave Background Radiation (CMBR) data. In QG, we found universes that can be attracted to the RR solution or recollapse toward the isotropic singularity attractor. If the Equation of State (EoS) parameter is ultra-radiative $w>1/3$, the tilt variable increases both in RR and Milne for QG or GR, respectively. In both cases, the fluid expansion and acceleration diverge, while the vorticity initially increases and then decreases to zero.
\PACS{
      {PACS-key}{}\and
      {PACS-key}{}\and
      {PACS-key}{}
     }
}

\maketitle

\global\long\def\imsize{0.85\columnwidth}
\global\long\def\halfsize{0.5\columnwidth}
\global\long\def\big{1.2\columnwidth}
\global\long\def\peq{0.65\columnwidth}

\section{Introduction}\label{intro}

A fluid component is said to be tilted if its four-velocity is not orthonormal to some spatially homogeneous surfaces \cite{King:1972td, ellisking}. In this case, a nonzero vorticity, shear and anisotropies arise.

Tilted cosmological Bianchi models have been explored in the literature in the context of GR; see Refs. \cite{Coley_2005,Coley_2006, Lim_2006, Coley:2008zz, Coley_2009, Krishnan:2022qbv, Krishnan:2022uar,Ebrahimian:2023svi,Allahyari:2023kfm}. Within GR, Coley, Hervik, and Lim demonstrated that for an EoS $p = w\rho$ with $1/3 < w < 1$, for the dynamic time evolution toward the future, the Bianchi V solution degenerates into the Milne solution, with a steadily increasing tilt \cite{Coley_2005,Coley_2006, Coley:2008zz}. Their analysis shows that this increase in the tilt leads to a divergence in both the acceleration and the expansion of the fluid source. Moreover, still within GR, Krishnan, Mondol, and Sheikh-Jabbari found results that agree well with the aforementioned tilt increase for the EoS parameter $1/3 < w < 1$ \cite{Krishnan:2022qbv, Krishnan:2022uar}. Furthermore, they showed that the tilt can increase even when the anisotropic Bianchi V model becomes isotropic (for example, when the shear goes to zero). In addition, the authors Ebrahimian, Krishnan, Mondol, and Sheikh-Jabbari discussed the significance of the sign of the tilt parameter in cases where the cosmic fluid has more than one fluid \cite{Ebrahimian:2023svi}. The effects of the tilt near the big bang past singularity were addressed in \cite{Allahyari:2023kfm}.

More recently, tilted source solutions were explored in both GR and QG for the anisotropic Bianchi V model in \cite{deMedeiros:2024pmc}. The time evolution was investigated toward the past singularity. In \cite{deMedeiros:2024pmc}, it is found that in QG, universes with higher and smaller matter density are attracted to the Kasner solution or to the asymptotic isotropic singularity, respectively. It is also shown in \cite{deMedeiros:2024pmc} that the Kasner singularity is a past attractor with zero vorticity for both GR and QG, while the isotropic singularity attractor may have divergent vorticity for QG. In this sense, only QG allows a geometric singularity with divergences in all the kinematic variables of the substance, which decrease to finite values for the future when time is reversed. The authors also found that for the EoS parameter $1/3 < w < 1$, all kinematic variables decrease to zero, both in GR and QG theories. According to their results, for the ultra-radiative regime, any small initial perturbation in the kinematic variables could increase for future evolution.

This work is a continuation of a previous one, \cite{deMedeiros:2024pmc}. Now only evolution to the future is considered. Here also, it is supposed a tilted fluid component on Bianchi V geometry both in GR and QG. We confirm all the aforementioned findings in \cite{Coley_2005,Coley_2006, Coley:2008zz, Krishnan:2022qbv, Krishnan:2022uar, deMedeiros:2024pmc} for GR and in \cite{deMedeiros:2024pmc} for QG.

The effective quadratic gravity theory arises in a scenario where quantized fields are considered on a classical gravitational background. The necessary counter-terms to the effective action for the quadratic gravity be renormalized are \cite{dewitt1965dynamical, article, Christensen:1976vb, Christensen:1978yd, Birrell:1982ix,Grib:330376}
\begin{align}
    S=&\int \mathrm{d}^4x\sqrt{-g}\frac{m^2_{p}}{2}\left [ \left(R-2\,\Lambda\right)+\beta R^2+\alpha  \left ( R_{ab}R^{ab} \right . \right.\nonumber\\&\left.\left.-{1}/{3}\,R^2\right ) \right ]\,.\label{S}
\end{align} 
This action is the basis for the Starobinsky inflationary model \cite{Starobinsky:1980te}. Here, $g$ is the determinant of the metric $g_{ab}$; $\Lambda$ is the cosmological constant, $\alpha$ and $\beta$ are the renormalization parameters; and $m_p=(\hbar c^5/G)^{1/2}$ is the Planck mass, $c$ is the speed of light, $\hbar$ is the Planck constant and $G$ is the gravitational constant. $R_{ab}$ is the Ricci tensor and $R$ is the Ricci scalar. Einstein-Hilbert GR is recovered when both $\beta=0$ and $\alpha =0$. 

Starobinsky inflation \cite{Starobinsky:1980te} is currently considered the best inflationary model due to its excellent correspondence with CMBR measurements \cite{Akrami:2018odb}. In four dimensions and isotropic space-times, quadratic gravity \eqref{S} reduces to a particular type of $f(R) \sim R+R^2$, where the action of the $f(R)$ gravity theories depends only on the Ricci scalar; see Refs. \cite{Sotiriou:2008rp, DeFelice2010, Nojiri_2011, Nojiri_2017}. Since this formulation is equivalent to $f(R)$, Starobinsky inflation can be converted by the appropriate conformal metric transformation to Einstein-Hilbert gravity with an additional scalar field coupled to matter fields \cite{Maeda:1988ab, Cotsakis_2008, Sotiriou:2008rp, DeFelice2010, Nojiri_2011, Nojiri_2017}.

Since Ostrogradsky's time, it has been well known that higher-order theories might develop instabilities due to the increased degrees of freedom associated with the higher-order derivative terms present in these theories \cite{Ostrogradsky}. In this context, the Lagrangian of a higher-order differential theory satisfies the non-degenerescence condition of Ostrogradsky, which leads to a Hamiltonian that is unbounded from below. 

A dynamical variable that suffers from Ostrogradsky's instability carries kinetic terms with opposite signs, which, through coupling, can result in infinite energy transfers between degrees of freedom, while the total energy of the system is conserved \cite{trodden2016theoretical}. This type of instability is known today as ghosts \cite{PhysRev.79.145,Stelle:1977ry,Woodard:2007,trodden2016theoretical}. It is well known that when the quadratic theory is linearized near Minkowski space, it shows a massless spin 2 field, a massive spin 0 field, and a massive spin 2 field that has energy with the wrong sign \cite{Stelle:1977ry}. 

Also, still in the linearization context near the Minkowski solution, the tachyon behavior in action \eqref{S} is eliminated for the following choice of the renormalization parameters $\beta>0$ and $\alpha<0$, see \cite{PhysRevD.32.379,MULLER:2014jaa}. However, ghosts in QG cannot be eliminated by an appropriate choice of theory parameters. Nevertheless, despite Ostrogradsky's ghost, the quadratic theory is healthy, in the sense that it does not suffer from causality problems \cite{Edelstein_2021}.

As discussed in \cite{Woodard:2007}, one of the main motivations of the theory $f(R)$ is that the Ostrogradsky ghosts are eliminated. In this case, the field equations of the gravitational theory $f(R)$ result in only a single higher-order derivative equation that carries the dynamics, while the lower-order derivative equation results in a constraint. This constraint limits the lower derivative degrees of freedom, which violate the main assumption of Ostrogradsky's instability, the non-degenerescence condition \cite{Ostrogradsky}. 

In this work, anisotropic metrics are addressed, and the renormalization parameter $\alpha$ cannot be disregarded in the effective quadratic action \eqref{S} since the metric is not isotropic. In this case, the effective action \eqref{S} depends on the Ricci tensor term $R_{ab}R^{ab}$. The evolution is toward the future for the anisotropic Bianchi V model for both GR and QG theories. Bianchi V model is chosen because of its non-trivial behavior of the kinematic variables such as vorticity, acceleration, and expansion of the tilted substance \cite{King:1972td}.

Despite the well-known asymptotic isotropic RR solution, we found universes that can be attracted to the isotropic singularity \cite{1969JETP...30..372R}. We also found that for all solutions attracted to RR or Milne for QG or GR, respectively, and for $1/3<w < 1$, all kinematic variables initially increase during the transient regime along with the tilt variable. However, we show that the acceleration and the expansion of this substance increase indefinitely, while the vorticity shows an increase followed by a decrease to zero. On the other hand, in QG, universes attracted for the isotropic singularity may have divergent vorticity along with all the other kinematic singularities \cite{deMedeiros:2024pmc}.

This work also explores how the tilted substance affects a basin plot for different initial conditions of the diagonal shear variables attracted to the RR solution or to the isotropic singularity. We identified initial conditions, both near and farther from Ruzmaikina's orbit, where the inflation occurs. For a specific set of renormalization parameters $\alpha$ and $\beta$, we show how the attraction regions for RR shift when the tilt is introduced. The farther regions of Ruzmaikina's attraction are moved to the left, while the region near the origin shifts to the right. Since the values considered to show a significant effect correspond to large values for the tilted source, its effects are small on the basin plots. Where for a more realistic set of the renormalization parameters with $\beta=1.305\times10^{9}\,m_p^{-2}$, as inferred by CMBR measurements \cite{Gorbunov_2015,Mishra:2018dtg,Akrami:2018odb}, the tilt has a negligible effect on the basin plot.

This paper is structured as follows. Section \ref{sec2} shows the field equations for a tilted source. In Sect. \ref{sec3}, the linear stability of some solutions in GR and QG is carefully analyzed. In Sect. \ref{sec4}, the numerical analysis of the field equations is presented. The behavior of kinematic variables is analyzed for both theories of gravity. We also show for QG how the tilted source changes a basin of initial conditions that are either attracted to the slow-roll inflationary solution of RR or recollapse toward the isotropic singularity attractor. Finally, the conclusions are presented in Sect. \ref{summary}. 

The results are expressed in terms of the expansion normalized variables (ENV), which are dimensionless and obtained by normalizing the time and the physical quantities with the appropriate powers of the Hubble parameter $H$, as defined in \cite{Barrow:2006xb}. However, we do not use the ENV to develop the numerical codes due to the singularity in the Hubble parameter when it passes through zero for solutions that evolve toward recollapse. The numerical codes were developed using the GNU/GSL Ordinary Differential Equation (ODE) package\footnote{For a detailed example of functions for solving initial value problems of ODEs, refer to the following link: \url{https://www.gnu.org/software/gsl/doc/html/ode-initval.html}.}, which contains the Runge-Kutta Prince-Dormand method (8, 9) on Linux. The codes were generated with the algebraic manipulator Maple 17. The following conventions and units are used: the metric signature is $-+++$, the Latin indices $a,\,b,\,...$ run from $0$ to $3$, and $G=\hbar=c=1$. The constraints fluctuate numerically, always less than $10^{-7}$. 

\section{Field equations}\label{sec2}

The field equations obtained from the metric variation of the effective action \eqref{S} for QG are given by
\begin{align}
     E_{ab} \equiv& \left( G_{ab}+g_{ab}\Lambda \right)+\left ( \beta -\frac{1}{3}\,\alpha  \right )H^{(1)}_{ab}\nonumber\\&+\alpha H^{(2)}_{ab}-\kappa T_{ab}=0\,,\label{eqdemov}
\end{align}
where:
\begin{align}
    G_{ab}=&\,R_{ab}-\frac{1}{2}\,g_{ab}R\,,\nonumber\\
    H^{(1)}_{ab}=&- \frac{1}{2}\,g_{ab}R^2+2\,RR_{ab}+2\,g_{ab}\square R-2\,{R}_{;ab}\,,\nonumber\\
    H^{(2)}_{ab}=&-\frac{1}{2}\,g_{ab}R_{cd}R^{cd}-R_{;ab}+2\,R^{cd}{{{{R_c}_{bda}}}}+\square R_{ab}\nonumber\\&+\frac{1}{2}\,\square Rg_{ab}\,.
\end{align}
Here, $G_{ab}$ is the Einstein-Hilbert tensor, $H^{(1)}_{ab}$ is the tensor resulting from metric variation on the term $R^2$, while $H^{(2)}_{ab}$ comes from metric variation on the term $R_{ab}R^{ab}$. As usual, $T_{ab}$ is the energy-momentum tensor and $\kappa = 8\pi G$. The field equations \eqref{eqdemov} contain partial differential equations of order $4$ in the metric.

It is well known that the covariant divergence of the field equations \eqref{eqdemov} vanishes, leading to the constraints $E_{00}$ and $E_{0i}=0$, with the dynamics governed by the spatial part $E_{ij}$; see, for example, \cite{Weinberg:1972kfs, Stephanibook, Stelle:1977ry}. Therefore, the temporal component of the field equations \eqref{eqdemov} corresponds to the lowest-order equation, which is a constraint dynamically preserved and used as a numerical check.

In this work, the proper time $t$, with the dimension of the Planck time $t_p$, is chosen. An orthogonal, nonrigid base $e_a^c$ is employed, where the index $a$ numerates the tetrad. This base results in the projected metric tensor $g_{ab}=-\delta_{a0}\delta_{b0} +\delta_{ij}\omega^i\otimes \omega^j$, with $\omega^i$ the dual of the spatial part of the basis $e_j$, where $\omega^i e_j=\delta^i_j$. In other words,
\begin{equation}
    {g_{ab}=\begin{pmatrix}
-1 &0  &0  &0 \\ 
 0&1  & 0 &0 \\ 
 0& 0 &1  &0 \\ 
 0& 0 & 0 &1 \label{mett}
\end{pmatrix}}\,.
\end{equation}
We choose a time-like vector $u^a=(1,0,0,0)$, with normalization $g_{ab}u^a u^b=-1$, and the projector in the orthogonal three-dimensional space to $u^a$ as $h_{ab}=u_au_b+g_{ab}$, with ${h^c}_bg_{ca}=h_{ab}$ and ${h^c}_au_c=0$. 

The covariant derivative $\nabla _au_b$ can be decomposed into its irreducible parts \cite{Stephanibook}: an anti symmetric part, called the rotation tensor $\omega_{ab}$, a symmetric traceless part, $\sigma_{ab}$, called the shear tensor, and another part containing only its trace $\Theta$:
\begin{align}
    &\nabla _au_b=\sigma _{ab}+\omega _{ab}+\frac{1}{3}\,\Theta \,\delta _{ab}-\dot{u}_au_b\,,\nonumber\\
    &\sigma _{ab}=u_{(a;b)}-\frac{1}{3}\,\Theta \,\delta _{ab}+\dot{u}(_{a}u_b)\,,\nonumber\\
    &\omega  _{ab}=u_{[a;b]}+\dot{u}[_{a}u_b]\,,\nonumber\\
    &\dot{u}_a=u^b\nabla _bu_a\,,\nonumber\\
    &\Theta =\nabla _au^a \label{cinva}\,,
\end{align}
the vector $\dot{u}_a$ is the acceleration vector.

The vorticity vector is related to the rotation tensor by
\begin{equation}
    \omega_a=\frac{1}{2}\,\epsilon_{abcd}\,u^b\,\omega^{cd}\,,\label{vortve}
\end{equation}
where $\epsilon^{abcd}$ is the Levi-Civita tensor. For spatially homogeneous geometries, it can be seen that the vector $u^a$ is geodesic and has zero vorticity.

Otherwise, any covariant derivative in the base $e_a^c$ can be written with the appropriate connection:
\begin{equation}
    \nabla _au_b=e^c_a\partial_c u_b-\Gamma^c_{ba}u_c\label{nabla}\,.
\end{equation}
For this specific vector $u_a=(-1,0,0,0)$, the above equation becomes: 
\begin{equation}
    {\nabla _au_b=\Gamma ^0_{ab}}\,,\label{tri}
\end{equation}
which, according to \eqref{cinva}, the temporal part of the connection is given by:
\begin{equation}
    {\Gamma ^0_{ab}=\left\{\begin{array}{ll}
&0,  \,  \mbox{if} \,  a=0 \,\mbox{ or} \, b=0\,,  \\
& \sigma _{ij}+H\delta _{ij}, \,\mbox{if} \, a \neq 0 \,  \mbox{and} \,  b\neq 0\,.  
\end{array}\right.} \label{Gamma_0ab}
\end{equation}
Also, considering \eqref{cinva}, \eqref{tri}, \eqref{Gamma_0ab}, and the shear $\sigma_{ij}$ with zero trace, it is immediately seen that using this connection the expansion is $\Theta = \nabla_a u^a=3H$. The shear in \eqref{Gamma_0ab} is chosen as:
\begin{equation}
      {{\sigma _{ij}}= \left( \begin {array}{ccc} \,- {2\,\sigma_{{+}}
}&0& {\phi_{{3}} }\\ \noalign{\medskip}0&{\sigma_{{+}}  +\sqrt {3}\,
\sigma_{{-}}  }&{0}\\ \noalign{\medskip} {\phi_{{3}} }&{0}& {\sigma_{{+
}} -\sqrt {3}\,\sigma_{{-}} }\end {array} \right) }\,.\label{sig1}
  \end{equation}
Here, zero shear describes the isotropic case, as discussed in Section \ref{subsec21}. 

Now we address the spatial part of the connection. Given that the basis $e_a^c$ are non-coordinate, the connection is uniquely determined by metricity and zero torsion
\begin{align}
   &\nabla _cg_{ab}=0 \implies e_c^d\partial_d g_{ab}-\Gamma^d_{ac}g_{db}-\Gamma^d_{bc}g_{ad}=0\,,\nonumber\\
   &\nabla _ae_{b}-\nabla _be_{a}=[e _{a},e_{b}]\implies (\Gamma^d_{ba}-\Gamma^d_{ab})e_d=[e _{a},e_{b}]\,, \label{mtr_torc}
\end{align}
with the metric given by \eqref{mett}.

For spatially homogeneous models, the commutator of the spatial part of the basis is given by
\begin{equation}
    [e_i,e_j]=-C^k_{ij}e_k\,,\label{sc1}
\end{equation}
for $C^k_{ij}=-C^k_{ji}$.

Since the spatial derivatives of the metric \eqref{mett} are zero, they do not contribute to the connection. In this case, the metricity condition implies for the spatial part that $\Gamma_{ijk}=-\Gamma_{jik}$ \eqref{mtr_torc} while zero torsion implies that $\Gamma_{c\,ba}-\Gamma_{c\,ab}=C_{c\,ab}$ \eqref{mtr_torc} which results in:
\begin{equation}
{\Gamma _{i\,jk}=\frac{1}{2}(C_{jki}+C_{kji}-C_{ikj})}\,,\label{cestru}
        \end{equation}
for the pure spatial part of the connection.
In this work, the Bianchi V geometry is selected because it is one of the simplest models that contains non-trivial behavior for tilted sources \cite{King:1972td}. The only non-null structure constants appropriate for Bianchi V geometries are \cite{Stephani:2003tm}
\begin{align}
    &C^2_{12}=C^3_{13}=b(t),\,C^2_{21}=C^3_{31}=-b(t)\,.\label{ce}
\end{align}
The time dependence implies that the structure constants are preserved at the slices of constant time homogeneous hypersurfaces. These structure constants \eqref{ce} for Bianchi V result in the only non-null components for the pure spatial part of the connection 
\begin{align}
    &\Gamma _{212}=\Gamma _{313}=b(t), &\Gamma _{122}=\Gamma _{133}=-b(t)\,. \label{Gamma_ijk_q}
\end{align}
The connection is not completely determined by \eqref{Gamma_0ab} with shear \eqref{sig1} and \eqref{Gamma_ijk_q}. This occurs because the connection is assumed to be the dynamical variable, instead of assuming the metric as such, as is usual. In this case, there is an additional condition that must be fulfilled for consistency, as we demonstrate below.

As it is well known, the Riemann tensor is derived from the commutator of the covariant derivatives:
\begin{equation}
    {\left [ \nabla _c,\nabla _d \right ]V^a={R^a}_{bcd}V^b}\,,
\end{equation}
which leads to the expression
        \begin{align}
            &{{R^{a}}_{bcd}=\Gamma ^{a}_{bd|c}-\Gamma ^{a}_{bc|d}+\Gamma ^{a}_{fc}\Gamma ^{f}_{bd}-\Gamma ^{a}_{fd}\Gamma ^{f}_{bc}+C^f_{cd}\Gamma ^{a}_{bf}}\,.
        \end{align}
The Riemann tensor must satisfy the Jacobi identity, which is given by $R_{abcd}+R_{acdb}+R_{adbc}=0$. This condition results in the following first-order differential equation:
\begin{equation}
     \dot{b} +bH -2b \sigma_{{+}}  =0\,.\label{jacobi}
\end{equation}
 
\subsection{Tilted source and Einstein-Hilbert gravity}\label{subsec21}

We begin by addressing the isotropic Bianchi V in GR for the non-tilted case, which reproduces the well-known open Friedmann model. The spatial isotropic line element $d\sigma^2=a(t)^2\delta_{ab}\omega^a\otimes \omega^b$ that for the appropriate Bianchi V 1-form basis $\omega^a$ \cite{Stephani:2003tm} results in:
\begin{equation}
 d\sigma ^2=a^2(t)\left(dx^2+e^{2x}dy^2+e^{2x}dz^2\right)\,.
\end{equation}
The coordinate transformation:
\begin{align}
&x=\ln\left ( \cosh\chi - \sinh\chi \cos\theta  \right )\,,\nonumber\\
&y=\frac{\sin\theta \cos\phi }{\coth \chi -\cos\theta }\,,\nonumber\\
&z=\frac{\sin\theta \sin\phi }{\coth \chi -\cos\theta }\,,
\end{align}
leads to the \textit{Friedmann-Lemaître-Robertson-Walker} (FLRW) line element: 
\begin{equation}
    d\sigma^2 = a^2(t)\left[d\chi^2+\sinh^2\chi\left ( d \theta^2+\sin^2\phi \, d \phi ^2\right )\right]\,,
\end{equation}
for negative spatial curvature.

As already mentioned, the field equations for GR are obtained by setting $\alpha=0$ and $\beta=0$ in Eq. \eqref{eqdemov}, resulting in
\begin{equation}
    E_{ab }\equiv G_{ab }-\kappa T_{ab }=0\,,
\end{equation}
with perfect fluid $T_{ab}=(\rho+p)u_au_b+pg_{ab}$ as source, where $p=w\rho$ with equation of state EoS parameter $w$, metric \eqref{mett} and fluid velocity $u_a=(-1,0,0,0)$. This results in Friedmann equations: 
\begin{align*}
   & -\frac{\kappa \rho}{3H^2}-\frac{b^2}{H^2}+1=0\,,\nonumber\\
   &-2\dot{H}-3H^2+b^2-w\kappa\rho=0\,,
\end{align*}
for the usual density and curvature parameters:
\begin{align*}
&\Omega_{m }=\frac{\kappa\rho}{3H^2}\,, & \Omega_K=b^2/H^2\,, 
 \end{align*}
we get the following Friedmann equations:
\begin{align}
   & \Omega_m+\Omega_K-1=0\,,\nonumber\\
   &2\frac{\dot{H}}{H^2}-\Omega_K+3(1+w\Omega_m)=0\,,\label{fried}
\end{align}
altogether with the additional equation, which must be satisfied by $\Omega_K$:
\begin{equation}
     \dot{\Omega}_{K} =- 2\,\Omega_{K}H +4\,\Omega_{{K}} \Sigma_{+} H-2\,\Omega_{{K}} \dot{H}/H\,,\label{deromegak}
\end{equation}
which is the Jacobi identity shown in \eqref{jacobi} in terms of the curvature density $\Omega_K$. This equation \eqref{deromegak} is one of the dynamical equations that must be satisfied in this work. 

Now the connection is completely defined by the temporal part \eqref{Gamma_0ab} with shear \eqref{sig1} and the spatial part given by
\begin{align}
    &\Gamma _{212}=\Gamma _{313}=H\sqrt{\Omega_K}, &\Gamma _{122}=\Gamma _{133}=-H\sqrt{\Omega_K}\,. \label{Gamma_ijk}
\end{align}
the expansion-normalized variables ENV, usually used to express the dynamical system, are dimensionless and are obtained by normalizing the shear components by the Hubble parameter $H$. The necessary ENV for describing the Einstein-Hilbert gravity are \cite{Barrow:2006xb}:
\begin{align}
&\Omega_{m }=\frac{\kappa\rho}{3H^2}\,,  &\Phi_3=\frac{\phi _{3 }}{H}\,,     \nonumber\\ 
      &\Sigma_{\pm}=\frac{\sigma _{\pm }}{H}\,,
      & \Omega_K\,.\label{ENV1}
 \end{align}
Now we address the non-isotropic tilted Bianchi V solutions in GR. First, remember the well-known relation for the Hubble parameter $H$, $H = \dot{a}/a$, with $a$ being the scale factor. As discussed in \cite{deMedeiros:2024pmc}, for the non-isotropic case, or the nonzero shear case, there is no scale factor; nevertheless, it can be defined as
\begin{equation}
H=\dot{a}/a \implies a=a_0e^{\int H d\tau}\,,
\end{equation}
with $a_0$ a constant of integration. The deceleration parameter can be rewritten as:
\begin{equation}
    q=-\frac{\ddot{a}a}{(\dot{a})^2}=-(\dot{H}/H^2+1)\,.\label{decpar}
\end{equation} 
Since the tilt is a source property, it is described by the energy-momentum tensor \cite{deMedeiros:2024pmc}:
\begin{equation}
    T_{ab}=\frac{1}{\kappa}\left [ (1+w)\,3\,H^2\Omega_m  \hat{u}_a \hat{u}_b +3\,wH^2\Omega_m\,g_{ab}\right]\,,\label{tem}
\end{equation}
with the metric given by \eqref{mett} and $w$ the EoS parameter with $p=w\rho$ and the time-like vector $\hat{u}^a$ is:
\begin{equation}    
\hat{u}^a=\left[\cosh(r),\sinh(r)\cos(\eta),0,\sinh(r)\sin(\eta)\right]\,,\label{timeve}
\end{equation}
with normalization $\hat{u}^a\hat{u}^bg_{ab}=-1$. Here, $r$ is the dimensionless tilt variable, and $\eta$ specifies the tilt direction, and it is given in radians. In this sense, $r$ describes the changes in tilt in time and $\eta$ the direction of this tilt. It can be seen that when the tilt is zero, $r=0$, the usual perfect fluid source is recovered, and the time-like vector $\hat{u}^a$ coincides with the time-like vector $u^a=(1,0,0,0)$. The tilt $r$ is with respect to the vorticity-free vector $u^a$.

Now, since the tilted substance moves according to $\hat{u}^a$ as defined in \eqref{timeve}, it is necessary to have a suitable set of kinematic variables instead of those shown in \eqref{cinva}-\eqref{vortve}. To derive these kinematic variables for the tilted matter, $\nabla_a\hat{u}_b$, the metric \eqref{mett} and the connection defined by \eqref{Gamma_0ab} with shear \eqref{sig1}, \eqref{Gamma_ijk}, and \eqref{deromegak} are used, which results in:
\begin{align}
\hat{\dot{u}}_0=&- {H}w\sinh \left( r  \right)  D\,,\nonumber\\
\hat{\dot{u}}_1=&- 3\,H w \cos \left( \eta \right) \cosh \left( r  \right)  D\,,\nonumber\\
\hat{\dot{u}}_3=&-3\,H w\cosh \left( r  \right) \sin \left( \eta \right) D\,,\nonumber\\
\hat{\Theta}=&-\left\{ \cosh \left( r  \right) H \left[ 2\,\sinh \left( r  \right) \cos \left( \eta \right) \cosh
 \left( r  \right) \sqrt {\Omega_{{K}}} \right.\right.\nonumber\\&\left.\left. +A
 \right]
\right\}/\left\{ B \right\}\,,\nonumber\\
   \hat{\omega}^2 =&-
{\frac { \left| H \right| \sinh \left( r   \right) 
\sin \left( \eta  \right) \sqrt {\Omega_{{K}}}}{\cosh
 \left( r  \right) }}\,,\label{cininc}
\end{align}
where $\hat{\dot{u}}_2$ and the other vorticity components are null. To simplify notation, the terms $A$, $B$, $C$, and $D$ are defined as follows: 
\begin{align}
   A  =  &-\Sigma_- \left(\left( \cos
 \left( \eta \right)\right)^2  -1 \right) \left( \left( \cosh \left( r  \right)\right)^2 -1 \right) \sqrt {3}\nonumber\\&+ \left( 3\, \left( \cos \left( \eta \right)  \right) ^{2}\Sigma_{+}-2\,\cos \left( \eta \right) \sin \left( \eta \right) \Phi_3+2\right. \nonumber\\&\left.-\Sigma_{+}\right) \left( \cosh \left( r  \right)  \right) ^{2}+2\,\cos \left( \eta \right) \sin \left( \eta \right)\Phi_3+1\nonumber\\&+\Sigma_{+}\left(1-3\, \left( \cos \left( \eta \right)  \right) ^{2}\right)\,,\nonumber\\
B = &\left( w-1 \right)  \left( \cosh \left( r  \right)  \right) ^{2}-w\,,\nonumber\\
C = & \,2\,\cosh \left( r  \right) \cos \left( \eta \right)\sqrt {\Omega_{{K}}}\left( \left( \cosh \left(r  \right)  \right) ^{2} -1\right)\,,\nonumber\\
D =&  \left[  \,C +A \sinh \left( r  \right)  \right] /\left(B\right)\,.
\end{align}
When the tilt is null $r=0$, $\hat{u}^a=(1,0,0,0)$ and the matter follows a geodesic $\hat{\dot{u}}^a=0$, with zero vorticity $\hat{\omega}^a=0$ and expansion $\hat{\Theta}=3H$. It can be seen from Eq. \eqref{cininc} that the vorticity vector $\hat{\omega}^a$ is zero when $\eta=0$, $\eta=\pi$, or whenever $\Omega_K$ is zero.

Given this specific time-like vector $\hat{u}^a$ in \eqref{timeve}, the tensor energy-momentum \eqref{tem}, the metric \eqref{mett}, and the connection defined by \eqref{Gamma_0ab} with \eqref{sig1}, \eqref{Gamma_ijk}, and \eqref{deromegak}, the field equations $E_{02}\equiv 0$, $E_{21}\equiv 0$, and $E_{23}\equiv 0$ in \eqref{eqdemov} are identically satisfied for any values of $\alpha$ and $\beta$, which means that $E_{02}\equiv 0$, $E_{21}\equiv 0$, and $E_{23}\equiv 0$ hold both in GR and QG.

Nevertheless, the energy-momentum covariant conservation $\nabla_bT^{ab} = 0$ must be obeyed independently from the gravitational theory. In the same setting of the time vector $\hat{u}^a$ given in \eqref{timeve}, the energy-momentum source in \eqref{tem}, the metric in \eqref{mett}, and the connection defined by \eqref{Gamma_0ab} with \eqref{sig1}, \eqref{Gamma_ijk}, and \eqref{deromegak} result in non-trivial differential equations for the variables $H$, $\Omega_m$, $\eta$, and $r$, which are shown in the Appendix \ref{appA}.

Also, considering the same settings, the non-trivially Einstein-Hilbert field equations with shear and tilt were shown in the Appendix \ref{appA}, including the constraints. The constraints must be satisfied by the initial conditions. Once fulfilled, they must hold throughout all the time evolution, and for this reason, they are used as a numerical check in this work.

The dynamical system is fully defined for GR by \eqref{deromegak} and the Appendix \ref{appA}.

\subsection{Quadratic gravity}\label{subsec22}

As previously mentioned, the field equations for quadratic gravity in \eqref{eqdemov} contain terms with fourth-order derivatives. Therefore, in addition to the variables introduced in \eqref{ENV1}, new variables are required. According to \cite{Barrow:2006xb}, they are defined as:
 \begin{align}
 &\Phi_{3 ,0}=\frac{\dot{\phi} _{3 }}{H^2},
 &\Phi_{3,1}=\frac{\ddot{\phi} _{3 }}{H^3}, \nonumber \\&\Sigma_{\pm1}=\frac{\dot{\sigma} _{\pm }}{H^2},
 &\Sigma_{\pm2}=\frac{\ddot{\sigma} _{\pm}}{H^3}\label{vnq1}\,.
 \end{align}
Based on their definition, the dimensionless variables \eqref{vnq1} must satisfy the following set of first-order differential equations:
\begin{align}
    &\dot{\Sigma }_{\pm}=\Sigma_{\pm1}H-\Sigma_{\pm}\,\dot{H}/H,
&\dot{\Sigma }_{\pm1}=\Sigma_{\pm2}H-2\Sigma_{\pm1}\,\dot{H}/H,\nonumber\\
 &\dot{\Phi }_{3}=\Phi_{3 ,0}H-\Phi_{3}\,\dot{H}/H,
        &\dot{\Phi }_{3, 0}=\Phi_{3,1}H-2\Phi_{3 ,0}\,\dot{H}/H\,.\label{senv}
\end{align}
The time evolution of the higher derivatives $\ddot{H}$, $\Phi_{3,1}$, and $\Sigma_{\pm2 }$ for quadratic gravity are derived from the field equations \eqref{eqdemov} and expressed in the ENV formalism \eqref{ENV1}, \eqref{vnq1} with $\alpha\neq 0$ and $\beta\neq 0$. Using the metric \eqref{mett} and the connection defined by \eqref{Gamma_0ab}, \eqref{Gamma_ijk}, the non-trivial components $E_{11}$, $E_{22}$, $E_{33}$, and $E_{31}$ from the field equations \eqref{eqdemov} are presented in the Appendix \ref{appB}. The covariant conservation of the source $\nabla_aT^{ab}=0$, shown in the Appendix \ref{appA}, results in the time evolution of $\Omega_m$, $\eta$, and $r$, which are given in the Appendix \ref{appB}. The constraints $E_{00}$, $E_{01}$, and $E_{03}$ obtained from the field equations \eqref{eqdemov} are also shown in the Appendix \ref{appB}. In this sense, the dynamical system is completely defined in the Appendix \ref{appB} and by the ENV first-order differential equations \eqref{deromegak}, \eqref{senv}. The initial conditions must always satisfy the constraints, and once initially satisfied, they must be maintained throughout the time.

It can be noted that the non-diagonal shear variable $\Phi_3=0$ is crucial for the component $E_{ 31}$ of \eqref{eqdemov} to exhibit non-trivial dynamics for tilt $r \neq 0$ in both GR and QG. In GR, this can be verified by inspection in the dynamical equation $E_{ 31}$ in the Appendix \ref{appA}. Additionally, the initial conditions must satisfy $\Omega_K\neq 0$, $\eta\neq0$, $\eta\neq\pi$, and $\eta\neq\pi/2$, as well as the EoS parameter $w\neq -1 $ for a non-trivial dynamics for the vorticity.  

The curvature invariants $R$, $R_{ab}R^{ab}$, and $R_{abcd}R^{abcd}$, expressed in terms of the ENV \eqref{ENV1} and \eqref{vnq1}, are given in the Appendix \ref{appc}.

\section{Fixed-points and their stability}\label{sec3}

The stability of some selected solutions for Einstein-Hilbert GR and QG models was analyzed in \cite{deMedeiros:2024pmc}. In this section, we extend these results with a well-known analysis of the stability of the Minkowski exact solution \cite{PhysRevD.32.379,MULLER:2014jaa}, and the RR asymptotic solution, both in QG.

The solutions of the field equations correspond to fixed points of the dynamical system. As shown in \cite{deMedeiros:2024pmc}, the dynamical equations are linearized near the static solutions, and the eigenvalue for each fixed point is found. Evolution to the past or the future near the solutions is discussed.

\subsection{Einstein-Hilbert gravity}\label{sec31}

The FLRW solution without spatial curvature in GR is an exact solution
\begin{align}
 &H =\frac{2}{3\,t(1+w)}, & \Omega_{m}= 1\,,\label{FLRW}
\end{align}
with zero tilt $r=0$ and all other variables being zero. For further details, see the textbooks \cite{Weinberg:1972kfs,Stephanibook}.

The Milne universe is an exact solution for both GR and QG, also with zero tilt:
\begin{align}
    &H=1/t, & \Omega_{K}= 1,\nonumber\\&\eta=0, &r=0\,,\label{Milne}
\end{align}
and the other variables are null. The Milne solution corresponds to Minkowski space, where the Riemann tensor vanishes \cite{Mukhanov:2005sc}.

The exact vacuum solution of Kasner is an exact solution with zero tilt for both GR and QG theories
\cite{Barrow:2006xb,Toporensky:2016kss}:
\begin{align}
    &H={1}/{3t}\,,\label{kasner}
\end{align}
where $\Sigma_+$ and $\Sigma_-$ are constants that must satisfy the $E_{00}$ field equation
\begin{align}  
&\Sigma_{+}^2+\Sigma_{-}^2=1\,.\label{kasner1}
\end{align}
This constraint connects the Kasner circle with the parameter $\phi$, which is given in radians as: 
\begin{align}    
      &\Sigma_{+} = \cos{\phi}, &\Sigma_{-}= \sin{\phi}\,.
\end{align}
The other variables are null. 

The Ref. \cite{deMedeiros:2024pmc} address the fixed points of FLRW, Milne, and Kasner for GR in terms of the dynamic time $\tau$, instead of the proper time $t$, with $dt=d\tau/H(\tau)$ \cite{Coley:2008zz,Barrow:2006xb,Toporensky:2016kss}. The dynamical equations are linearized near the static solutions, and the eigenvalues for each fixed point are found. It is shown in \cite{deMedeiros:2024pmc} that the Kasner orbit is an attractor toward the past singularity. Meanwhile, the FLRW and Milne orbits are attractors to the future for the EoS parameter $-1<w<-1/3$ and $-1/3<w < 1$, respectively, with $p=w\rho$. According to the eigenvalues of the Milne solution shown in \cite{deMedeiros:2024pmc}, the non-perfect source has an increase in its tilt variable for the EoS parameter in the ultra-radiative regime of $1/3<w<1$. This result is in good agreement with \cite{Coley_2005,Coley_2006,Coley:2008zz,Krishnan:2022qbv,Krishnan:2022uar}.

\subsection{Quadratic gravity}\label{sec32}
\begin{figure*}[htpb]
      \begin{centering}
     \begin{tabular}{c c}
            \resizebox{\imsize}{!}{\includegraphics[width=0.2\textwidth]{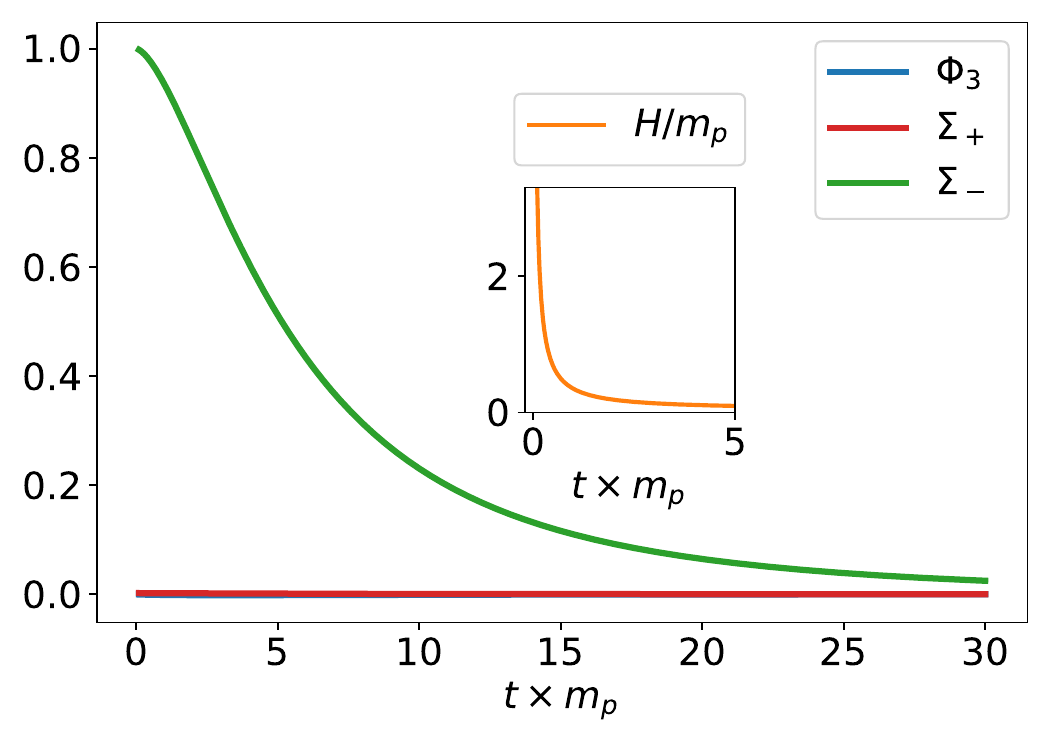}} &
            \resizebox{\imsize}{!}{\includegraphics[width=0.2\textwidth]{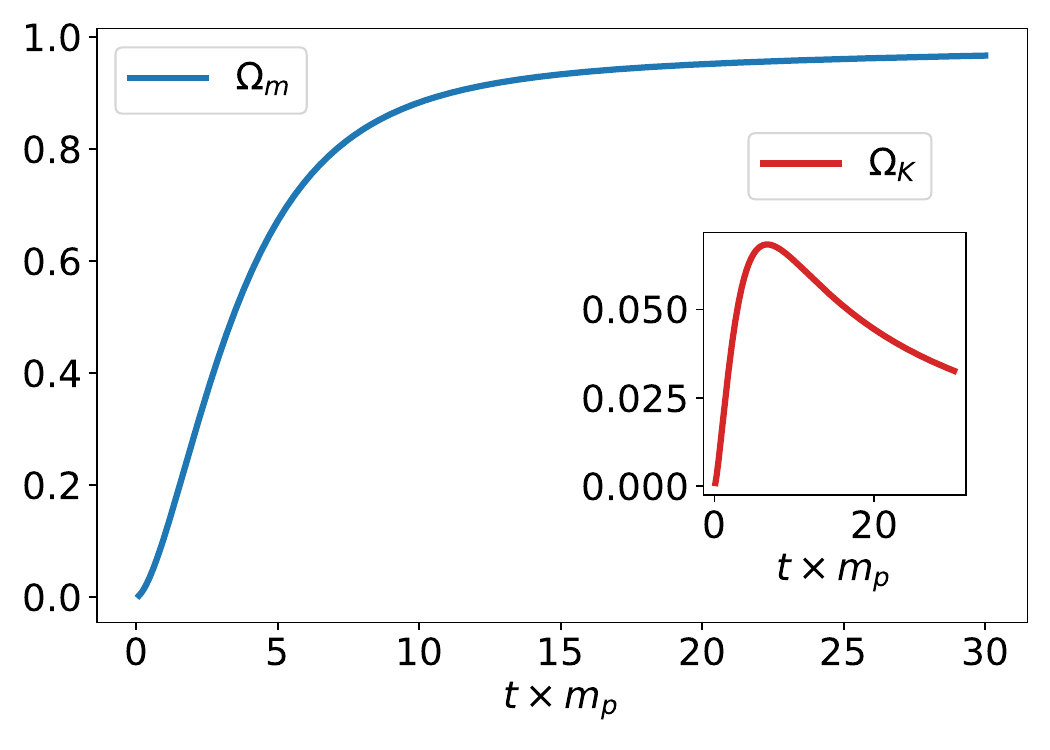}} \\
           \\
        (a) & (b)\\
         \resizebox{\imsize}{!}{\includegraphics[width=0.2\textwidth]{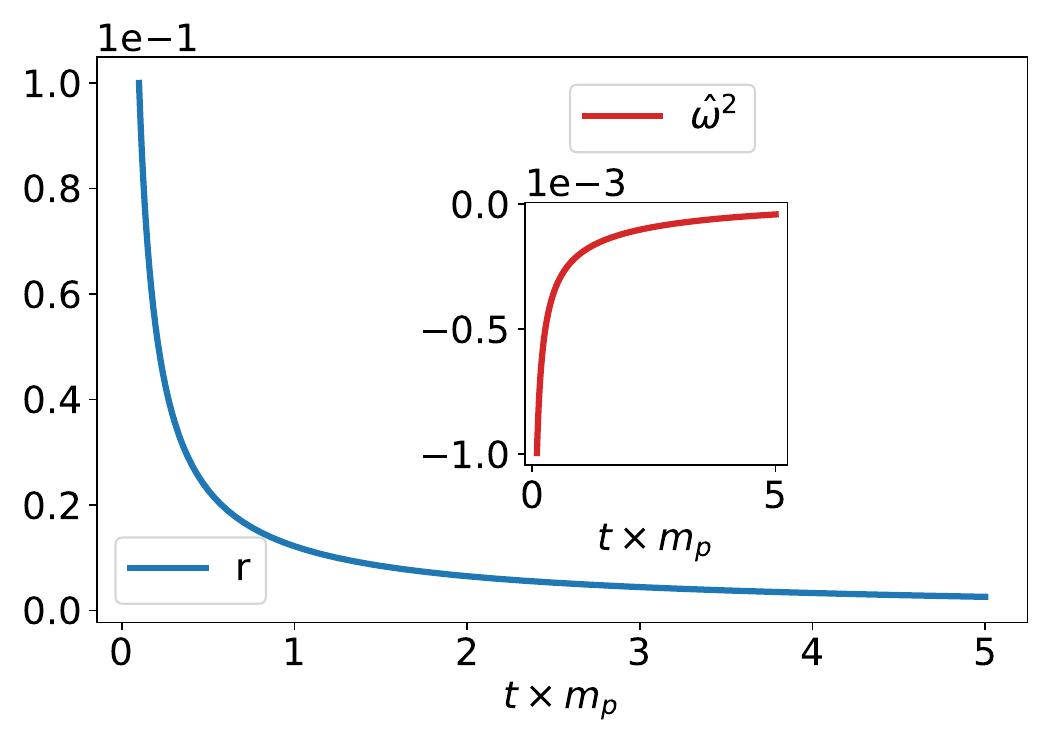}} &
            \resizebox{\imsize}{!}{\includegraphics[width=0.2\textwidth]{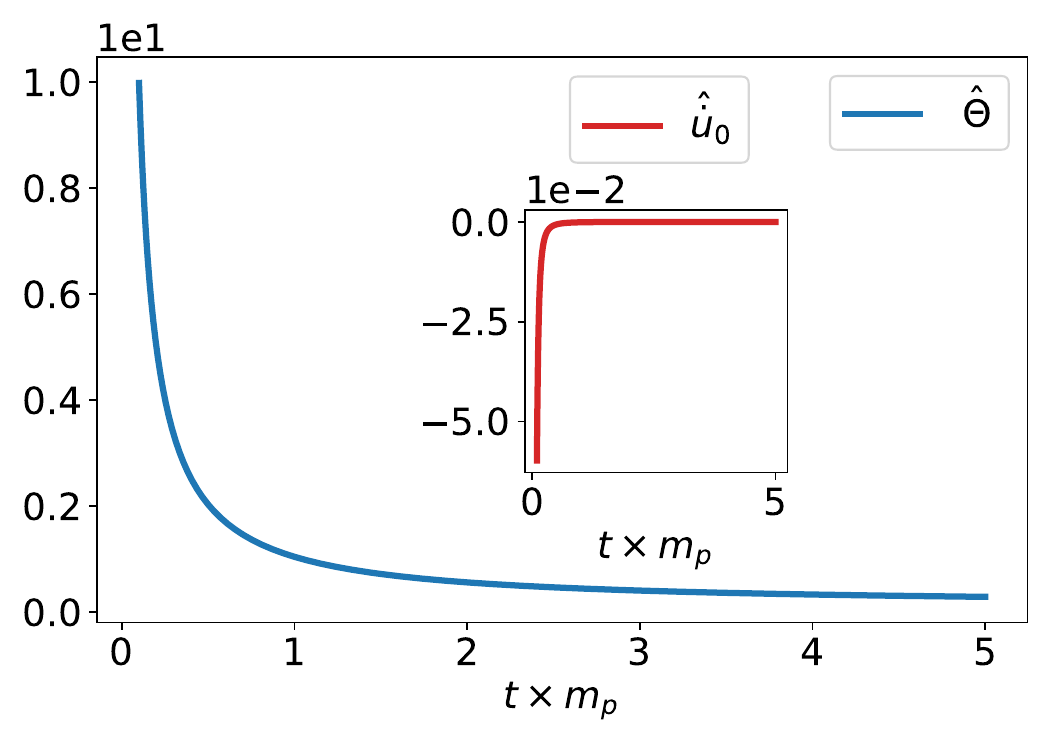}} \\
           \\
        (c) & (d)\\
           \end{tabular}
  \par\end{centering}\caption{The dynamic time evolution to the future for Einstein-Hilbert GR with an initial condition near the Kasner exact solution, Eqs. \eqref{kasner}-\eqref{kasner1}, is addressed. The initial conditions are given by $H=3.33333\,m_p$, $\Omega_K=9.0\times 10^{-4}$, $\Phi_3=-4.02001\times 10^{-4}$, $\Sigma_+=2.00330\times 10^{-3}$, $\Sigma_-=9.98040\times 10^{-1}$, and a small amount of tilted matter $\Omega_m=3.0\times 10^{-3}$ with $r=1.0\times 10^{-1}$ and $\eta=1.0\times 10^{-1}$. The EoS parameter $w$, defined by $p=w\rho$, is chosen as $w=-0.6$, and the Kasner parameter is set to $\phi=1.5$. The solution is attracted to the FLRW orbit, as described by Eq. \eqref{FLRW}. a) The graph plots the isotropization of the shear components $\Phi_3$, $\Sigma_+$, and $\Sigma_-$ in blue, red, and green, respectively. In the inset, it is shown in orange the Hubble parameter $H$, which approaches zero toward the universe expansion. b) The graph in blue shows the matter density $\Omega_m$, which approaches $1$, while the plot in the red inset shows the curvature density $\Omega_K$, which tends to zero, also in agreement with Eq. \eqref{FLRW}. c) The graph shows the decrease of the tilt variable $r$ and the vorticity, both approaching zero, with $r$ plotted in blue and the vorticity in red inset. d) The blue graph shows the decrease of the matter expansion $\hat{\Theta}$, while the red inset displays one of the matter components of the acceleration, $\hat{\dot{u}}_0$, which also decreases to zero}\label{f1}
\end{figure*}

\begin{figure*}[htpb]
      \begin{centering}
     \begin{tabular}{c c}
            \resizebox{\imsize}{!}{\includegraphics[width=0.2\textwidth]{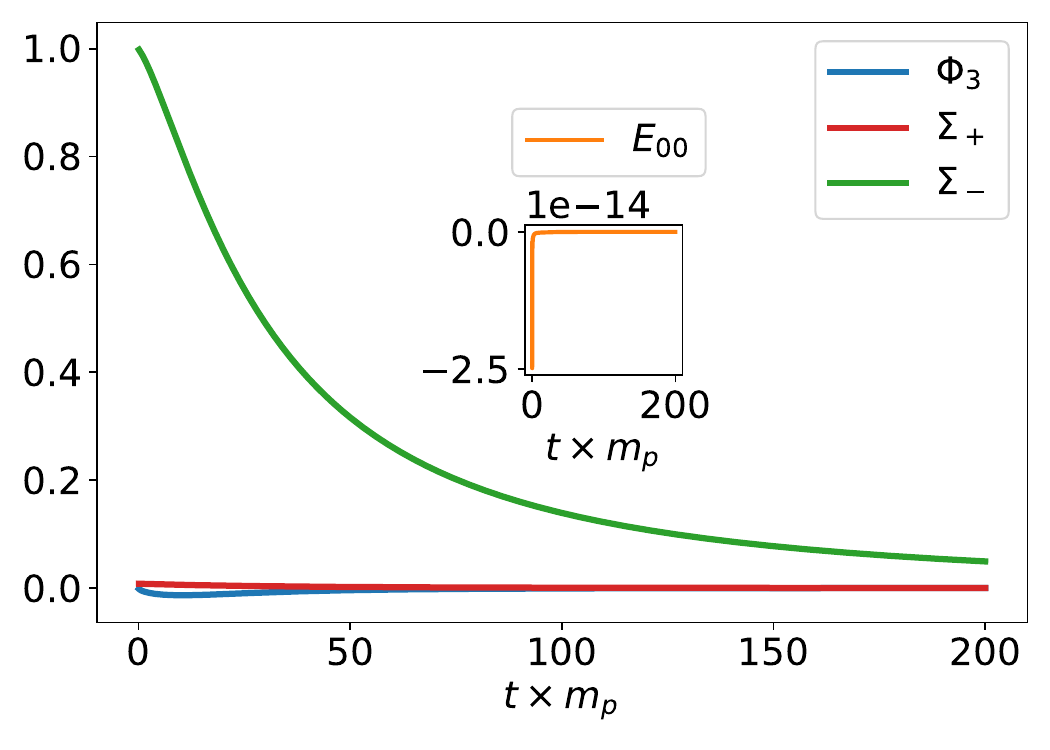}} &
            \resizebox{\imsize}{!}{\includegraphics[width=0.2\textwidth]{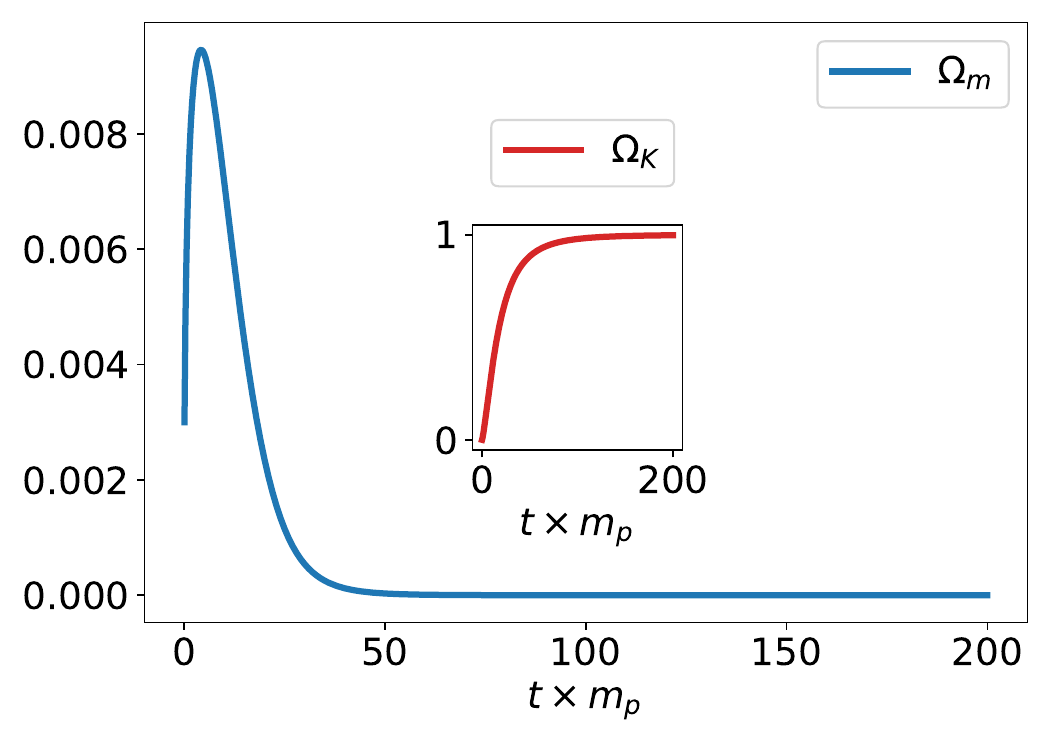}} \\
           \\
        (a) & (b)\\
         \resizebox{\imsize}{!}{\includegraphics[width=0.2\textwidth]{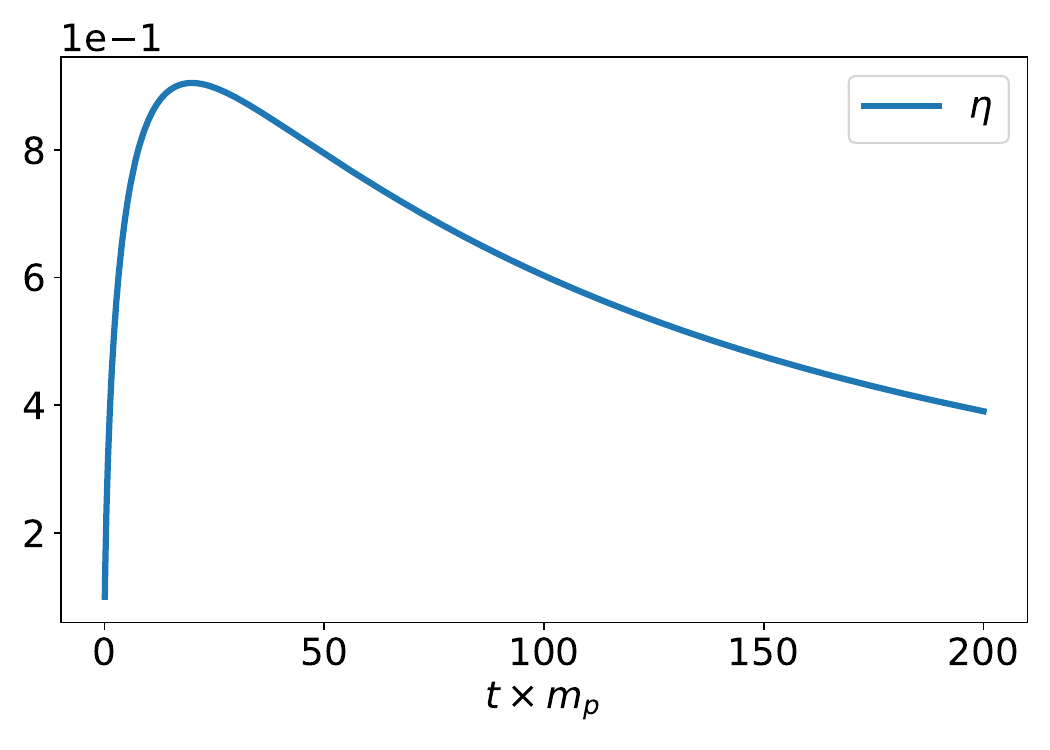}} &
            \resizebox{\imsize}{!}{\includegraphics[width=0.2\textwidth]{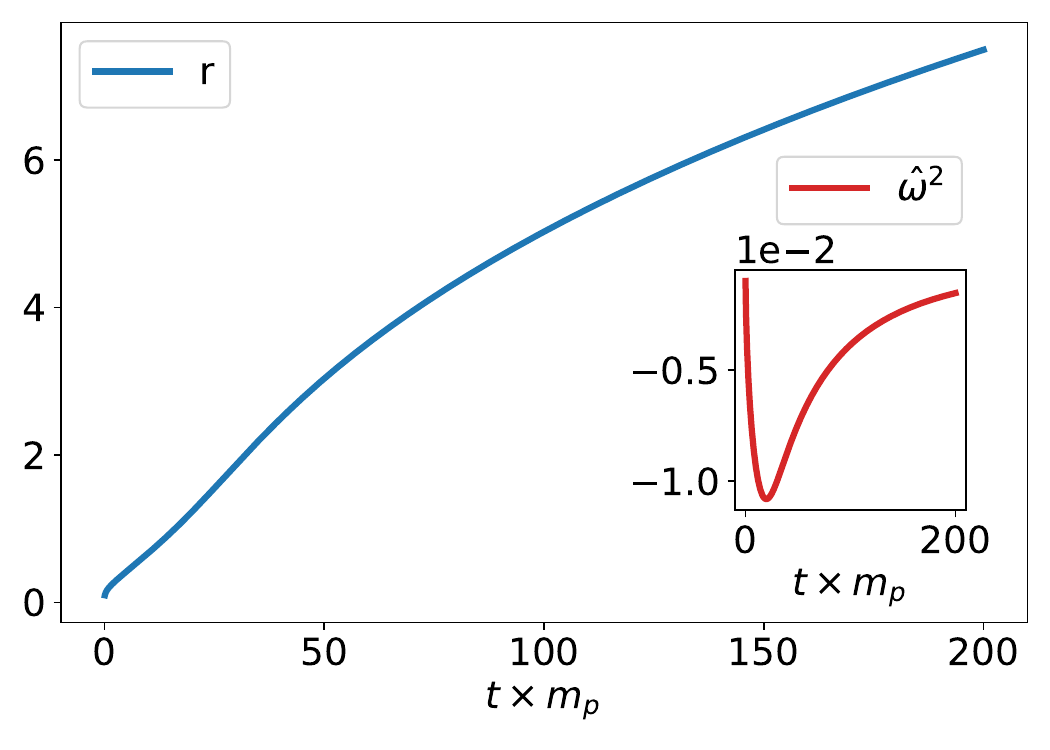}} \\
           \\
        (c) & (d)\\
        \resizebox{\imsize}{!}{\includegraphics[width=0.2\textwidth]{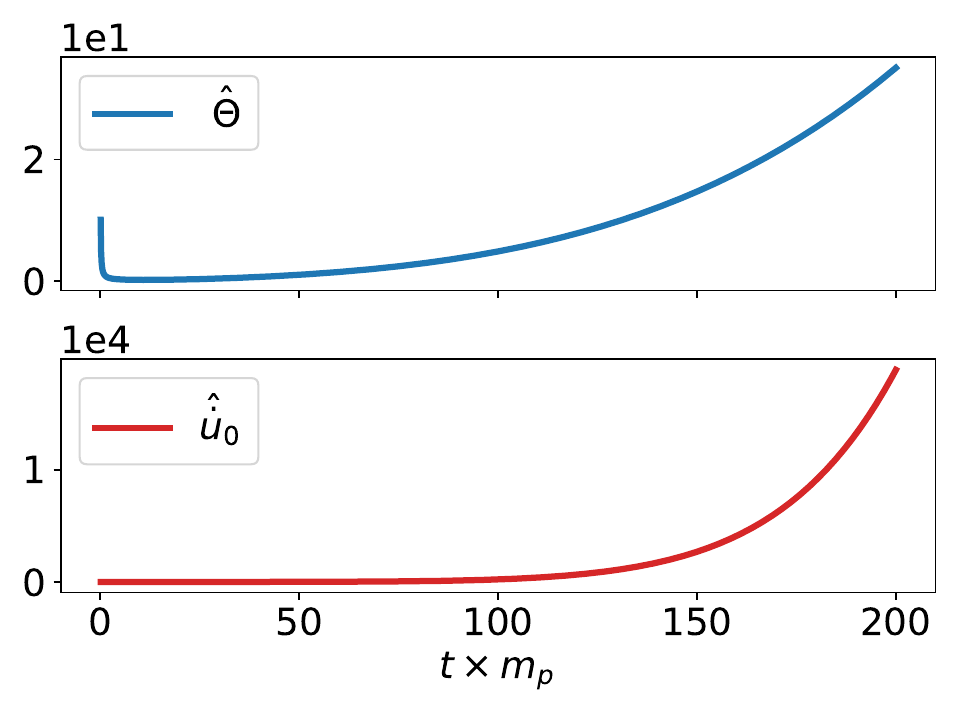}} &
        \resizebox{\imsize}{!}{\includegraphics[width=0.2\textwidth]{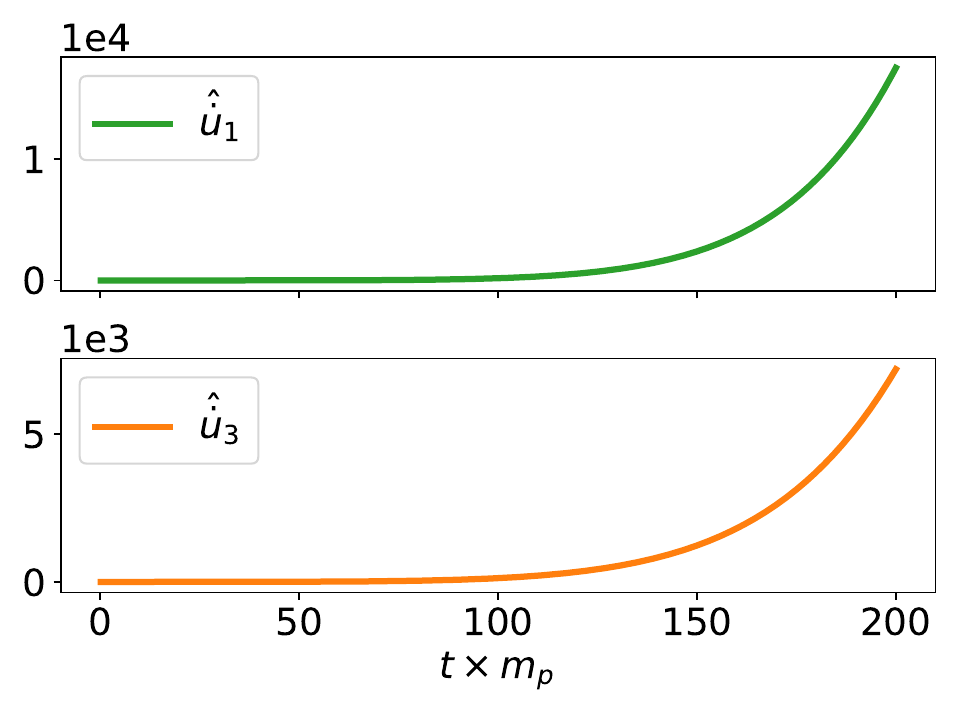}}\\
        (e) & (f) \\
           \end{tabular}
  \par\end{centering}\caption{The future dynamics evolution for GR, with initial conditions near the Kasner solution $H=3.33333\,m_p$, $\Omega_K=9.0\times 10^{-4}$, $\Phi_3=-1.60800\times 10^{-3}$, $\Sigma_+=8.01321\times 10^{-3}$, $\Sigma_-=9.97991\times 10^{-1}$, and a small amount of tilted matter $\Omega_m=3.0\times 10^{-3}$, $r=1.0\times 10^{-1}$, and $\eta=1.0\times 10^{-1}$, with the EoS parameter $w=0.6$ and Kasner parameter $\phi=1.5$, shows that the solution is attracted to the Milne orbit. a) The isotropization of the shear variables $\Phi_3$, $\Sigma_+$, and $\Sigma_-$ is illustrated in blue, red, and green, respectively. The plot inset in orange shows the constraint $E_{00}$ with fluctuations smaller than $10^{-13}$. b) The matter density is displayed in blue and approaches zero, while in red in the inset shows the curvature density $\Omega_K$, which tends to 1, in accordance with Eq. \eqref{Milne}. c) The evolution of the tilt direction $\eta$ is shown in blue, with an initial increase followed by a decrease. d) The plot in blue shows the increase in the tilt variable $r$. The inset in red shows the increase of the vorticity $\hat{\omega}^2$ in absolute value, which initially increases along with the tilt $r$ in the transient regime and then decreases to zero as the solution is attracted to the Milne universe. e) The matter expansion $\hat{\Theta}$ plotted in blue shows it decreasing followed by an increase during the transient regime. After this regime, $\hat{\Theta}$ continues to increase and diverges. While the matter component of acceleration $\hat{\dot{u}}_0$ increase followed by a divergence is displayed in red. f) The remaining matter acceleration components $\hat{\dot{u}}_1$ and $\hat{\dot{u}}_3$, which also increase indefinitely, are shown in green and in orange, respectively
  }\label{f2}
\end{figure*} 

The isotropic singularity is a non-tilted vacuum asymptotic $t\to -\infty$ solution for QG theory. The non-null variables are 
\begin{align}
    &H=1/2t, &\dot{H}=-1/2t^2,\nonumber\\ &\ddot{H}=1/t^3\,.
\end{align}
This solution is also called false radiation because it is a vacuum solution for QG. 

The linear analysis done in \cite{deMedeiros:2024pmc} shows that both the isotropic singularity and Kasner solutions are past singularity attractors for QG. It is also shown in \cite{deMedeiros:2024pmc} that the Kasner singularity is a past attractor with zero vorticity, while the isotropic singularity attractor may have divergent vorticity. 

The Minkowski solution is an exact solution for QG with zero tilt and all the other variables null. The well-known eigenvalues for the Minkowski fixed point are: \[\left[\pm\frac{1}{\sqrt{\alpha}}\right]_3,\pm \frac{1}{\sqrt{-6\beta}},\left[0\right]_8\,.\] 
Here the dynamic system is not expressed in terms of the ENV since $H=0$ for the Minkowski fixed point. To avoid the tachyon behavior, the renormalization parameters should be chosen such that $\beta>0$ and $\alpha<0$ \cite{PhysRevD.32.379,MULLER:2014jaa}. The eigenvalue $\pm {1}/{\sqrt{-6\beta}}$ corresponds to the damped oscillations of the Hubble parameter, exhibiting the asymptotic scalaron behavior $e^{i\omega t}$, where $\omega=1/\sqrt{6\beta}$. The eigenvalue $\pm{1}/{\sqrt{\alpha}}$ is associated with the shear variables $\phi_3$, $\sigma_+$, and $\sigma_-$, also showing damped oscillations and characterized by the Ostrogradsky ghost behavior $e^{i\omega t}$, with $\omega=1/\sqrt{-\alpha}$.

In this subsection, we also analyze the stability of the asymptotic Ruzmaikina and Ruzmaikin RR solution for QG \cite{1969JETP...30..372R}. This solution describes the slow-roll regime of the Starobinsky inflationary model \cite{Starobinsky:1980te}. The nonzero variables are
\begin{align}
    &H>\dot{H}, &\dot{H}=-{1}/{36\beta}\,,\label{RR}
\end{align}
with eigenvalues \[-H, \left[ -3H\right]_5,\left[0\right]_2,-3H\left(w+1\right),H\left(3w-1\right),\] \[\left[-\frac{3}{2} H\pm \frac{H}{2\alpha}\sqrt{\alpha^2+96\alpha\beta}\right]_3.\]
An analysis of the eigenvalue shows that the matter density approaches zero for the EoS parameter $w<1$, while for $w>1/3$, the tilt variable increases.

\section{Numeric results}\label{sec4}

The remainder of the article presents numerical results for initial conditions near the exact solution of Kasner for GR. It also addresses both initial conditions near and farther from the RR solution and near the isotropic singularity for QG. A basin plot for different initial conditions of the diagonal shear variables attracted to the RR solution or to the isotropic singularity are shown. This work is going to investigate only the time evolution to the future. The time evolution toward the past singularity was previously discussed in \cite{deMedeiros:2024pmc}. 

As mentioned earlier, the dynamical system for GR is defined in the Appendix \ref{appA} and by equation \eqref{deromegak}, while for QG, it is described by equations \eqref{deromegak} and \eqref{senv} and in the Appendix \ref{appB}, as detailed in Section \ref{subsec22}. The initial condition must satisfy the constraints; for GR, these are discussed in Section \ref{subsec21} and the Appendix \ref{appA}, and for QG, in Section \ref{subsec22} and the Appendix \ref{appB}. Since the constraints are maintained in time with good accuracy, they are used as a numerical check of the results.

\subsection{Einstein-Hilbert gravity}\label{sec41}

This subsection addresses the numerical time evolution to the future of the tilted Bianchi V geometry for the Einstein-Hilbert GR theory. The dynamical system is defined by equation \eqref{deromegak} and in the Appendix \ref{appA} for the ENV defined in \eqref{ENV1}. The variables $\Phi_{3}$, $\Sigma_{+}$, and $\Sigma_{-}$ are fixed by the constraints given in the Appendix \ref{appA}, and the numerical code is check, with fluctuations always smaller than $10^{-13}$.

As pointed out in \cite{deMedeiros:2024pmc}, for the EoS $p=w\rho$ with $-1<w<-1/3$, the FLRW orbit is an attractor. While for $-1/3<w < 1$, the solution is attracted to the Milne orbit with a tilt steadily increasing for $1/3<w < 1$. In this section, we show that for the ultra-radiative regime, $1/3<w < 1$, all the kinematic variables initially increase together with the tilt variable during the transient regime. The kinematic variables, such as the expansion and acceleration, continue to increase and diverge as the solution is attracted to the Milne orbit, while the vorticity initially increases and then decreases to zero.

It can be seen in Fig. \ref{f1}, the initial condition near the Kasner exact solution \eqref{kasner}-\eqref{kasner1}, for the EoS parameter $w=-0.6$. In well agree with the results above mentioned, this solution is attracted to the FLRW orbit, as described by Eq. \eqref{FLRW}. In Fig. \ref{f1}a, the isotropization of the shear variables $\Phi_3$, $\Sigma_+$, and $\Sigma_-$ are shown in blue, red, and green, respectively. The decrease of the Hubble parameter $H$ toward zero is shown in the orange inset of Fig. \ref{f1}a. In Fig. \ref{f1}b the matter density is plotted in blue, approaching $1$, while the red inset shows the curvature density tending to zero. The decrease of the tilt variable $r$ is shown in blue in Fig. \ref{f1}c, where the kinematic variables decrease together with the tilt. In the red inset of Fig. \ref{f1}c, the vorticity is shown to decrease to zero. Finally, Fig. \ref{f1}d shows the graph of the matter expansion in blue, approaching zero $\hat{\Theta}\rightarrow0$. In the inset of Fig. \ref{f1}d, the matter component of the acceleration $\hat{\dot{u}}_0$ is plotted in red, which also tends to zero. The other components of the matter acceleration have the same behavior. In this manner, the non-perfect fluid approaches a perfect fluid that follows a geodesic and is vorticity-free.

Figure \ref{f2} depicts the initial condition near the Kasner solution for the EoS parameter in the ultra-radiative regime chosen as $w=0.6$. This solution is attracted to the Milne orbit; see Eq. \eqref{Milne}. Fig. \ref{f2}a shows the isotropization of the shear variables $\Phi_3$, $\Sigma_+$, and $\Sigma_-$ plotted in blue, red, and green, respectively. The constraint $E_{00}$ is plotted in the red inset of Fig. \ref{f2}a, with fluctuations always smaller than $10^{-13}$. In Fig. \ref{f2}b, the matter density, shown in blue, approaches zero, while the matter curvature approaches $1$, as shown in red in the inset. In Fig. \ref{f2}c, the tilt direction $\eta$ is shown in blue, with an initial increase followed by a decrease. While the increase of the tilt variable $r$ is plotted in blue in Fig. \ref{f2}d. In the red inset of Fig. \ref{f2}d, the initially increasing vorticity in absolute value is shown to occur during the transient regime. After that, the vorticity decreases to zero as the solution is attracted to the Milne orbit. Fig. \ref{f2}e shows in blue the initial decrease followed by an increase during the transient regime of the matter expansion $\hat{\Theta}$ and then its indefinite increase toward the Milne universe. The matter component acceleration $\hat{\dot{u}}_0$ increase followed by a divergence is plotted in red. The remaining matter acceleration components, $\hat{\dot{u}}_1$ and $\hat{\dot{u}}_3$, also increase indefinitely, as shown in green and orange, respectively, in Fig. \ref{f2}f.

In the situation described above, the contribution of the energy-momentum tensor to the field equations becomes zero and decouples from the field equations. In this sense, the tilt effects no longer influence the dynamics dominated by the Milne solution.

\subsection{Quadratic gravity}

In this subsection, we consider the future evolution of the anisotropic Bianchi V model for the QG, using the same tilted source described in GR. The variables fixed by the constraints shown in the Appendix \ref{appB} are $\ddot{H}$, $\Phi_{3,1}$, and $\Sigma_{+2}$. The constraints used to check the numerical code have fluctuations always smaller than $10^{-7}$.

\begin{figure*}[htpb]
      \begin{centering}
     \begin{tabular}{c c}
            \resizebox{\imsize}{!}{\includegraphics[width=0.2\textwidth]{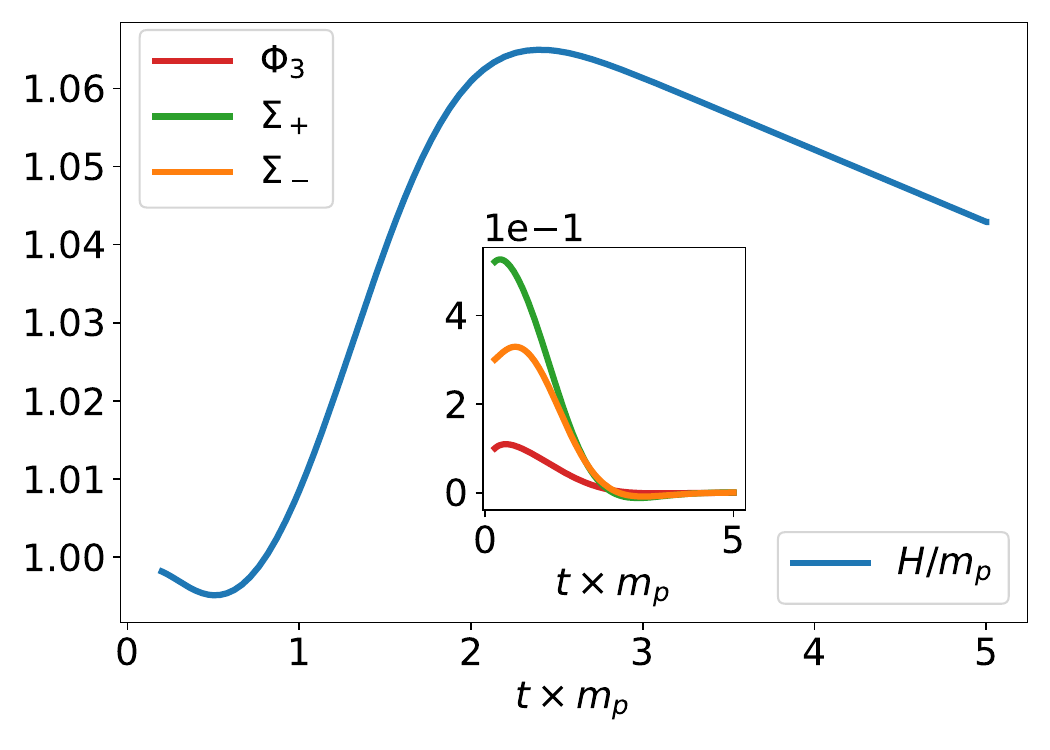}} &
            \resizebox{\imsize}{!}{\includegraphics[width=0.2\textwidth]{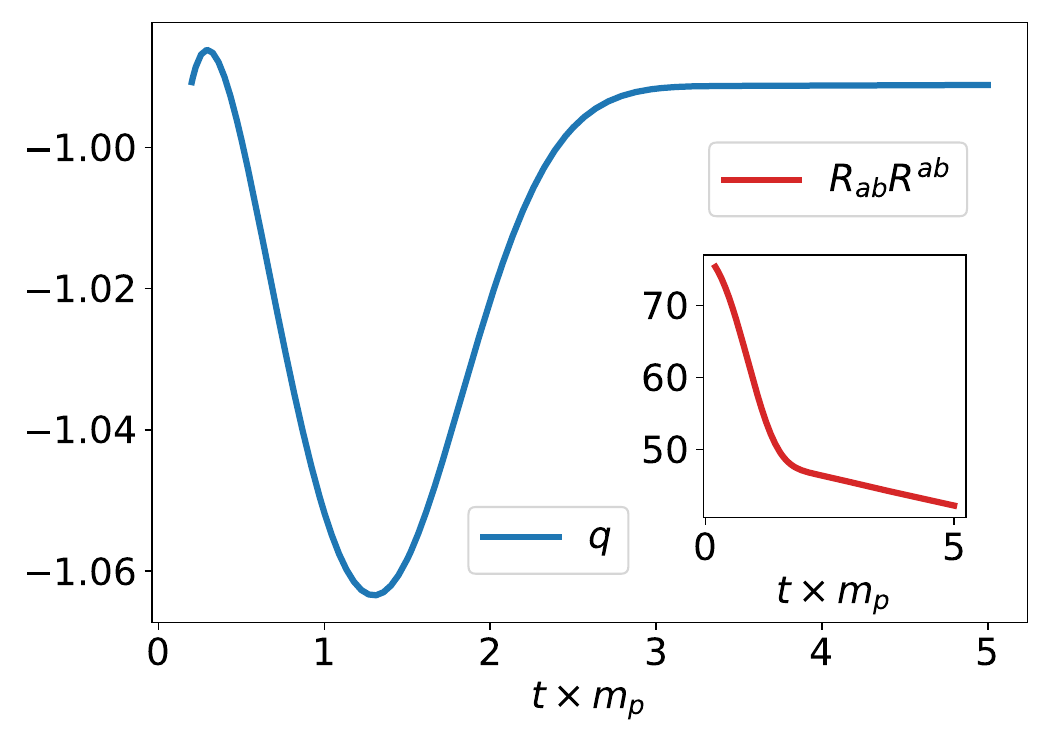}} \\
           \\
        (a) & (b)\\
         \resizebox{\imsize}{!}{\includegraphics[width=0.2\textwidth]{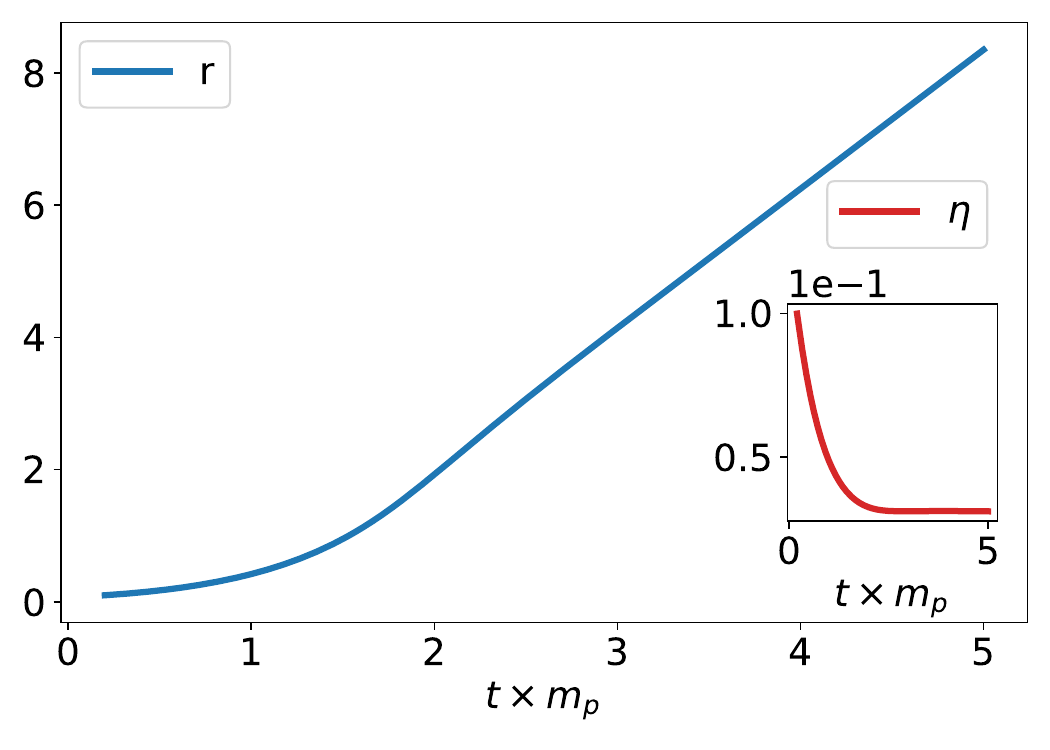}} &
            \resizebox{\imsize}{!}{\includegraphics[width=0.2\textwidth]{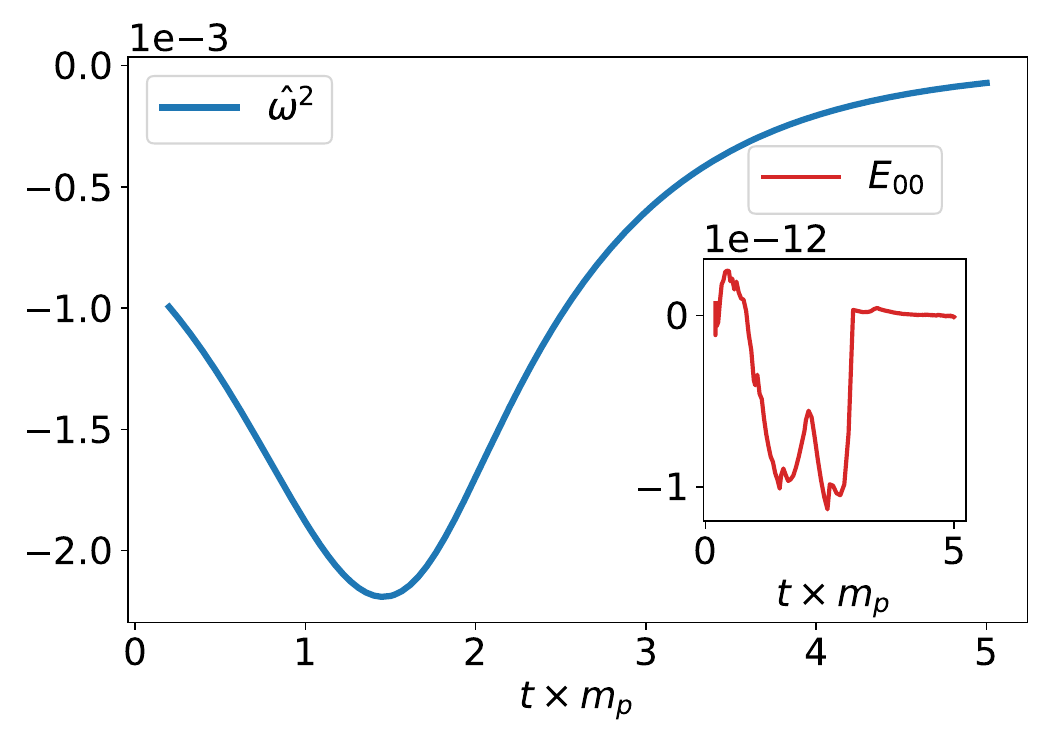}} \\
           \\
        (c) & (d)\\
          \resizebox{\imsize}{!}{\includegraphics[width=0.2\textwidth]{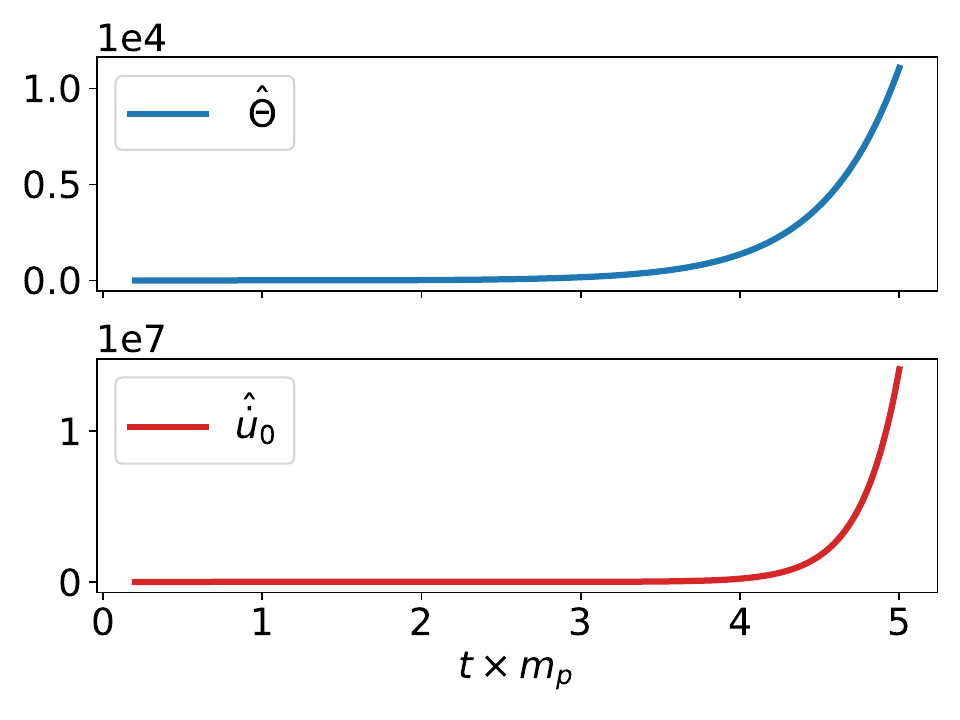}} &
            \resizebox{\imsize}{!}{\includegraphics[width=0.2\textwidth]{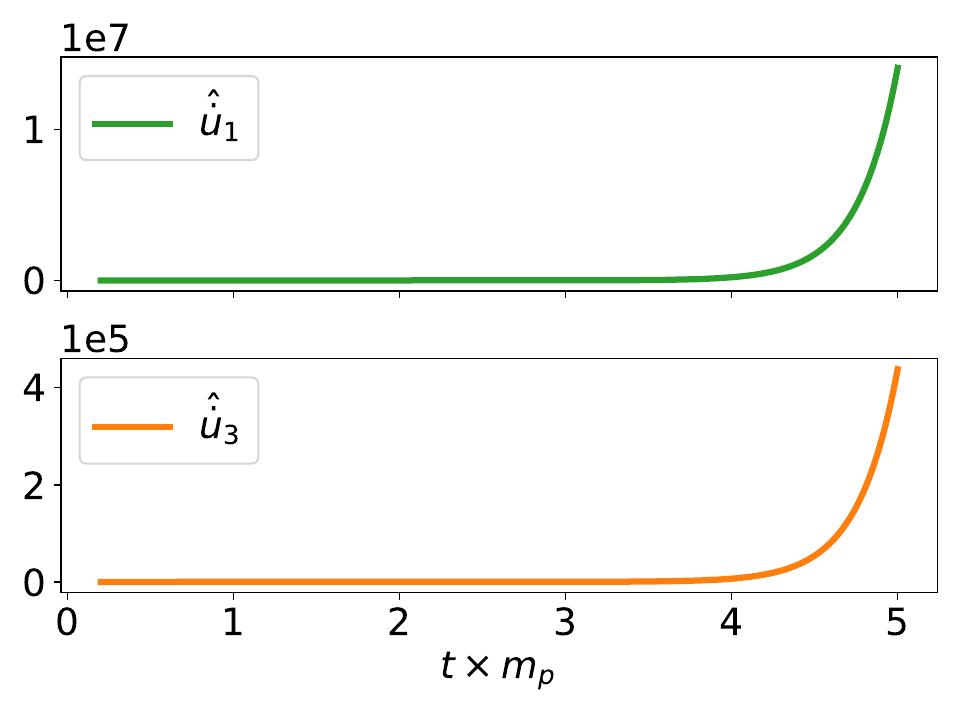}}  \\
        (e) & (f)\\
           \end{tabular}
  \par\end{centering}\caption{QG shows the solution with the initial condition: $H=1.0\,m_p$, $\dot{H}=-9.25926\times 10^{-3}\,m_p^2$, $\ddot{H}=-8.18805\times 10^{-2}\, m_p^3$, $\Omega_K=1.0\times 10^{-2}$, $\Phi_3=1.0\times 10^{-1}$, $\Phi_{3,0}=1.0\times 10^{-1}$, $\Phi_{3,1}=-7.33847\times 10^{-1}$, $\Sigma_+=5.0\times 10^{-1}$, $\Sigma_{+1}=1.0\times 10^{-1}$, $\Sigma_{+2}=-1.26245$, $\Sigma_-=3.0\times 10^{-1}$, $\Sigma_{-1}=1.0\times 10^{-1}$, $\Sigma_{-2}=1.0\times 10^{-1}$, near the RR asymptotic solution \cite{1969JETP...30..372R} with a small amount of tilted matter $\Omega_m=3.33333\times 10^{-2}$, $r=1.0\times 10^{-1}$ and $\eta=1.0\times 10^{-1}$. The numeric evolution is to the future, and the solution is attracted to the Ruzmaikina's regime. The EoS parameter is chosen as $w=0.6$, and the renormalization parameters are $\alpha=-30\, m_p^{-2}$ and $\beta=3 \,m_p^{-2}$. a) The blue plot shows the transient behavior of the Hubble parameter $H$, followed by the slow-roll inflationary regime. The inset graph shows the isotropization of the shear variables: $\Phi_3$ in red, $\Sigma_+$ in green, and $\Sigma_-$ in orange. b) The deceleration parameter $q$ is shown in blue, approaching $-1$, as expected. In the inset, the curvature scalar $R_{ab}R^{ab}$ is plotted in red, tending to zero as the universe expands. c) The blue plot shows the increase in the tilt variable $r$, while the inset in red shows the decrease of the tilt direction $\eta$. d) The blue graph shows the initial increase in the absolute value of the vorticity during the transient regime and its decrease to zero during the onset of the slow-roll inflationary regime. The inset in red displays the numerical checks for the constraint $E_{00}$, with fluctuations smaller than $10^{-11}$. e) The increase, followed by divergence, in the matter expansion $\hat{\Theta}$ and in the matter acceleration component $\hat{\dot{u}}_0$ are shown in blue and in red, respectively. f) The remaining matter acceleration components, $\hat{\dot{u}}_1$ and $\hat{\dot{u}}_3$, are plotted in green and in orange, respectively}\label{f3}
\end{figure*} 

 \begin{figure*}[htpb]
      \begin{centering}
     \begin{tabular}{c c}
            \resizebox{\imsize}{!}{\includegraphics[width=0.2\textwidth]{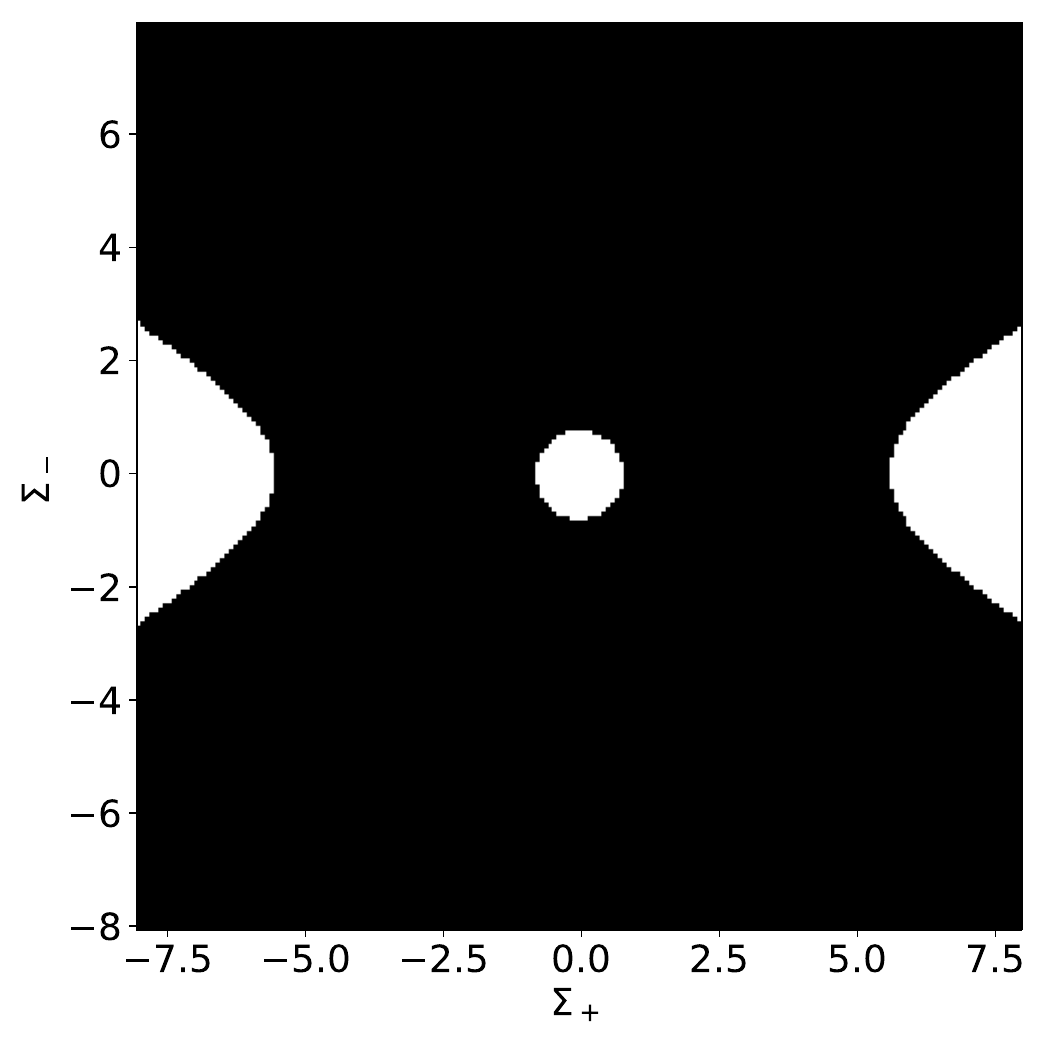}} &
            \resizebox{\imsize}{!}{\includegraphics[width=0.2\textwidth]{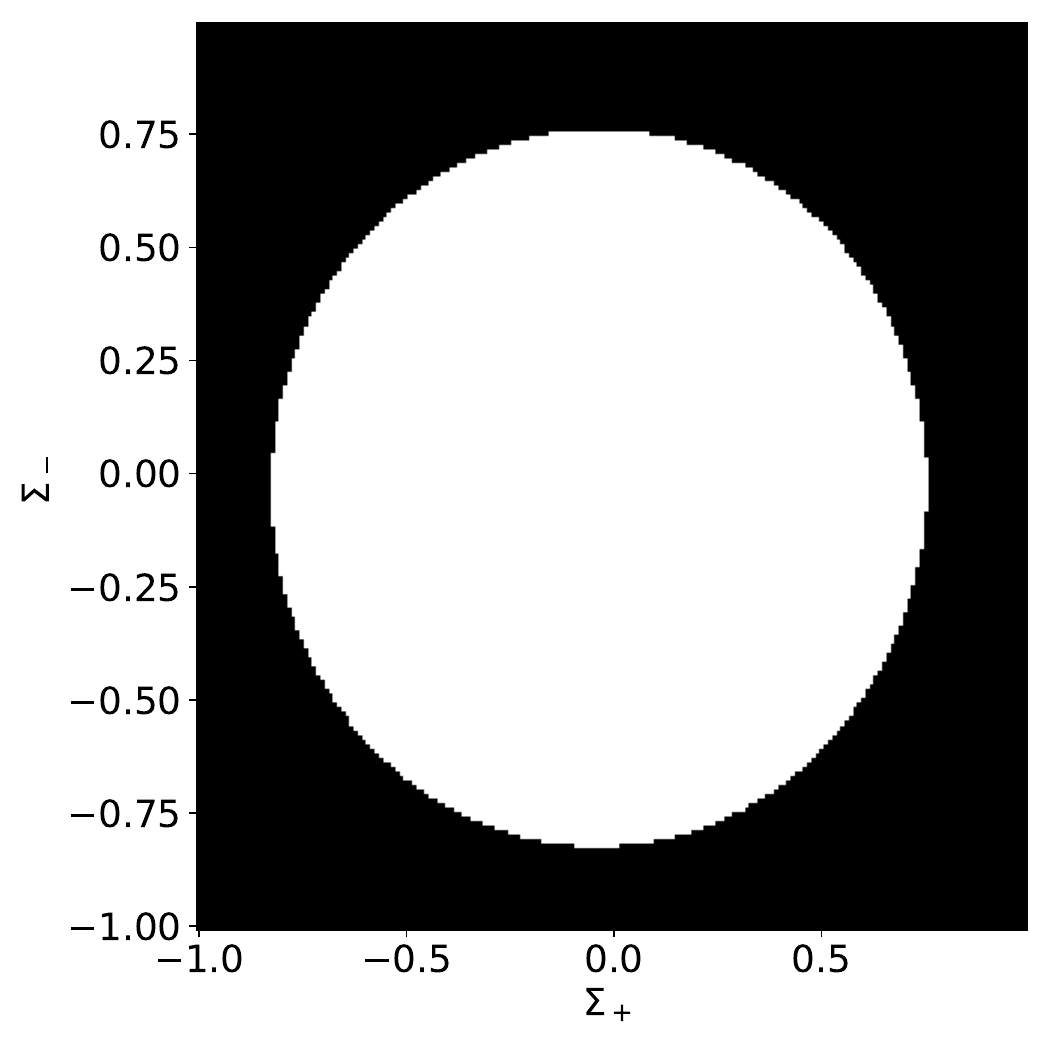}} \\
           \\
        (a) & (b)\\
        \resizebox{\imsize}{!}{\includegraphics[width=0.2\textwidth]{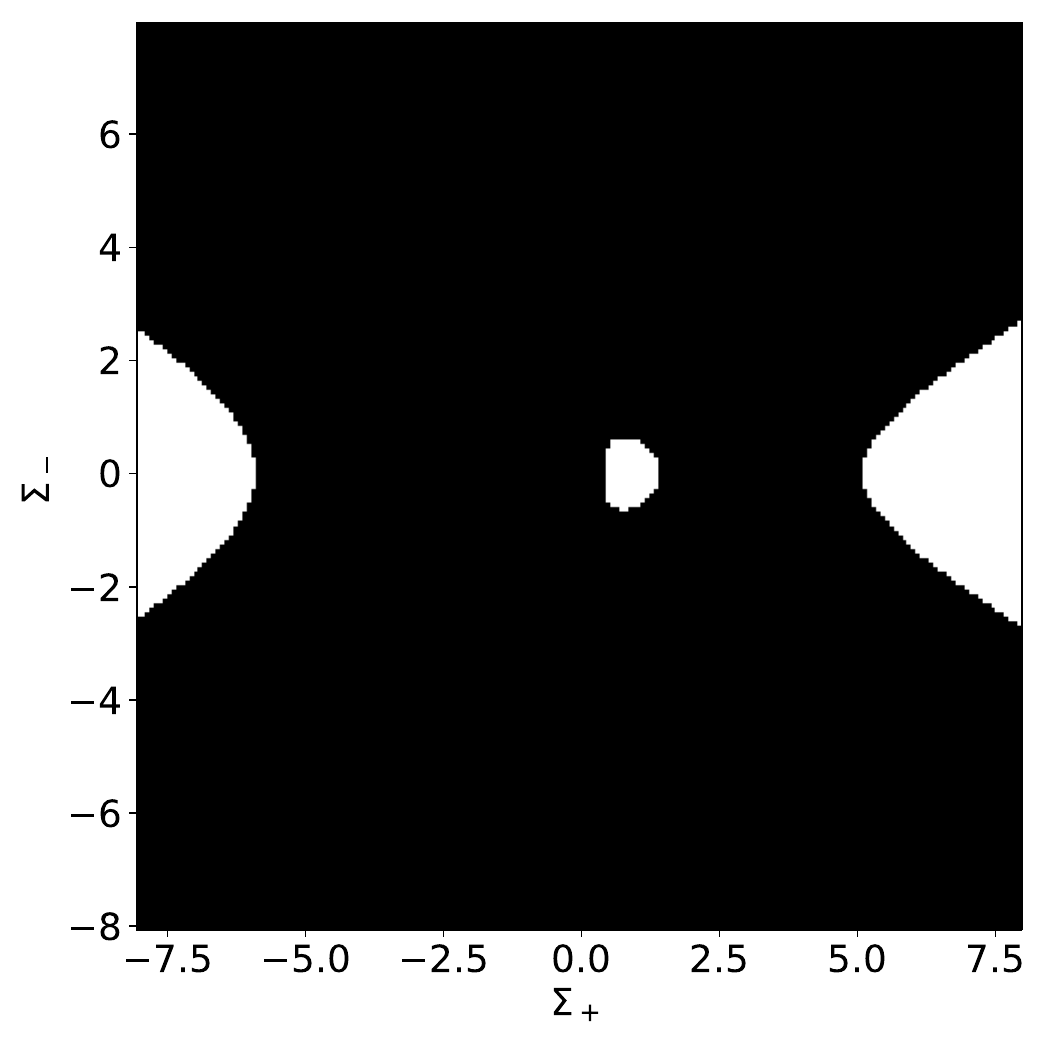}} &
            \resizebox{\imsize}{!}{\includegraphics[width=0.2\textwidth]{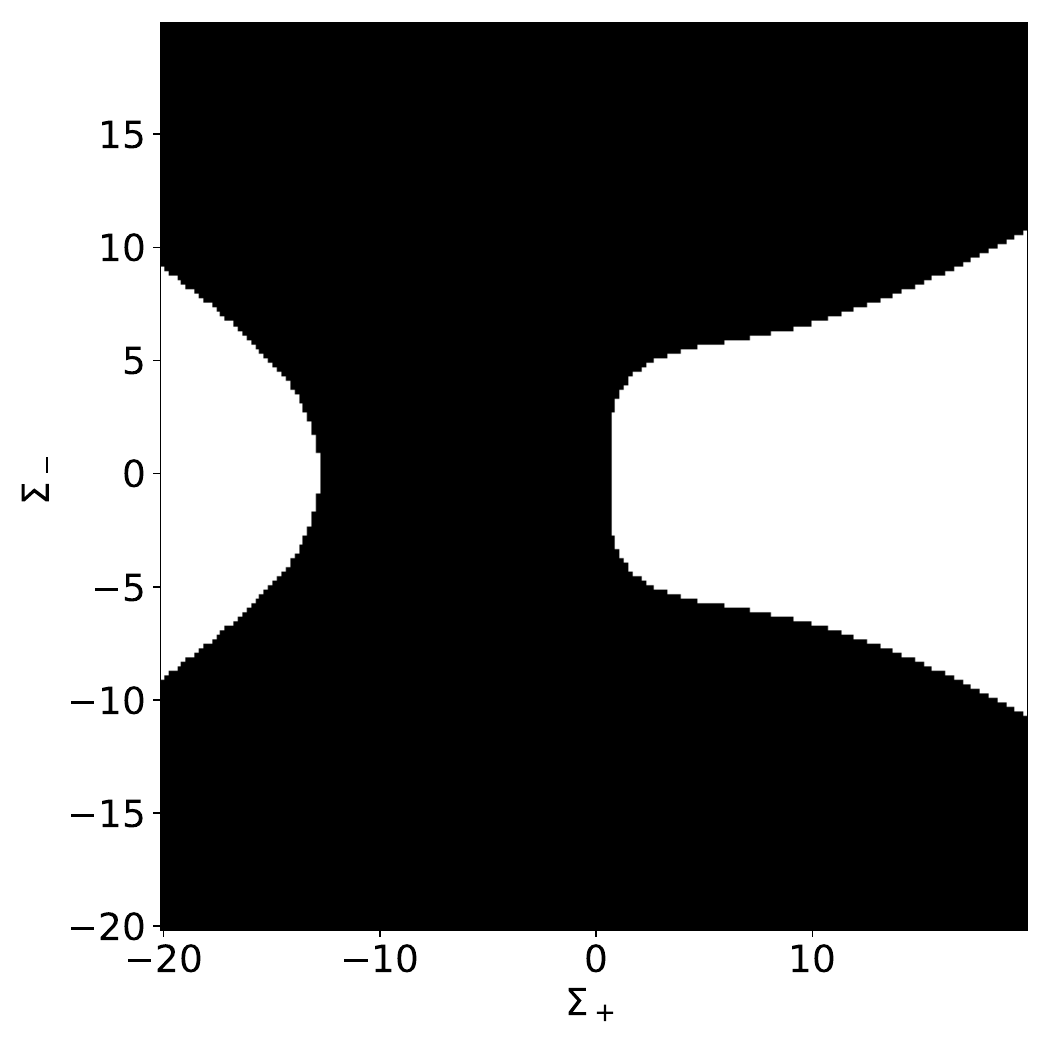}} \\
           \\
        (c) & (d)\\
           \end{tabular}
  \par\end{centering}\caption{A basin plot is shown for different initial conditions of the diagonal shear components $\Sigma_+$ and $ \Sigma_-$, with the white points showing initial conditions attracted to the RR orbit, while the black points represent solutions attracted, which evolve to a recollapse and are asymptotically attracted to the isotropic singularity. The dynamical evolution is towards the future for QG. The renormalization parameters are set as $\alpha=-30\, m_p^{-2}$ and $\beta=3 \,m_p^{-2}$ and the EoS parameter $w=1/3$ for the radiation fluid. The initial condition are given by $H=9.98148\times 10^{-1}\,m_p$, $\dot{H}=-9.25926\times 10^{-3}\,m_p^2$, $\Omega_K=1.00371\times 10^{-2}$, $\Phi_3=1.00186\times 10^{-1}$, $\Phi_{3,0}=1.00371\times 10^{-1}$, $\Sigma_{+1}=1.00371\times 10^{-1}$, $\Sigma_{-1}=1.00371\times 10^{-1}$, $\Sigma_{-2}=1.00558\times 10^{-1}$, and a small amount of matter $\Omega_m=3.34571\times 10^{-2}$. a) The graph plots a basin for a non-tilted source with $r=0$ and $\eta=0$. b) The plot shows a zoom near the origin shown in the left panel. c) Here, a small amount of tilted matter with $r=4.8$ and $\eta=1.0\times 10^{-1}$ is considered. The plot shows how the attraction regions for RR shift when the tilt is introduced. The farther regions of Ruzmaikina's attraction are moved to the left, while the region near the origin shifts to the right. d) The plot shows a not small initial condition for the tilt, $r=7.0$ and $\eta=1.0\times 10^{-1}$. In this case, the farther regions of attraction for RR are mostly relocated to the left, leading to a superposition with the attraction region near the origin. The plot is generated considering $200\times200$ points}\label{f4}
\end{figure*} 

 \begin{figure*}[htpb]
      \begin{centering}
     \begin{tabular}{c c}
       \resizebox{\imsize}{!}{\includegraphics[width=0.2\textwidth]{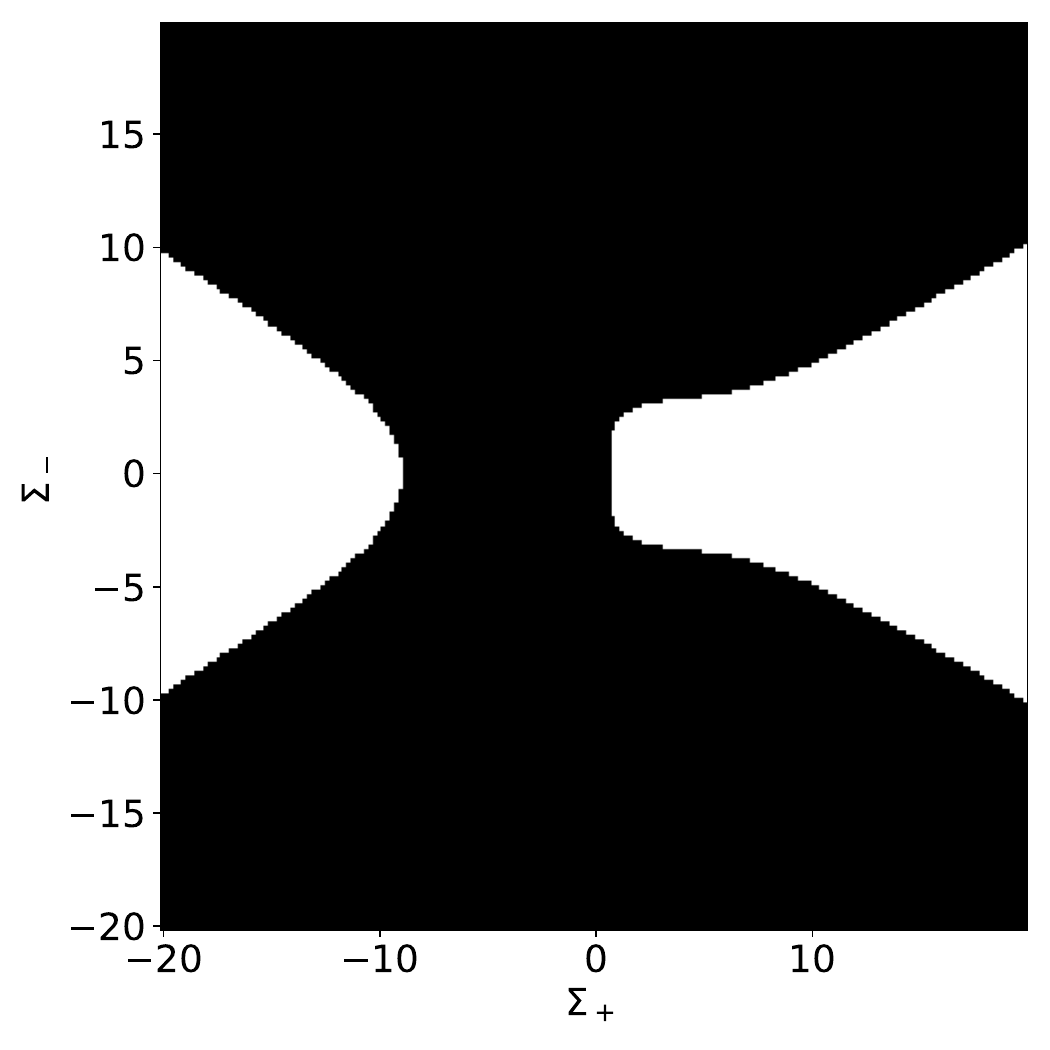}} &
            \resizebox{\imsize}{!}{\includegraphics[width=0.2\textwidth]{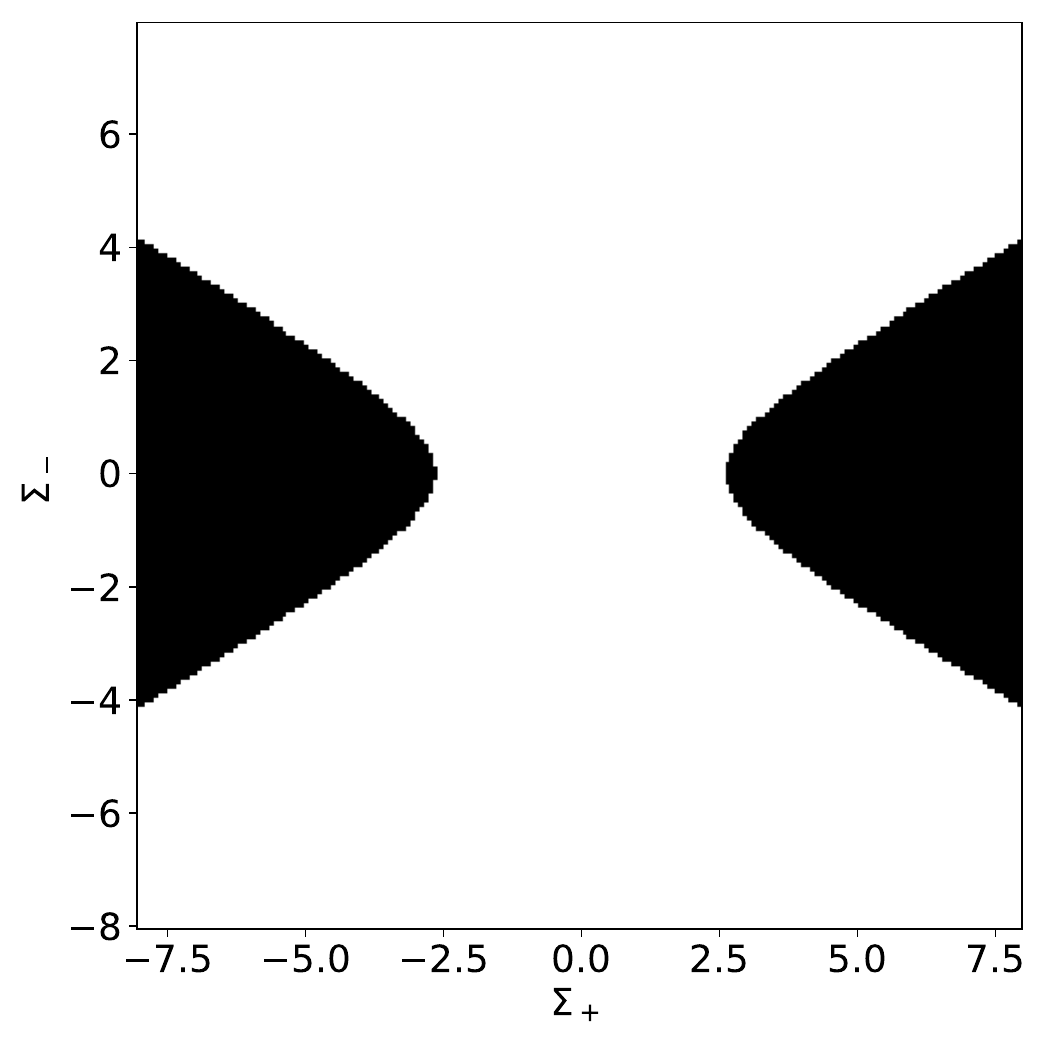}}   \\
            \\
            (a) & (b)
           \end{tabular}
  \par\end{centering}\caption{a) A basin of attraction is plotted for the same initial conditions shown in the previous figure: $H=9.98148\times 10^{-1}\,m_p$, $\dot{H}=-9.25926\times 10^{-3}\,m_p^2$, $\Omega_K=1.00371\times 10^{-2}$, $\Phi_3=1.00186\times 10^{-1}$, $\Phi_{3,0}=1.00371\times 10^{-1}$, $\Sigma_{+1}=1.00371\times 10^{-1}$, $\Sigma_{-1}=1.00371\times 10^{-1}$, $\Sigma_{-2}=1.00558\times 10^{-1}$, but with different values of tilted matter $\Omega_m=6.69143\times 10^{-1}$ for $r=4.8$ and $\eta=1.0\times 10^{-1}$. The time numerical evolution is towards the future for QG, and the EoS parameter is $w=1/3$ for the radiation fluid. The renormalization parameters are $\alpha=-30\, m_p^{-2}$ and $\beta=3 \,m_p^{-2}$. The graph shows that an increase in the tilted matter density $\Omega_m$ behaves similarly to an increase in the tilt variable $r$. b) Now, the plot considers the value for the renormalization parameter $\beta=1.305\times 10^{9}\,m_p^{-2}$, which is inferred by CMBR measurements. The other renormalization parameter is set to $\alpha=-3.0\times 10^{8}\,m_p^{-2}$, with the EoS parameter $w=1/3$ for the radiation source. The initial conditions are given by $H=1.0\, m_p$, $\dot{H}=-2.12857\times 10^{-11}\, m_p^2$, $\Omega_K=1.0\times 10^{-2}$, $\Phi_3=1.0\times 10^{-1}$, $\Phi_{3,0}=1.0\times 10^{-1}$, $\Omega_m=3.33333\times 10^{-2}$, $\Sigma_{+1}=1.0\times 10^{-1}$, $\Sigma_{-1}=1.0\times 10^{-1}$, $\Sigma_{-2}=1.0\times 10^{-1}$ with $r=4.8$ and $\eta=1.0\times 10^{-1}$. The basin to RR is changed when compared to the arbitrary values of the renormalization parameters shown in the previous figure and in the left panel. Plot made considering $200\times200$ points}\label{f5}
\end{figure*}

The eigenvalue analyses shown in Sect. \ref{sec3} for the RR solution show that the tilt increases for the EoS parameter $1/3<w<1$, $p=w \rho$. As a numerical example, Figure \ref{f3} shows the initial condition attracted to the RR solution \eqref{RR}, with EoS parameter $w=0.6$. The renormalization parameters are set to $\alpha=-30\, m_p^{-2}$ and $\beta=3 \,m_p^{-2}$. Fig. \ref{f3}a shows in blue the transient regime of the Hubble parameter $H$ followed by the slow-roll inflationary regime. The inset shows the isotropization of the shear variables $\Phi_3$, $\Sigma_+$, and $\Sigma_-$ plotted in red, grenn, and orange, respectively. The deceleration parameter $q$, which approaches $-1$, is plotted in blue in Fig. \ref{f3}b, as expected. In the red inset of Fig. \ref{f3}b, the decrease and approach to zero of the curvature scalar $R_{ab}R^{ab}$ is shown. The increase of the tilt variable $r$ is shown in blue in Fig. \ref{f3}c, while the red inset shows the tilt direction decreases. Fig. \ref{f3}d displays the evolution of the vorticity, shown in blue, which increases during the transient regime and decreases to zero during the onset of the slow-roll inflationary regime. In the red inset of Fig. \ref{f3}d, the constraint $E_{00}$ is plotted, with fluctuations always smaller than $10^{-11}$. Fig. \ref{f3}e shows the increase and subsequent divergence of the matter expansion $\hat{\Theta}$ and the matter acceleration component $\hat{\dot{u}}_0$ displayed in blue and in red, respectively. The remaining divergence in the non-null matter acceleration components, $\hat{\dot{u}}_1$ and $\hat{\dot{u}}_2$, are plotted in green and orange in Fig. \ref{f3}f, respectively. The solution presents a graceful end for inflation while the kinematic variables continue with the behavior described above. The solution is then attracted to the Minkowski orbit.

In the above case, the fluid becomes non-perfect with the tilt increase, and the matter source does not follow a geodesic, with divergence in its expansion and acceleration components. However, the energy-momentum tensor decreases to zero, and the tilt effects decouple from the field equations dominated by inflation.

\begin{figure*}[htpb]
      \begin{centering}
     \begin{tabular}{c c}
            \resizebox{\imsize}{!}{\includegraphics[width=0.2\textwidth]{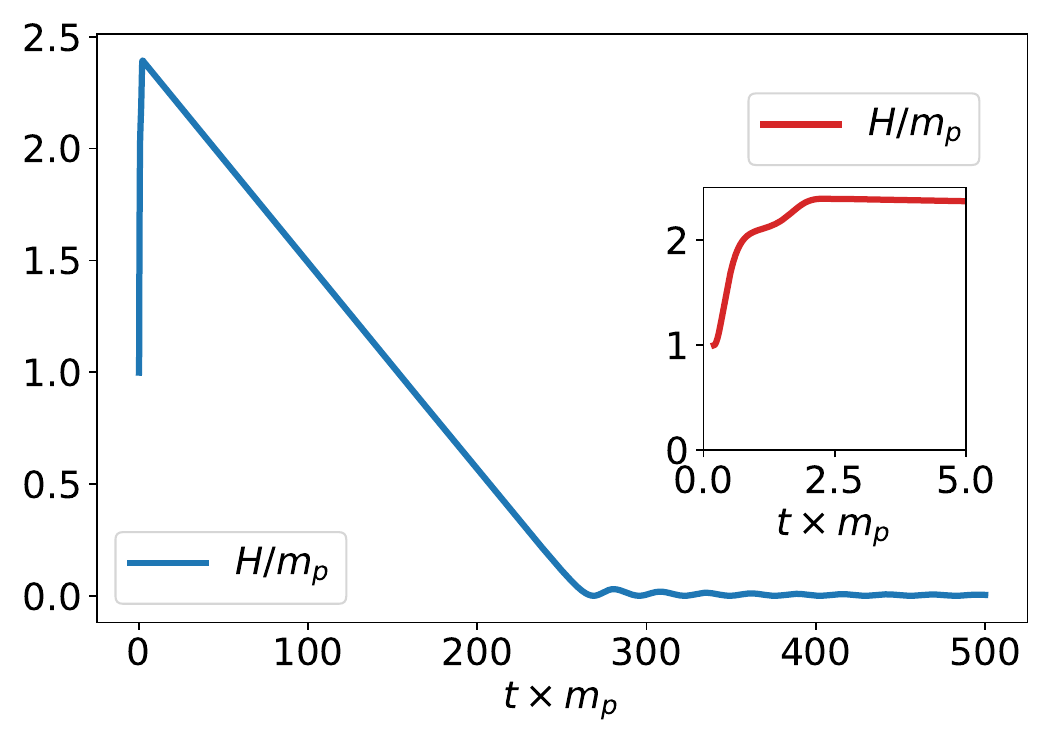}} &
            \resizebox{\imsize}{!}{\includegraphics[width=0.2\textwidth]{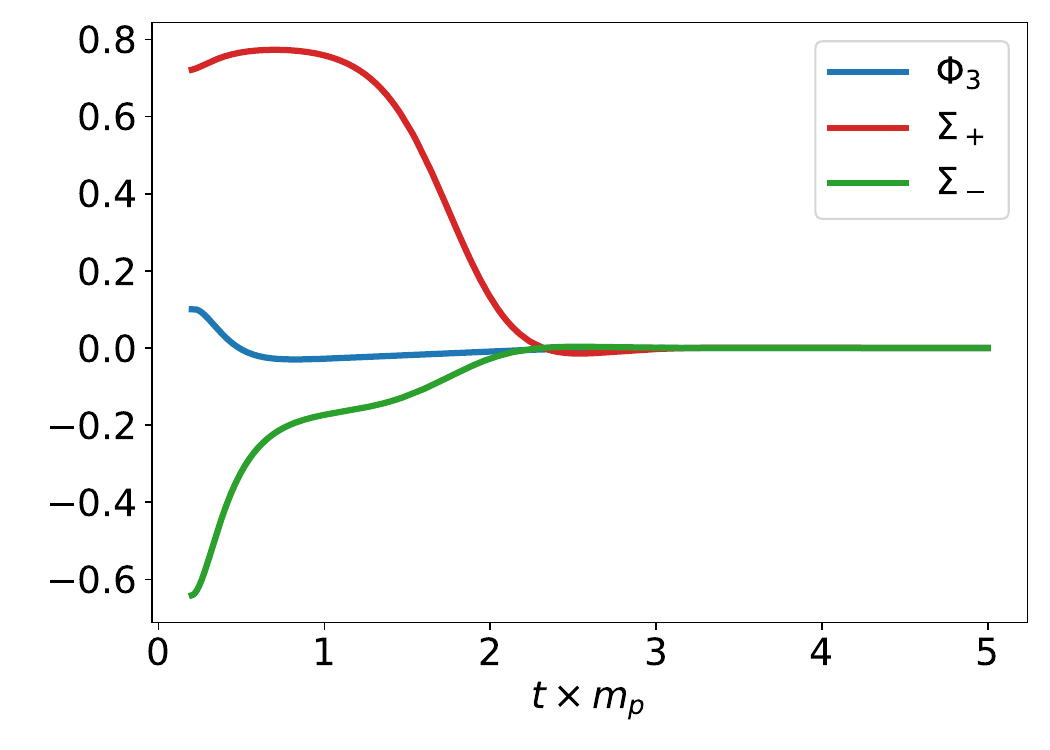}} \\
           \\
        (a) & (b)\\
         \resizebox{\imsize}{!}{\includegraphics[width=0.2\textwidth]{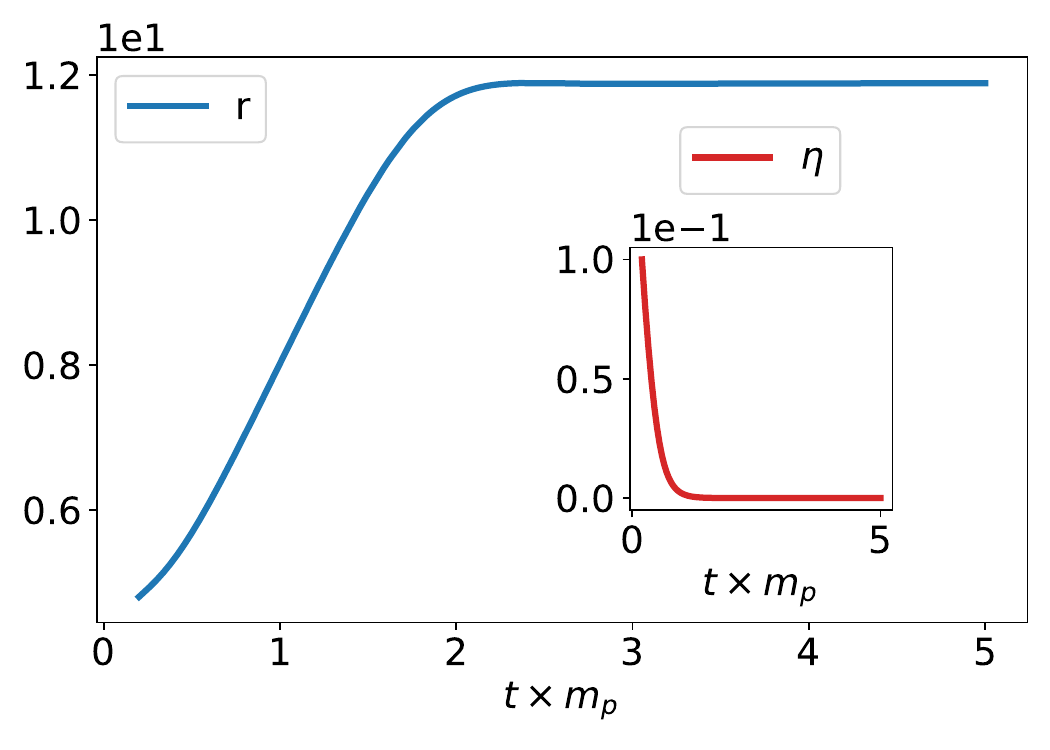}} &
            \resizebox{\imsize}{!}{\includegraphics[width=0.2\textwidth]{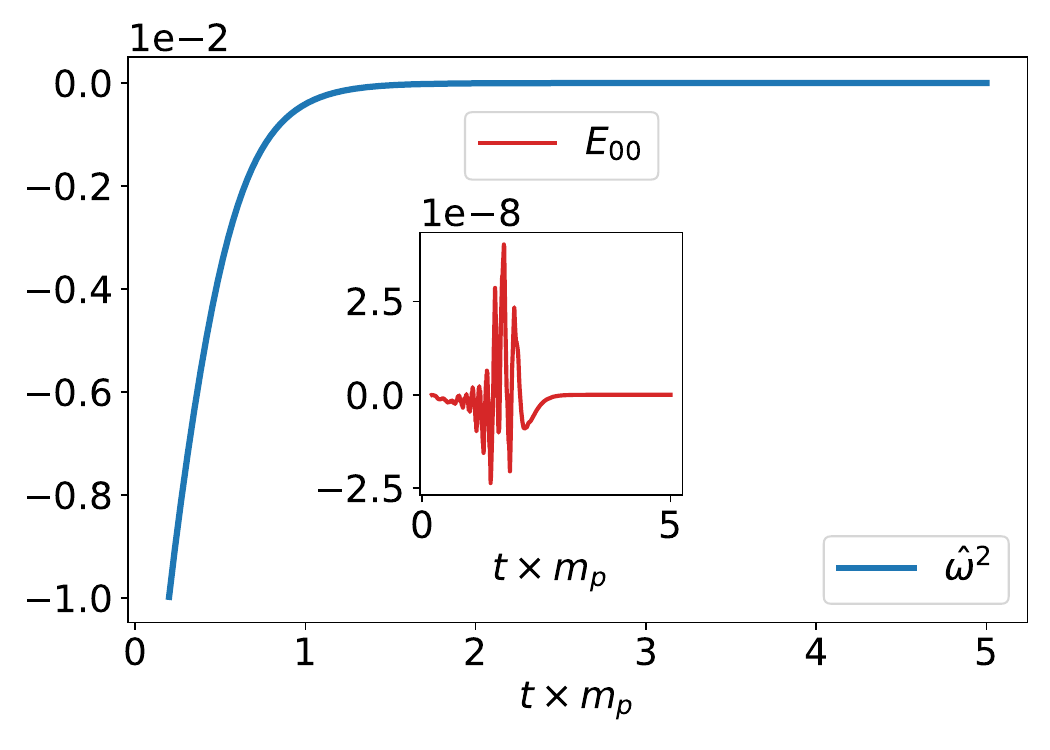}} \\
           \\
        (c) & (d)\\
            \resizebox{\imsize}{!}{\includegraphics[width=0.2\textwidth]{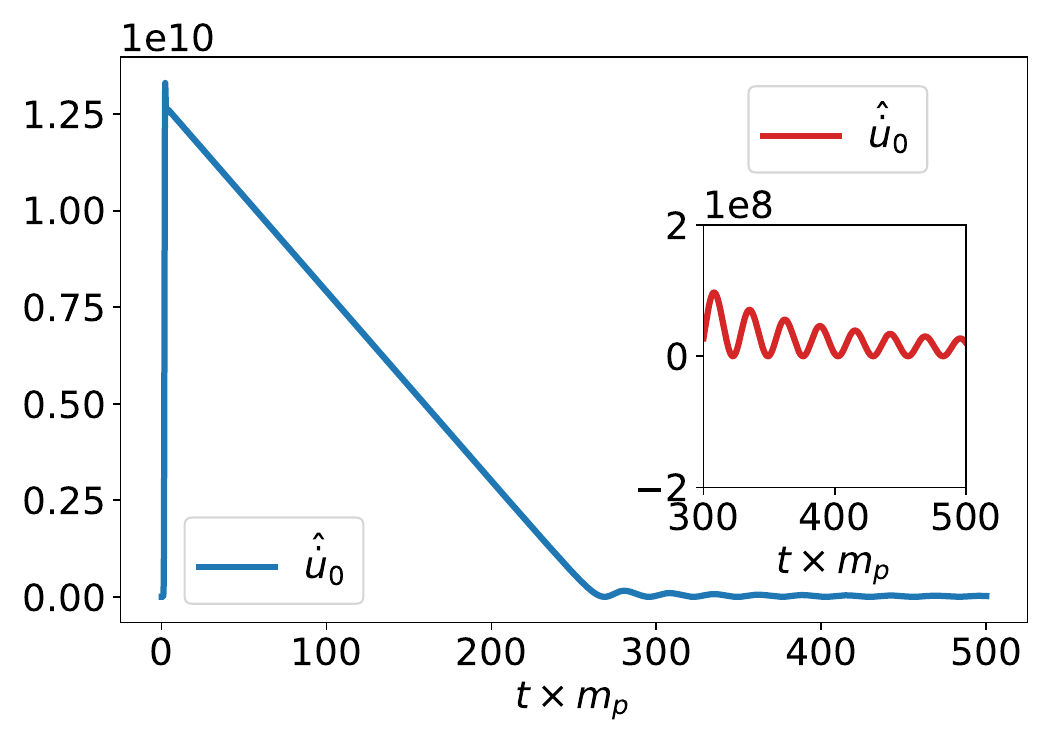}} &
            \resizebox{\imsize}{!}{\includegraphics[width=0.2\textwidth]{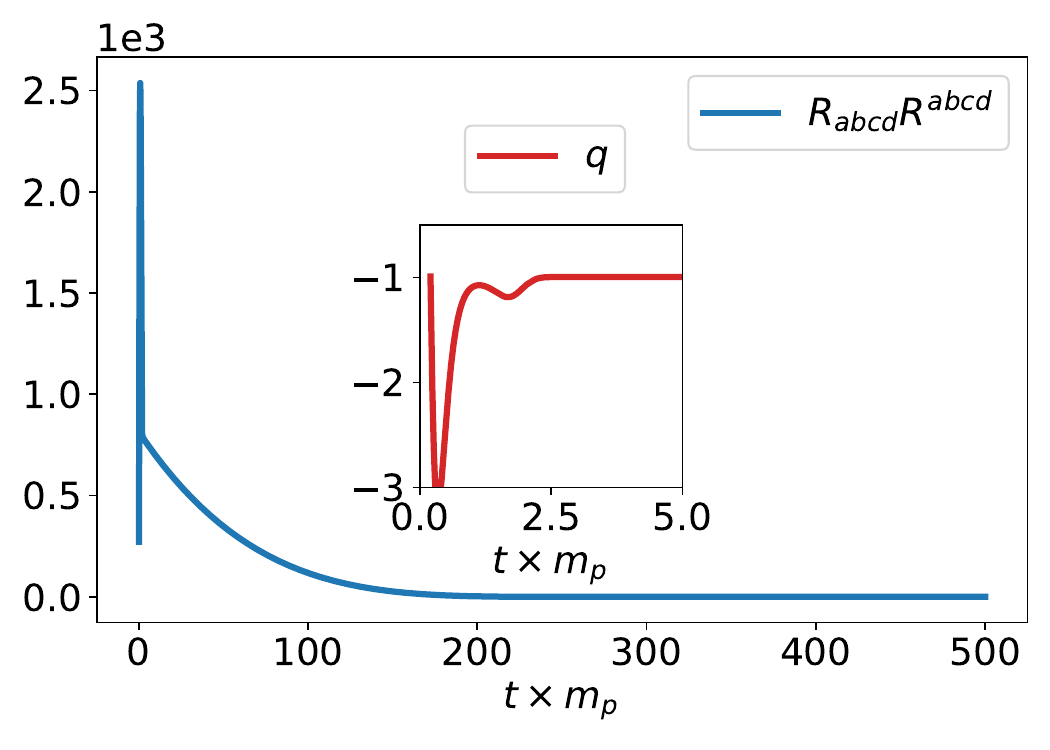}}  \\
           \\
        (e) & (f)\\
           \end{tabular}
  \par\end{centering}\caption{The dynamical evolution toward the future for QG with radiation source EoS parameter $w=1/3$ is addressed. The renormalization parameters are $\beta=3\,m_p^{-2}$ and $\alpha=-30\,m_p^{-2}$. The solution with initial conditions $H=9.98148\times 10^{-1}\,m_p$, $\dot{H}=-9.25926\times 10^{-3}\,m_p^2$, $\ddot{H}=3.58604\times 10^{1}\,m_p^3$, $\Omega_K=1.00371\times 10^{-2}$, $\Phi_3=1.00186\times 10^{-1}$, $\Phi_{3,0}=1.00371\times 10^{-1}$, $\Phi_{3,1}=-5.53408$, $\Omega_m=3.34571\times 10^{-2}$, $\Sigma_+=7.21336\times 10^{-1}$, $\Sigma_{+1}=1.00371\times 10^{-1}$, $\Sigma_{+2}=2.85600\times 10^{1}$, $\Sigma_-=-6.41187\times 10^{-1}$, $\Sigma_{-1}=1.00371\times 10^{-1}$, $\Sigma_{-2}=1.00558\times 10^{-1}$ with tilt $r=4.8$ and $\eta=1.0\times 10^{-1}$ is attracted to the RR orbit. This initial condition corresponds to a white point near the center of the basin of attraction to RR shown in Fig. \ref{f4}c. a) The graph, plotted in blue, shows the Hubble parameter $H$, which starts in a transient regime, followed by the slow-roll regime, and a graceful exit of inflation with damped oscillations. In red in the inset, it is shown a zoom on the Hubble parameter, with a focus on the transient and the beginning of the slow-roll regime. b) The isotropization of the shear variables $\Phi_3$, $\Sigma_+$, and $\Sigma_-$ is shown in blue, red, and green, respectively. c) The blue plot shows that the tilt variable $r$ initially increases and then stabilizes at a constant value, while the red inset plot shows that the tilt direction $\eta$ decreases. d) The graph in blue shows that the vorticity $\hat{\omega}^2$ approaches zero in the slow-roll phase. Inset in red, it shows the constraints $E_{00}$, with fluctuations smaller than $10^{-7}$. e) The plot in blue shows the matter component $\hat{\dot{u}}_0$, which, due to the tilt constant approaches, evolves in accordance with the Hubble parameter. In red in the inset, a zoom of the matter component $\hat{\dot{u}}_0$ is plotted, showing the high-damped amplitudes of the oscillations. f) The graph in blue shows the decrease to zero of the curvature scalar $R_{abcd}R^{abcd}$. The inset in red shows the deceleration parameter $q$, which approaches $-1$, as expected}\label{f6}
\end{figure*} 

\begin{figure*}[htpb]
      \begin{centering}
     \begin{tabular}{c c}
            \resizebox{\imsize}{!}{\includegraphics[width=0.2\textwidth]{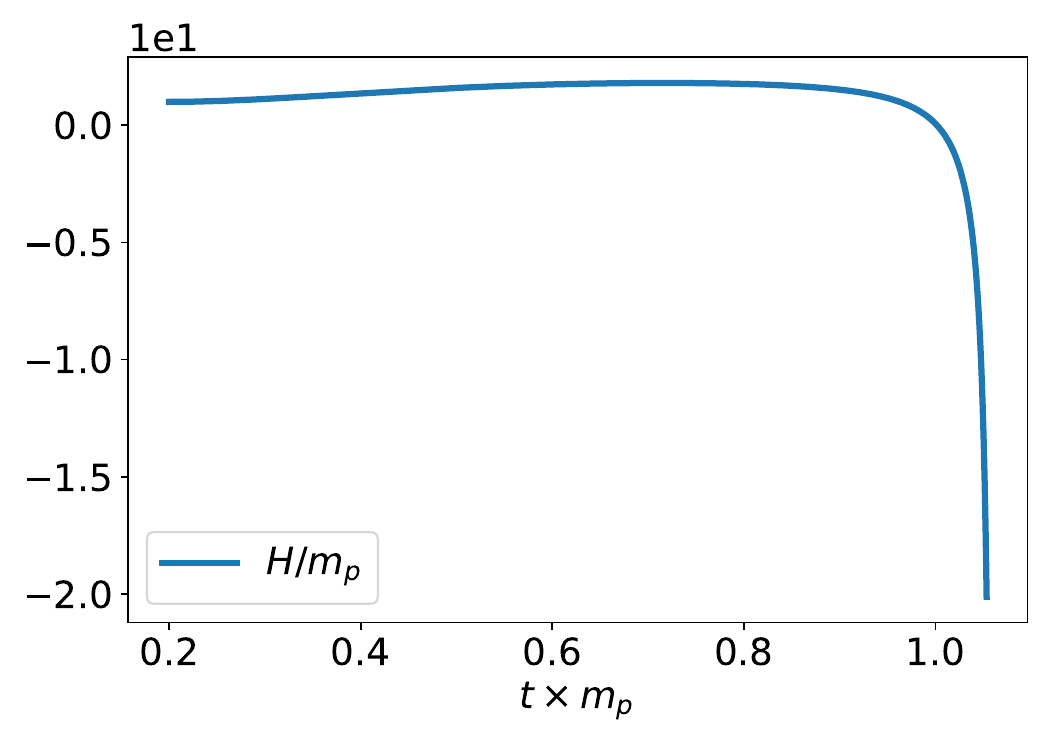}} &
            \resizebox{\imsize}{!}{\includegraphics[width=0.2\textwidth]{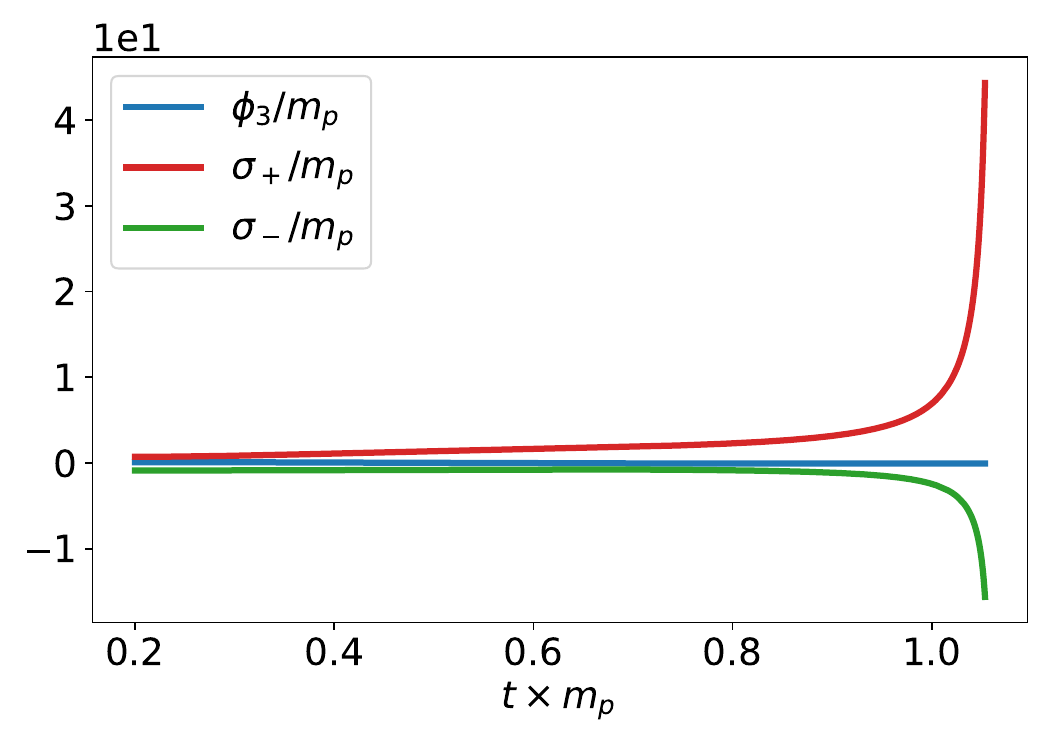}} \\
           \\
        (a) & (b)\\
         \resizebox{\imsize}{!}{\includegraphics[width=0.2\textwidth]{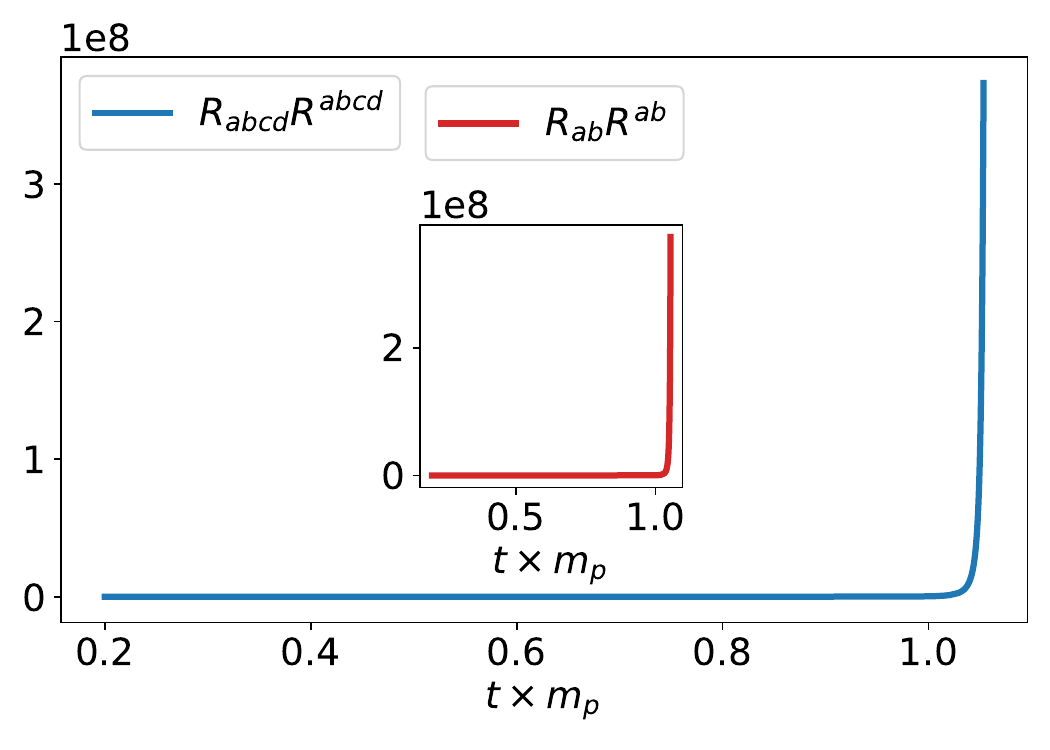}} &
            \resizebox{\imsize}{!}{\includegraphics[width=0.2\textwidth]{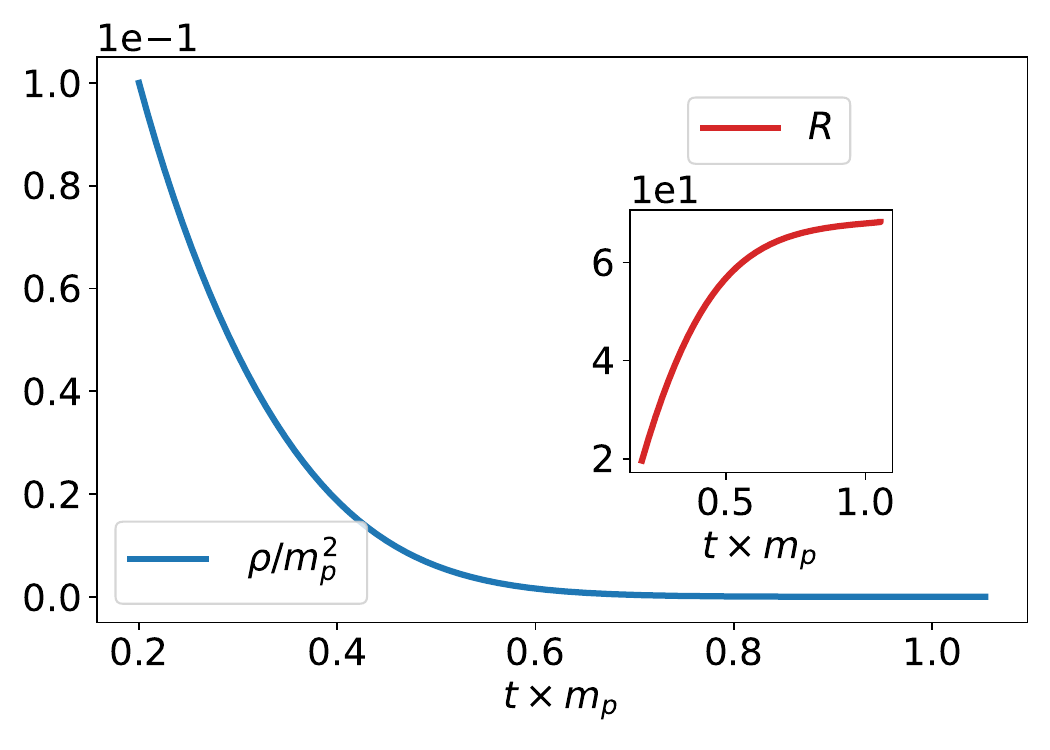}} \\
           \\
        (c) & (d)\\
        \resizebox{\imsize}{!}{\includegraphics[width=0.2\textwidth]{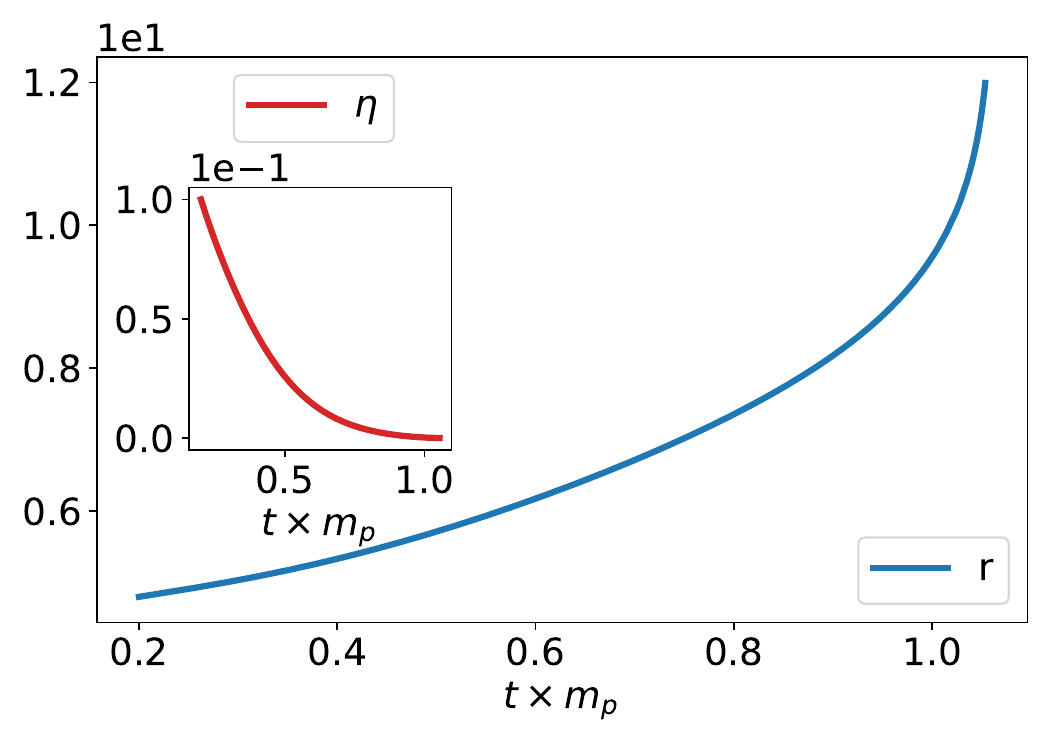}} 
           &
            \resizebox{\imsize}{!}{\includegraphics[width=0.2\textwidth]{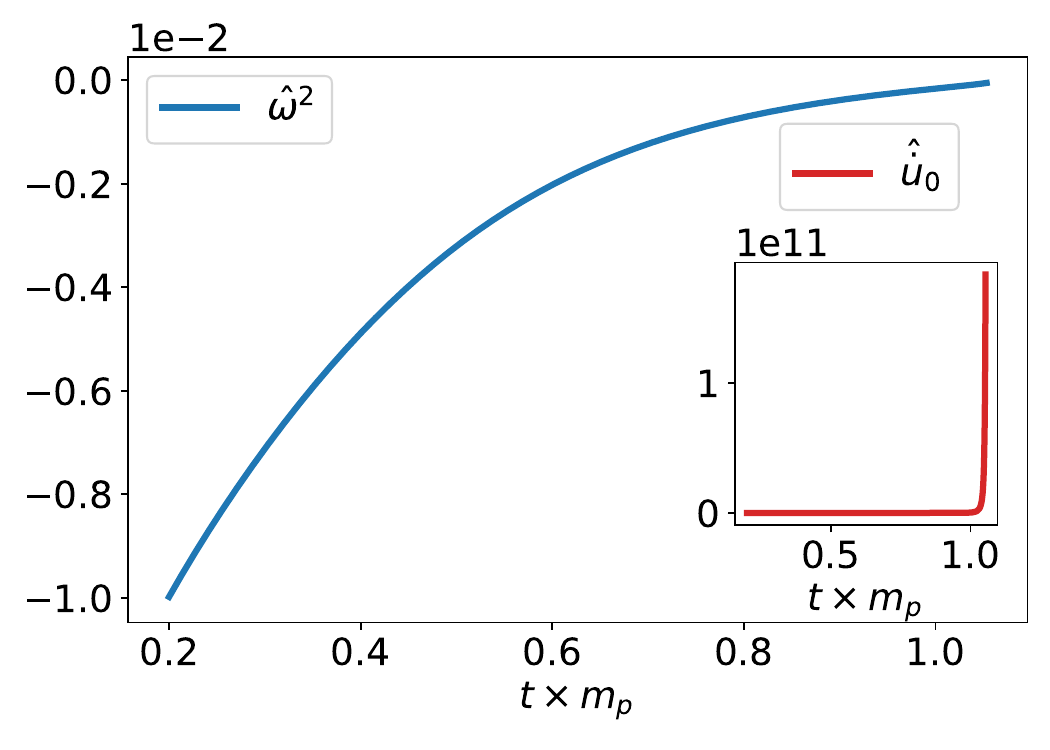}} \\
           \\
        (e) & (f)
           \end{tabular}
  \par\end{centering}\caption{Now, QG is addressed for the radiation source with the EoS parameter $w=1/3$. The initial condition contains a significant amount of tilt $r=4.8$ with tilt direction $\eta=1.0\times 10^{-1}$ along with $H=9.98148\times 10^{-1}\,m_p$, $\dot{H}=-9.25926\times 10^{-3}\,m_p^2$, $\ddot{H}=3.29973\times 10^{1}\,m_p^3$, $\Omega_K=1.00371\times 10^{-2}$, $\Phi_3=1.00186\times 10^{-1}$, $\Phi_{3,0}=1.00371\times 10^{-1}$, $\Phi_{3,1}=-5.11917$, $\Omega_m=3.34571\times 10^{-2}$, $\Sigma_+=7.21336\times 10^{-1}$, $\Sigma_{+1}=1.00371\times 10^{-1}$, $\Sigma_{+2}=3.03602\times 10^{1}$, $\Sigma_-=-8.81633\times 10^{-1}$, $\Sigma_{-1}=1.00371\times 10^{-1}$, $\Sigma_{-2}=1.00558\times 10^{-1}$. The renormalization parameters are $\beta=3\,m_p^{-2}$ and $\alpha=-30\,m_p^{-2}$. The numerical evolution is to the future. The solution evolves to a recollapse and then is attracted to the isotropic singularity. The initial condition corresponds to a black point located outside the center of the basin of attraction to RR shown in Fig. \ref{f4}c. a) The plot shows in blue the Hubble parameter decrease, passing through zero, and subsequently increasing towards negative values. b) The graph shows an increase in the diagonal shear components $\sigma_+$ and $\sigma_-$ plotted in red and green, respectively. The plot in blue shows the non-diagonal shear component $\phi_3$, which does not show any increase. c) It is shown the increase and then a divergence in the scalars of curvature $R_{abcd}R^{abcd}$ and $R_{ab}R^{ab}$, plotted in blue and in red inset, respectively. d) The blue plot shows that the energy density decreases to zero. In the inset, it is shown in red the scalar of curvature $R$, which evolves without reaching a divergence. e) The increase in the tilt variable $r$ is shown in blue. In the inset it is plotted in red the decrease of the tilt direction $\eta$. f) The graph shown in blue shows the vorticity decreases to zero. The plot inset is shown in red, the increase indefinitely in the matter acceleration component $\hat{\dot{u}}_0$}\label{f7}
\end{figure*} 

 \begin{figure*}[htpb]
      \begin{centering}
     \begin{tabular}{c c}
       \resizebox{\imsize}{!}{\includegraphics[width=0.2\textwidth]{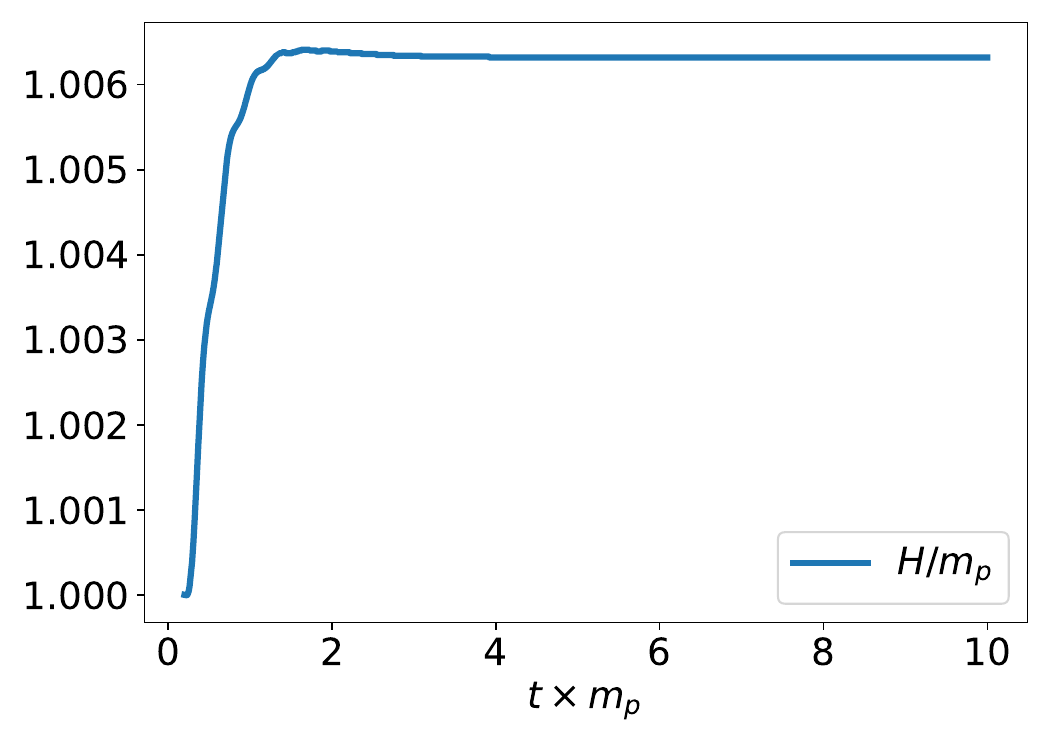}} &
            \resizebox{\imsize}{!}{\includegraphics[width=0.2\textwidth]{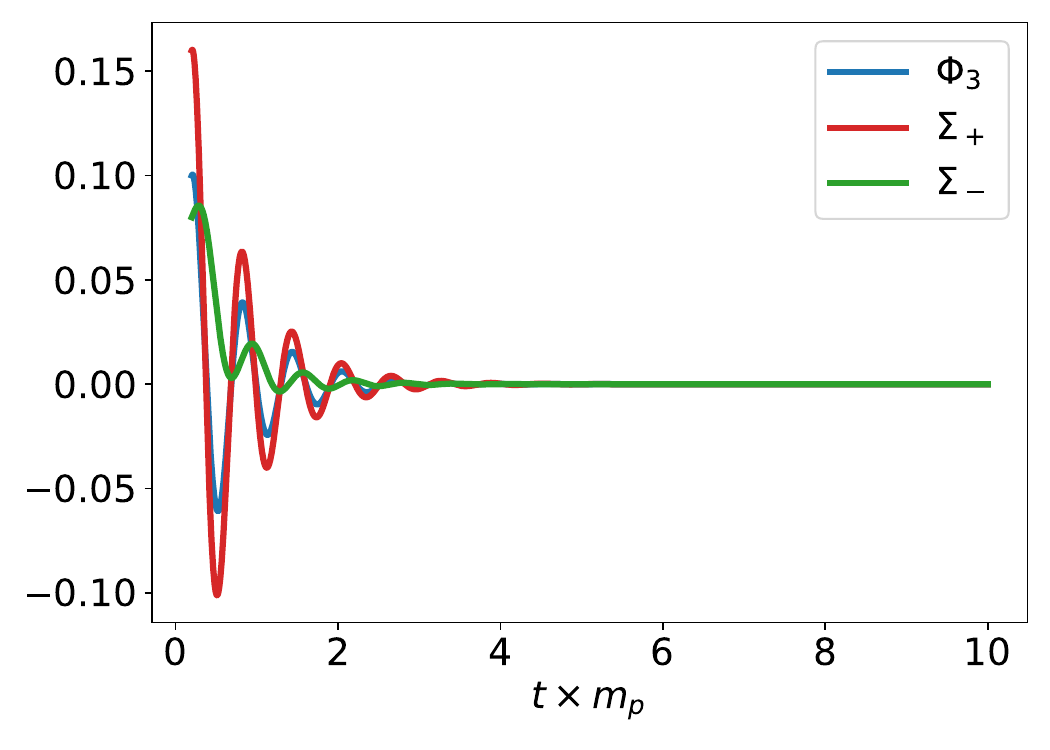}}   \\
            \\
            (a) & (b)\\
             \resizebox{\imsize}{!}{\includegraphics[width=0.2\textwidth]{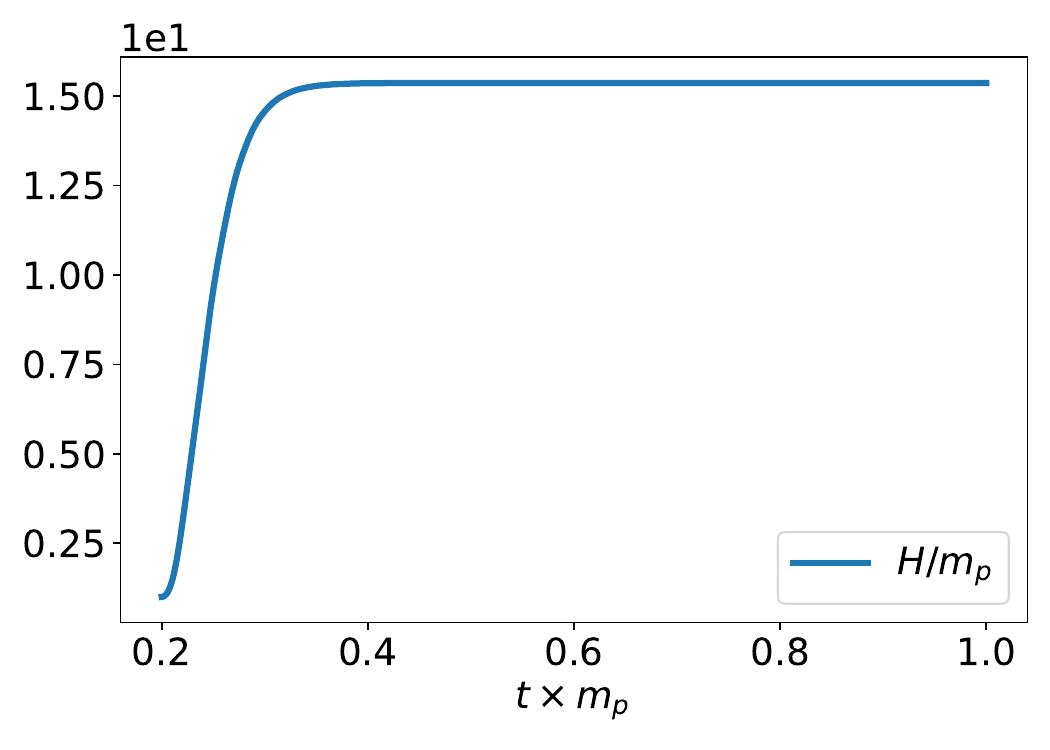}} &
            \resizebox{\imsize}{!}{\includegraphics[width=0.2\textwidth]{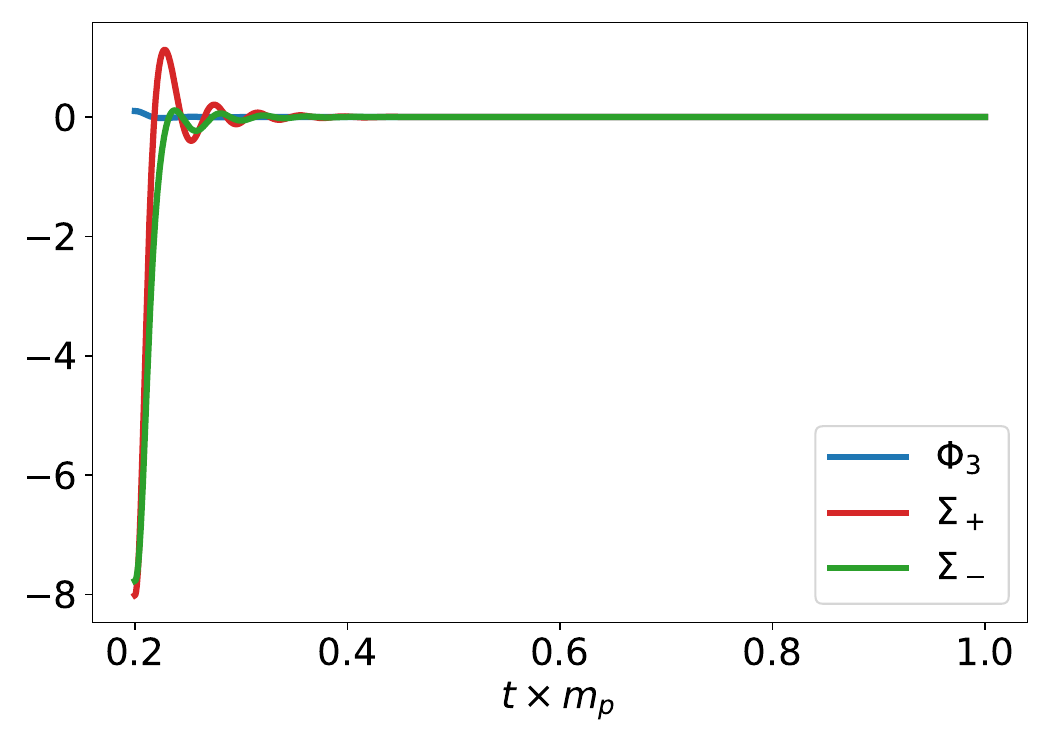}}   \\
            \\
            (c) & (d)
           \end{tabular}
  \par\end{centering}\caption{It addresses the orbit with renormalization parameters $\alpha=-3.0\times 10^{8}\,m_p^{-2}$ and $\beta=1.305\times 10^{9}\,m_p^{-2}$ for the radiation fluid with EoS parameter $w=1/3$. The numeric evolution is to the future, and both solutions are attracted to RR orbit. a) and b) The initial conditions are set near the RR solution and represented by a white point near the origin in the basin shown in Fig. \ref{f5}b. The initial conditions are: $H=1.0\,m_p$, $\dot{H}=-2.12857\times 10^{-11}\,m_p^2$, $\ddot{H}=-6.82592\times 10^{-2}\,m_p^3$, $\Omega_K=1.0\times 10^{-2}$, $\Phi_3=1.0\times 10^{-1}$, $\Phi_{3,0}=1.0\times 10^{-1}$, $\Phi_{3,1}=-1.11645\times 10^{1}$, $\Omega_m=3.33333\times 10^{-2}$, $\Sigma_+=1.60297\times 10^{-1}$, $\Sigma_{+1}=1.0\times 10^{-1}$, $\Sigma_{+2}=-1.74597\times 10^{1}$, $\Sigma_-=8.01484\times 10^{-2}$, $\Sigma_{-1}=1.0\times 10^{-1}$, $\Sigma_{-2}=1.0\times 10^{-1}$ with $r=4.8$ and $\eta=1.0\times 10^{-1}$. c) and d) The initial conditions are given by $H=1.0\,m_p$, $\dot{H}=-2.12857\times 10^{-11}\,m_p^2$, $\ddot{H}=6.30244\times 10^{3}\,m_p^3$, $\Omega_K=1.0\times 10^{-2}$, $\Phi_3=1.0\times 10^{-1}$, $\Phi_{3,0}=1.0\times 10^{-1}$, $\Phi_{3,1}=-5.47648\times 10^{2}$, $\Omega_m=3.33333\times 10^{-2}$, $\Sigma_+=-8.01484$, $\Sigma_{+1}=1.0\times 10^{-1}$, $\Sigma_{+2}=4.50177\times 10^{4}$, $\Sigma_-=-7.77440$, $\Sigma_{-1}=1.0\times 10^{-1}$, $\Sigma_{-2}=1.0\times 10^{-1}$ with $r=4.8$ and $\eta=1.0\times 10^{-1}$. This solution corresponds to a white point located farther from the origin in the basin shown in Fig. \ref{f5}b. The left panel in a) and c) shows the dynamics for the Hubble parameter $H$ as the orbit is attracted to the RR asymptotic regime. In both right panels on b) and d), the isotropization of the shear variables $\Phi_3$, $\Sigma_+$, and $\Sigma_-$ is plotted in blue, in red, and in green, respectively}\label{f8}
\end{figure*}

\begin{figure}[htpb]
      \begin{centering}
       \resizebox{\imsize}{!}{\includegraphics[width=0.2\textwidth]{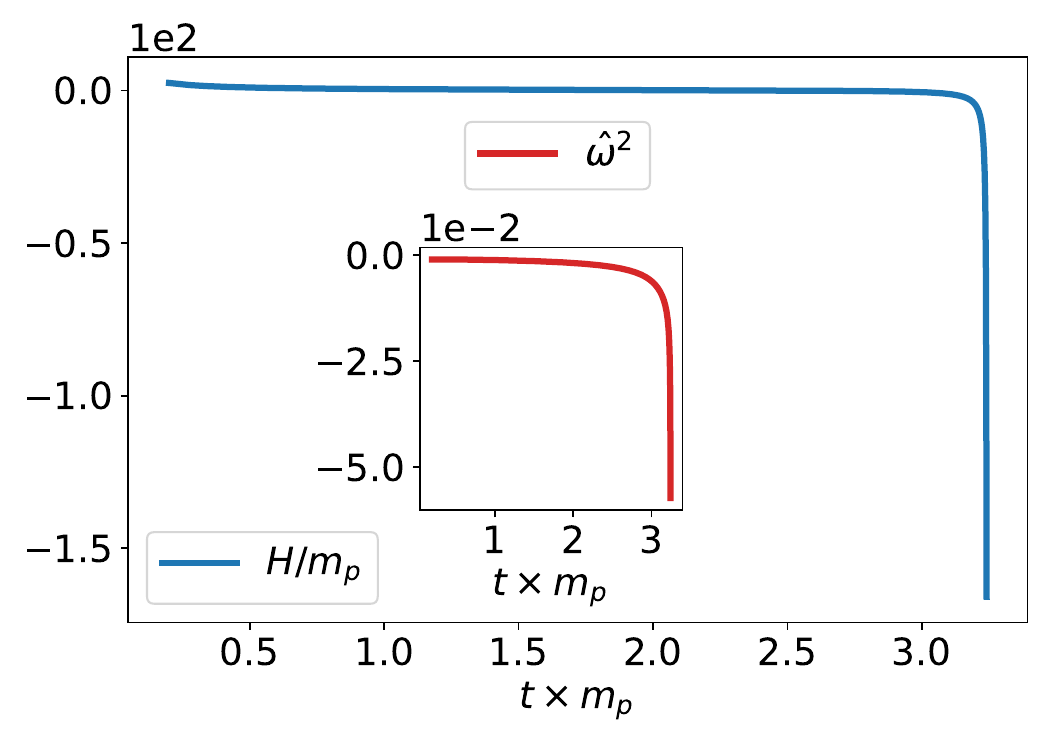}}   \par\end{centering}\caption{The dynamical time evolution to the future for QG with an EoS parameter $w=0.6$ and renormalization parameters $\beta=3\,m_p^{-2}$ and $\alpha=-30\,m_p^{-2}$ is addressed. The initial conditions near the isotropic singularity are given by $H=2.5\,m_p$, $\dot{H}=-1.25\times 10^{1}\,m_p^2$, $\ddot{H}=1.24774\times 10^{2}\,m_p^3$, $\Omega_K=1.6\times 10^{-3}$, $\Phi_3=4.0\times 10^{-2}$, $\Phi_{3,0}=1.6\times 10^{-2}$, $\Phi_{3,1}=-5.49742\times 10^{-2}$, $\Omega_m=5.33333\times 10^{-3}$, $\Sigma_+=4.0\times 10^{-2}$, $\Sigma_{+1}=1.6\times 10^{-2}$, $\Sigma_{+2}=-4.02629\times 10^{-2}$, $\Sigma_-=4.0\times 10^{-2}$, $\Sigma_{-1}=1.6\times 10^{-2}$, $\Sigma_{-2}=6.4\times 10^{-3}$ with $r=1.0\times 10^{-1}$ and $\eta=1.0\times 10^{-1}$. The solution reaches a recollapse and then is attracted to the isotropic singularity. The blue plot shows the Hubble parameter decreasing through zero and diverging toward negative values. The red inset displays the increase in module toward negative values of the vorticity $\hat{\omega}^2$}\label{f9}
\end{figure}

Despite the well-known RR solution for QG, we find universes that after a recollapse are asymptotically attracted to the isotropic singularity attractor. As above mentioned, RR is an asymptotic attractor with zero vorticity. However, solutions attracted to the isotropic singularity may have divergent vorticity \cite{deMedeiros:2024pmc}.

A basin plot for different initial conditions of the diagonal shear components $\Sigma_+$ and $ \Sigma_-$, with the white points showing initial conditions attracted to RR orbit, while the black points represent solutions attracted, which evolve to a recollapse and are asymptotically attracted to the isotropic singularity, is shown in Figs. \ref{f4} and \ref{f5}. In both cases, the EoS parameter is chosen as $w=1/3$ for the radiation fluid. However, the analyzed regions of attraction in the basin plots considered in this work do not change with the value of the EoS parameter.

Now, we analyze the tilt effect on the basin plot above mentioned. Figure \ref{f4} considers a set of arbitrary values for the renormalization parameters $\alpha=-30\, m_p^{-2}$ and $\beta=3 \,m_p^{-2}$, with a small matter density $\Omega_m=3.34571\times 10^{-2}$. Figures \ref{f4}a and b correspond to the case of zero tilt $r=0$, while Figures \ref{f4}c and d show the basin for nonzero tilt values $r=4.8$ and $r=7$, respectively. It can be seen that in Figures \ref{f4}c and d, increasing the tilt variable shifts the attraction region for RR near the origin in Fig. \ref{f4}a and b for $r=0$ to the right, while the farther regions of attraction to RR shift to the left.

Figure \ref{f6} shows the initial condition, which corresponds to a white point in the region near the center of the basin for the RR solution shown in Fig. \ref{f4}c. The tilt variable is $r=4.8$ for the radiation fluid with EoS parameter $w=1/3$. Figure \ref{f6}a shows, in blue, the dynamic evolution of the Hubble parameter. It starts in a transient regime and then enters the slow-roll inflationary regime of RR, where it decreases linearly slowly. After a graceful exit for the inflation, the Hubble parameter enters the phase of damped oscillations, characterized by the asymptotic scalaron behavior $e^{i\omega t}$, with $\omega=1/\sqrt{6\beta}$. A zoom of the Hubble parameter showing the transient regime and the start of the slow-roll regime is plotted in the red inset in Fig. \ref{f6}a.

Fig. \ref{f6}b shows the isotropization of the shear variables $\Phi_3$, $\Sigma_+$, and $\Sigma_-$, plotted in blue, red, and green, respectively.  The initial damped oscillations shown in Fig. \ref{f6}b correspond to the transient regime before the solution enters the slow-roll phase. In this work, it is not addressed the followed damped oscillation in the shear variables characterized by the asymptotic Ostrogradsky ghost behavior $e^{i\omega t}$, with $\omega=1/\sqrt{-\alpha}$. Fig. \ref{f6}c shows, in blue, the non-small initial condition for the tilt, which, after an increase, approaches a constant value. This result is in good agreement with the eigenvalue found in the Subsection \ref{sec32}. In the red inset Fig. \ref{f6}c, the graph of the tilt direction $\eta$ is shown. Figure \ref{f6}d plot, in blue, the vorticity, which decreases toward zero during the slow-roll regime. It is plotted in the red inset the constraint $E_{00}$, shown in the Appendix \ref{appB}, with fluctuations always smaller than $10^{-7}$. Fig. \ref{f6}e plots in blue the component of the matter acceleration $\hat{\dot{u}}_0$, which, due to the tilt constant approaches, evolves in accordance with the Hubble parameter. In the red inset, a zoom of the matter component $\hat{\dot{u}}_0$ is plotted, showing the high-damped amplitudes of the oscillations. For an initial condition with a smaller value for the tilt variable $r$ and for the radiation fluid $w = 1/3$, the oscillations shown in Fig. \ref{f6}e in the matter acceleration have a smaller amplitude. As expected, the scalars of curvature decrease and approach zero as the universe expands. For example, Figure \ref{f6}f plots, in blue, the scalar of curvature $R_{abcd}R^{abcd}$, which decreases to zero. Inset of Fig. \ref{f6}f is plotted in red the deceleration parameter, which approaches $-1$ during the slow-roll regime, indicating the accelerated expansion characteristic of this phase.

Fig. \ref{f7} shows the solution with the initial condition represented by a black point outside the center of the basin for the RR solution shown in Fig. \ref{f4}. The solution evolves to a recollapse and then is asymptotically attracted to the isotropic singularity. The EoS parameter is $w=1/3$, and the tilt variable is $r=4.8$. In Fig. \ref{f7}a, the blue plot shows the Hubble parameter $H$ reaching a recollapse. It can be seen that the Hubble parameter decreases, passes through zero, and diverges toward negative values. Figure \ref{f7}b shows the increase in the diagonal shear variables $\sigma_+$ and $\sigma_-$, plotted in red and green, respectively. Furthermore, the non-diagonal shear component $\phi_3$ is plotted in blue without showing any increase. However, in terms of the ENV, the shear variables $\Phi_3$, $\Sigma_+$, and $\Sigma_-$ decrease to zero. In Fig. \ref{f7}c, an increase indefinitely in the geometric variables $R_{ab}R^{ab}$ and $R_{abcd}R^{abcd}$ is shown, plotted in blue and in red inset, respectively. The Ricci scalar $R$ approaches a constant, as shown in the red inset of Fig. \ref{f7}d. Fig. \ref{f7}d shows in blue the energy density decreases to zero. Fig. \ref{f7}e shows the tilt increasing in blue, with the tilt direction decreasing in the red inset. The expansion and the acceleration of the source diverge along with the tilt increase, while the vorticity approaches zero. Fig. \ref{f7}f shows in blue the decrease of the vorticity and in red inset the divergence of the matter acceleration component $\hat{\dot{u}}_0$.

For a perfect fluid with zero tilt, $r=0$, the value of $\Omega_m$ does not affect the basin of attraction to RR for different initial conditions of $\Sigma_+$ and $\Sigma_-$. However, for a non-perfect fluid, as shown in Figure \ref{f5}a with $\Omega_m=6.69143\times 10^{-1}$ and $r=4.8$, compared to Fig. \ref{f4}c, where $\Omega_m=3.34571\times 10^{-2}$ and $r=4.8$, the increase in the matter density $\Omega_m$ shifts to the right the attraction points to RR located near the origin. Similarly, the farther white points are shifted to the left, resulting in their superposition, as shown in the basin plot of Fig. \ref{f5}a. Therefore, the basins with initial conditions considered in Figs. \ref{f4}c and d and Fig. \ref{f5}a show that a change in the initial condition of $\Omega_m$ is similar to a change in the initial condition of the tilt variable $r$.

Figure \ref{f5}b shows the basin plot to RR solution for different initial conditions of $\Sigma_-$ and $\Sigma_+$. Nevertheless, we address the renormalization parameter $\beta=1.305\times10^{9}\,m_p^{-2}$, as inferred by CMBR observations \cite{Gorbunov_2015,Mishra:2018dtg,Akrami:2018odb}. The tilt is set to $r=4.8$ and $\alpha=-3.0\times10^{8}\,m_p^{-2}$. The basin of attraction to RR inflation is modified in relation to Figures \ref{f4} and \ref{f5}a, which are based on an arbitrary set for the renormalization parameter $\alpha=-30\, m_p^{-2}$ and $\beta=3 \,m_p^{-2}$. In Figures \ref{f4} and \ref{f5}, both basins show an attraction region to RR solution for initial conditions farther from its orbit. As expected, the tilt has a small influence on the RR attractor, while for a realistic value of $\beta=1.305\times10^{9}\,m_p^{-2}$, it is insignificant.

Figures \ref{f8}a and b show a solution with an initial condition near the RR regime, represented by a white point near the origin in the basin of Fig. \ref{f5}b. Figure \ref{f8}a shows the Hubble parameter $H$ in blue, illustrating its evolution from the initial transient phase to the slow-roll regime, where it approaches a constant value. Fig. \ref{f8}b plots the shear variables $\Phi_3$, $\Sigma_+$, and $\Sigma_-$ in blue, red, and green, respectively. Figures \ref{f8}c and d show the orbit with the initial condition that is not near the Ruzmaikina and Rukmaikin asymptotic solution, represented by a white point farther from the origin in the basin of Fig. \ref{f5}b. Fig. \ref{f8}c shows in blue the evolution of the Hubble parameter, starting with the transient regime, followed by the slow-roll inflationary phase. Fig. \ref{f8}d shows the isotropization of the shear variables $\Phi_3$, $\Sigma_+$, and $\Sigma_-$ plotted in blue, red, and green, respectively.

We conclude our analysis by showing that universes with initial conditions attracted to recollapse toward the isotropic singularity may have divergent vorticity \cite{deMedeiros:2024pmc}. As an example, Fig. \ref{f9} shows an orbit with initial conditions near the isotropic singularity solution for QG. The solution is attracted to the recollapse and then is asymptotically attracted to the
isotropic singularity. The evolution of the Hubble parameter is shown in blue, while in red in the inset it is plotted that the vorticity increases toward negative values. In this case, the vorticity increase is accompanied by divergence in the other kinematic variables, such as the matter acceleration and its expansion. In the previous work \cite{deMedeiros:2024pmc}, it is shown, from the dynamical system, that the matter density $\Omega_m$ decouples from the dynamics as the solution approaches the physical singularity of the isotropic singularity. However, since the tilt and its effects are properties of the matter, they no longer influence the dynamics once the isotropic singularity is reached.

\section{Conclusions}\label{summary}
This work explores the effects of the kinematic variables: acceleration, vorticity, and expansion of the tilted component models for both GR and QG theories. The anisotropic Bianchi V model is addressed with numerical time evolution to the future. The main objective is the study of how these properties of matter can modify the future evolution of the attractors. This work also intends to explore the consequences of these modifications on the regions of initial conditions of the solutions. A previous work \cite{deMedeiros:2024pmc} aimed to better understand past singularities in the presence of these kinematic properties of the matter, such as tilt and vorticity. In both works the approach is non-perturbative, in the sense that the solution depends only on machine precision. A tilted source is defined by the energy-momentum tensor in \eqref{tem}, with a tilted time-like vector \eqref{timeve} and normalization $\hat{u}^a\hat{u}^bg_{ab}=-1$. In the presence of tilt, nonzero vorticity, shear and anisotropies arise \cite{King:1972td,ellisking,Coley_2005,Coley_2006,Lim_2006,Coley:2008zz,Coley_2009,Krishnan:2022qbv,Krishnan:2022uar,Ebrahimian:2023svi,Allahyari:2023kfm}. On the other hand, as pointed out in textbooks, a non-tilted source is a perfect fluid in which point particles follow geodesics, with zero acceleration and null vorticity \cite{Weinberg:1972kfs,Stephanibook}.

The same set of assumptions considered in this work was used in \cite{deMedeiros:2024pmc} for a numeric time evolution to the past singularity. This set of assumptions results in equations \eqref{Gamma_0ab} with \eqref{sig1}, \eqref{Gamma_ijk}, and \eqref{deromegak} to define the connection. The dynamical system for GR is defined by \eqref{deromegak} and in the Appendix \ref{appA}, while the dynamical system for QG is defined in the Appendix \ref{appB} and by \eqref{deromegak}, \eqref{senv}. The constraints for GR and QG are shown in the Appendix \ref{appA} and \ref{appB}, respectively. The constraints must be satisfied by the initial conditions. Once these constraints are initially satisfied, they must be maintained during time evolution. For this reason, they are used in this work as a numerical check. It was verified that all the constraints for GR show numerical fluctuations, always smaller than $10^{-13}$, while for QG they fluctuate numerically at $10^{-7}$.

The numeric time evolution to the past, done in \cite{deMedeiros:2024pmc}, shows that universes with higher and smaller matter density are attracted to the Kasner singularity or to the isotropic singularity, respectively. In \cite{deMedeiros:2024pmc}, it is found that for GR, all solutions are attracted to the Kasner past singularity with zero vorticity. While, in QG, the isotropic singularity attractor may have divergent vorticity. In this sense, only QG allows a geometric singularity with divergences in all the kinematic variables of the substance. 

It is also shown in \cite{deMedeiros:2024pmc} that for the EoS parameter $1/3 < w < 1$, all kinematic variables decrease to zero, both in GR and QG theories in backward evolution. According to this result, for the ultra-radiative regime, any small initial values in the kinematic variables increase up to the future when the time is reversed.

The present work is consistent with previous results in \cite{deMedeiros:2024pmc}. According to GR, the solutions are attracted to the FLRW orbit for the EoS $p=w\rho$ parameter $-1<w<-1/3$, with both the tilt and the kinematic variables decreasing to zero \cite{deMedeiros:2024pmc}. This behavior is described numerically by Fig. \ref{f1}. However, for solutions with $-1/3<w<1$, the Milne orbit is the future attractor, with an increasing tilt for $1/3< w<1$ \cite{Coley_2005,Coley_2006, Coley:2008zz,Krishnan:2022qbv, Krishnan:2022uar,deMedeiros:2024pmc}. This increase in the tilt occurs even when the model becomes isotropic, as shown by the shear variables decreasing to zero \cite{Krishnan:2022qbv, Krishnan:2022uar}. During the transient regime to the Milne orbit, we show that the expansion of the source has an initial decrease, followed by an increase. While still in the transient regime, the acceleration of the substance increases. After the transient regime, the kinematic variables, the expansion of the source, and its acceleration continue to increase and then diverge, also noted in \cite{Coley_2006, Coley:2008zz,deMedeiros:2024pmc}. Furthermore, for the vorticity, it is shown that after its initial increase in absolute value during the transient regime, it decreases to zero as the solution is attracted to the Milne orbit. Fig. \ref{f2} describes the aforementioned result for the initial condition near the Kasner solution and attracted to the Milne orbit with the EoS parameter set to $w=0.6$.

In the aforementioned situation, we show that, as the dynamical system evolves in time, the contribution of the energy-momentum tensor to the field equations decreases to zero and then decouples from the field equations. In this sense, the tilt effects no longer influence the dynamics dominated by the Milne solution.

Despite the well-known RR solution for QG, we find universes that after a recollapse are asymptotically attracted to the isotropic singularity attractor. This work intends to describe the behavior of the kinematic variables for these two attractors.

The linear stability for the RR solution done in Sect. \ref{sec32} shows that for the EoS parameter $1/3<w<1$, the tilt variable increases. As an example, Fig. \ref{f3} describes the solutions attracted to RR with the EoS parameter set to $w=0.6$. This tilt increase results in an initial increase in all the kinematic variables during the transient regime. After that, the expansion and acceleration of the substance increase indefinitely, and the vorticity decreases to zero during the slow-roll inflationary regime. The solution presents a graceful end for inflation while the kinematic variables continue with the behavior described above. The solution is then attracted to the Minkowski orbit. We also mention here that the tilt effects no longer influence the inflation once it has begun.

Nevertheless, universes attracted to the isotropic singularity may have divergent vorticity along with all the other kinematic singularities \cite{deMedeiros:2024pmc}. As an example, Fig. \ref{f7} shows an initial condition attracted to the isotropic singularity with zero vorticity, while Fig. \ref{f9} displays the solution with the initial condition near the isotropic singularity that also is attracted to the isotropic singularity, but with an increase in the module of the vorticity. In both cases, there is a singularity in the other kinematic variables, the expansion of the substances and its acceleration. 

However, since the matter decouples from the dynamic as the solution approaches the physical singularity of the isotropic singularity \cite{deMedeiros:2024pmc}, the tilt effects no longer influence the dynamics when the isotropic singularity is reached. In this sense, despite the increase of the matter proprieties for some cosmological models considered in this work, their effects no longer influence the future dynamical evolution of the universe once the solutions are settled.

This work also intends to investigate the tilt effect on a basin plot for different initial conditions of the diagonal shear components $\Sigma_+$ and $ \Sigma_-$, with the white points showing initial conditions attracted to the RR orbit, while the black points represent solutions attracted, which evolve to a recollapse and are asymptotically attracted to the isotropic singularity. Figures \ref{f4} and \ref{f5} describe the basin for the EoS parameter $w=1/3$. We identified initial conditions both near and farther from Ruzmaikina's orbit where the inflation occurs. As expected, the tilt has a small influence on the RR attractor, while for a realistic value of $\beta=1.305\times10^{9}\,m_p^{-2}$, as inferred from CMBR data \cite{Gorbunov_2015,Mishra:2018dtg,Akrami:2018odb}, it is insignificant. We also mention that no fractal structure was observed in these basins, which should mean that the dynamic system considered in this work is not chaotic for the diagonal shear variables $\Sigma_+$ and $\Sigma_-$. This result is predicted from the Bianchi types hierarchy mentioned in \cite{Coley_2005}.

\section*{Acknowledgements}

\begin{acknowledgement}

W. P. F. de Medeiros gratefully acknowledges the financial assistance provided by the Brazilian agency \textit{Coordena{\c c}{\~a}o de Aperfei{\c c}oamento de Pessoal de N{\'i}vel Superior} (CAPES) project number $88887.803891/2023-00$ for their financial support. D. A. Sales acknowledges the \textit{Conselho Nacional de Desenvolvimento Cient{\'i}fico e Tecnol{\'o}gico} (CNPq) and the \textit{Funda{\c c}{\~a}o de Amparo {\`a} Pesquisa do Estado do Rio Grande do Sul} (FAPERGS). 
\end{acknowledgement}

\appendix

\section{Appendix}\label{appA}
For the setting addressed in this work, namely the time-like vector $\hat{u}^a$ given in \eqref{timeve}, the energy-momentum source in \eqref{tem}, the metric in \eqref{mett}, and the connection defined by its temporal part in \eqref{Gamma_0ab} with the shear \eqref{sig1}, and the spatial part \eqref{Gamma_ijk} with the differential equation from the Jacobi identity \eqref{deromegak}, the energy-momentum covariant conservation
\begin{align*}
&\nabla_bT^{ab} = 0\,,
\end{align*} 
results in the following non-trivial differential equations for the variables $H$, $\Omega_m$, $\eta$, and $r$:
\begin{align*}
\nabla_bT^{0b}=&\, 3\,H  \left\{ -2\,\cos \left( \eta  \right) \Omega_{{m}}H^2  \cosh\left( r  \right) \sinh \left( r  \right)  \left( 1+w \right) \right.\nonumber\\&\left.\sqrt {\Omega_{{K}}  }-H  \left(  \left( 1+w \right)  \left( \cosh \left( r  \right)  \right) ^{2}-w \right) \dot{\Omega}_{{m}} + \left[2\,\right.\right.\nonumber\\&\left.\left.\left( w -\left( 1+w \right)  \left( \cosh \left( r  \right)  \right) ^{2}\right) \dot{H} + \left( 1+w \right) H  \left( -2\, \right.\right.\right.\nonumber\\&\left.\left.\left. \dot{r}\cosh \left( r  \right) \sinh \left( r  \right) + \left(  \left( - \Sigma_-  \sqrt {3}+ 3\,\Sigma_{+}   \right) \right.\right.\right.\right.\nonumber\\&\left.\left.\left.\left. \left( \cos \left( \eta  \right)  \right) ^{2}-2\,\cos \left( \eta   \right) \sin \left( \eta  \right)  \Phi_3   -  \Sigma_{+}   -4\,\right. \right.\right.\right.\nonumber\\&\left.\left.\left.\left. +\Sigma_-   \sqrt {3} \right)\left( \cosh \left( r \right)  \right) ^{2}+ \left( -  3\,\Sigma_{+}   +  \Sigma_-   \sqrt {3} \right)  \right.\right.\right.\nonumber\\&\left.\left.\left.\left( \cos \left( \eta  \right)  \right) ^{2}+2\,\cos \left( \eta \right) \sin \left( \eta   \right)  \Phi_3    +\left( \Sigma_{+}    +1\right.\right.\right.\right.\nonumber\\&\left.\left.\left.\left.- \Sigma_-   \sqrt {3} \right)\right) H \right]\Omega_{{m}}  \right\}\,,\\
\nabla_bT^{1b}=&-3\, H  \left( 1+w \right)  \left\{ -\Omega_{{m}}   \left( 3\, \left( \cos \left( \eta  \right)  \right) ^{2}-1 \right)  \sqrt {\Omega_{{K}}  }\right.\nonumber\\&\left. H^2\left(\left( \cosh \left( r   \right)\right)^2 -1 \right)   -\Omega_{{m}}  \left( 2\, \left( \cosh \left( r   \right)  \right) ^{2}-1 \right) \right.\nonumber\\&\left.\cos \left( \eta  \right) H \dot{r}  +\cosh \left( r  \right)  \left[ -H  \cos \left( \eta \right) \dot{\Omega}_{{m}}   +\Omega_{{m}}   \right.\right.\nonumber\\&\left.\left. \left( \sin \left( \eta   \right) \dot{\eta} H-2\,\cos \left( \eta   \right) \dot{H}  -2\,H^2   \left(  \left( -  \Sigma_{+}    +2 \right)\right.  \right. \right.\right.\nonumber\\&\left.\left.\left.\left.\cos \left( \eta  \right)+\sin \left( \eta \right)  \Phi_{3}  \right) \right)  \right] \sinh \left( r  \right)  \right\}\,,\\
    \nabla_bT^{3b}=&\,3\,H \left( 1+w \right)  \left\{ 3\,\sin \left( \eta  \right) \Omega_{{m}} H^2  \left( \cosh \left( r  \right) -1 \right) \right.\nonumber\\&\left. \left( \cosh \left( r  \right) +1 \right) \cos \left( \eta \right)\sqrt {\Omega_{{K}}  }+ \Omega_{{m}}  \left( 2\, \left( \cosh \left( r   \right)  \right) ^{2}\right.  \right.\nonumber\\&\left.\left.-1 \right) \sin \left( \eta  \right) H \dot{r}+\cosh \left( r  \right)  \left[ \sin \left( \eta  \right) H \dot{\Omega}_{{m}}  +\Omega_{{m}}     \right.\right.\nonumber\\&\left.\left.\left( H \cos \left( \eta  \right)  \dot{\eta}+\sin \left( \eta   \right)\left( 2\,\dot{H} +H^2  \left( - \Sigma_- \sqrt {3}  \right.\right.\right.\right.\right.\nonumber\\&\left.\left.\left.\left.\left.  +  \Sigma_{+} +4 \right)  \right)  \right)  \right] \sinh \left( r\right)  \right\}\,,
\end{align*}
while $\nabla_bT^{2b}\equiv 0$ is identically null. Considering the same setting, the non-trivially satisfied Einstein-Hilbert field equations with shear and tilt result in
\begin{align*}
E_{11}=&- \left\{  \left( 2\, \Sigma_{+}+2 \right) \dot{H} + \left[ 2\, \dot{\Sigma}_{+}+H\left(3\,\Omega_{{m}}\left( 1+w \right)   \right.\right.\right.\nonumber\\&\left.\left. \left.\left(\left( \cosh \left( r   \right)\right)^2 -1 \right)    \left( \cos \left( \eta   \right)  \right) ^{2}+3\,w\,\Omega_{{m}}  +3\,  \left( \Sigma_{+}    \right) ^{2}\right.\right.\right.\nonumber\\&\left.\left.\left.+3\, \left(  \Sigma_-  \right) ^{2}- \left(  \Phi_3   \right) ^{2}+6\,  \Sigma_{+}   -\Omega_{{K}}+3 \right)\right] H  \right\} ,\\
E_{22}=&- \left\{  \left( -  \Sigma_- \sqrt {3}-  \Sigma_{+} +2 \right) \dot{H} + \left[ - \dot{\Sigma}_{-}  \sqrt {3}-\dot{\Sigma}_{+}\right.\right.\nonumber\\&\left.\left.+\left(3\, \left( \Sigma_{-} \right)   ^{2}-3\,  \Sigma_{-}  \sqrt {3}+3\,w\,\Omega_{{m}}  + \left( \Phi_3   \right) ^{2}\right.\right.\right.\nonumber\\&\left.\left.\left.+3\, \left(  \Sigma_{+}   \right) ^{2}  -\Omega_{{K}} -3\,  \Sigma_{+}    +3 \right)H\right] H   \right\}\,,\\
  E_{33}=&  \left( -  \Sigma_- \sqrt {3}+ H \Sigma_{+}  -2 \right) \dot{H} + \left\{\dot{ \Sigma}_{+}   -  \dot{ \Sigma}_{-}  \sqrt {3}\right.\nonumber\\&\left.+\left[3\,\Omega_{{m}}   \left( \cosh \left( r  \right) -1 \right)  \left( \cosh \left( r   \right) +1 \right)  \left( 1+w \right)  \right.\right.\nonumber\\&\left.\left.\left( \cos \left( \eta   \right)  \right) ^{2}-3\,\Omega_{{m}}  \left( 1+w \right)  \left( \cosh \left( r  \right)  \right) ^{2}-3\, \left(   \Sigma_{+} \right)  ^{2}\right.\right.\nonumber\\&\left.\left.-3\, \left(   \Sigma_{-} \right)  ^{2}-3\,  \Sigma_-  \sqrt {3}-3\, \left(   \Phi_3 \right)   ^{2}+\Omega_{{K}}    \right.\right.\nonumber\\&\left.\left.  +3\,\Omega_{{m}}+3\,  \Sigma_{+}-3 \right]H\right\}H\,,\end{align*}\begin{align*}
  E_{31}=&\, H  \dot{\Phi}_3   + \dot{H}    \Phi_3   -3\, \left[ \sin \left( \eta   \right) \Omega_{{m}}  \left( \cosh \left( r   \right) -1 \right) \right.\nonumber\\&\left. \left( \cosh \left( r   \right) +1 \right)  \left( 1+w \right) \cos \left( \eta   \right) + \left( - \Sigma_-   \sqrt {3}  \right.\right.\nonumber\\&\left.\left. +3\, \Sigma_{+}  +3 \right)  \Phi_3   \right] H^2\,,
\end{align*}
where
\begin{align*}
E_{00}=&\,3\,H^2\left[-\Omega_{{m}}   \left( 1+w \right)  \left( \cosh \left( r   \right)  \right) ^{2}+w\Omega_{{m}}  -   \left( \Sigma_{+} \right)  ^{2}\right.\nonumber\\&\left.- \left( \Sigma_- \right)   ^{2}-\Omega_{{K}}  +1- \left(   \Phi_3  \right) ^{2}/3\right]\,,\nonumber\\
 E_{01}=&\,3\,H^2   \left[ -2\,\sqrt {\Omega_{{K}}  }  \Sigma_{+}    +\Omega_{{m}}  \cosh \left( r   \right) \sinh \left( r   \right) \cos \left( \eta   \right)  \right.\nonumber\\&\left.\left( 1+w \right)  \right]\,,\nonumber
\\
E_{03}=&\,3\,H^2  \left[ \sqrt {\Omega_{{K}}  }  \Phi_3  +\Omega_{{m}} \cosh \left( r  \right) \sinh \left( r   \right) \sin \left( \eta  \right) \right.\nonumber\\&\left. \left( 1+w \right)  \right]\,.
\end{align*}
are the constraints that must be satisfied by the initial conditions.

\section{Appendix}\label{appB}
The dynamical system for quadratic gravity is defined as follows and by the ENV first-order differential equations \eqref{deromegak} and \eqref{senv}.
\onecolumn

\begin{align*}
&y_1 = H(t), y_2 = \dot{H}(t), y_3 = \ddot{H}(t), y_4 = \Omega_K(t), y_5 = \Omega_m(t), y_6 = \Phi_{3}(t), y_7 = \Phi_{3, 0}(t), y_8 = \Phi_{3, 1}(t), y_9 = \Sigma_{+}(t),\\& y_{10} = \Sigma_{+ 1}(t), y_{11} = \Sigma_{+ 2}(t), y_{12} = \Sigma_{-}(t), y_{13} = \Sigma_{- 1}(t), y_{14} = \Sigma_{- 2}(t), y_{15} = \eta(t), y_{16} = r(t).
\end{align*}
With the first-order differential equations:

\begin{align*}
\dddot{H}=& \,1/36\beta \left\{6\,y_{{6}} \left(  \left( 3\,y_{{9}}y_{{12}}-y_{{13}}
 \right) y_{{6}}+y_{{7}}y_{{12}} \right) {y_{{1}}}^{4}\alpha\,\sqrt {3
}-3\,{y_{{1}}}^{2}y_{{5}} \left( 1+w \right)  \left( \cosh \left( y_{{
16}} \right)  \right) ^{2}+ \left(  \left( -6\,{y_{{6}}}^{4}+ \left( -
36\,{y_{{9}}}^{2}-36\,{y_{{12}}}^{2}\right.\right.\right.\right.\\&\left.\left.\left.\left.+12\,y_{{4}}-36 \right) {y_{{6}}}^
{2}+ \left( -48\,y_{{7}}-24\,y_{{8}} \right) y_{{6}}-54\,{y_{{9}}}^{4}
+ \left( -108\,{y_{{12}}}^{2}+612\,y_{{4}}-108 \right) {y_{{9}}}^{2}+
 \left( -144\,y_{{10}}-72\,y_{{11}}-288\,y_{{4}} \right) y_{{9}}\right.\right.\right.\\&\left.\left.\left.-54\,{
y_{{12}}}^{4}+ \left( 36\,y_{{4}}-108 \right) {y_{{12}}}^{2}+ \left( -
144\,y_{{13}}-72\,y_{{14}} \right) y_{{12}}+18\,{y_{{4}}}^{2}+ \left( 
144\,y_{{10}}-36 \right) y_{{4}}-24\,{y_{{7}}}^{2}-72\,{y_{{10}}}^{2}-
72\,{y_{{13}}}^{2} \right) \beta\right.\right.\\&\left.\left.- \left( 16\,{y_{{6}}}^{4}+ \left( 51
\,{y_{{9}}}^{2}+33\,{y_{{12}}}^{2}+19\,y_{{4}}-18\,y_{{10}}-3 \right) 
{y_{{6}}}^{2}+ \left( 18\,y_{{7}}y_{{9}}-4\,y_{{7}}-2\,y_{{8}}
 \right) y_{{6}}+36\,{y_{{9}}}^{4}+ \left( 72\,{y_{{12}}}^{2}+24\,y_{{
4}}-9 \right) {y_{{9}}}^{2}\right.\right.\right.\\&\left.\left.\left.+ \left( -12\,y_{{10}}-6\,y_{{11}}-12\,y_{{
4}} \right) y_{{9}}+36\,{y_{{12}}}^{4}+ \left( 12\,y_{{4}}-9 \right) {
y_{{12}}}^{2}+ \left( -12\,y_{{13}}-6\,y_{{14}} \right) y_{{12}}+{y_{{
7}}}^{2}+3\,{y_{{10}}}^{2}-12\,y_{{10}}y_{{4}}+3\,{y_{{13}}}^{2}
 \right)\right.\right.\\&\left.\left. \alpha \right) {y_{{1}}}^{4}+ \left( -72\, \left( {y_{{9}}}^{
2}+{y_{{12}}}^{2}+1/3\,{y_{{6}}}^{2}+y_{{4}}+9/2 \right) y_{{2}}\beta+
6\, \left( {y_{{9}}}^{2}+{y_{{12}}}^{2}+1/3\,{y_{{6}}}^{2} \right) y_{
{2}}\alpha-3\,{y_{{6}}}^{2}-9\,{y_{{9}}}^{2}-9\,{y_{{12}}}^{2}+3\,y_{{
4}}\right.\right.\\&\left.\left.-9+ \left( -6\,w+3 \right) y_{{5}} \right) {y_{{1}}}^{2}-216\,y_{{3
    }}y_{{1}}\beta-162\,{y_{{2}}}^{2}\beta-6\,y_{{2}}\right\}
\,,\\
 \dot{\Omega}_m=& -\left\{ -1/2\, \left( -2\,\cos \left( y_{{15}} \right) \cosh \left( y_{{16}}
 \right) \sinh \left( y_{{16}} \right) {y_{{1}}}^{2} \left( 1+w
 \right) \sqrt {y_{{4}}}+{y_{{1}}}^{2}y_{{12}} \left(  \left( \cosh
 \left( y_{{16}} \right)  \right) ^{2}-1 \right)  \left(  \left( \cos
 \left( y_{{15}} \right)  \right) ^{2}-1 \right)  \right.\right.\\&\left.\left. \left( 1+w \right) 
\sqrt {3}+ \left( -3\,{y_{{1}}}^{2}y_{{9}} \left( 1+w \right)  \left( 
\cos \left( y_{{15}} \right)  \right) ^{2}+2\,\sin \left( y_{{15}}
 \right) {y_{{1}}}^{2}y_{{6}} \left( 1+w \right) \cos \left( y_{{15}}
 \right) + \left( y_{{9}}-2 \right)  \left( 1+w \right) {y_{{1}}}^{2}+
2\,y_{{2}} \left( w-1 \right)  \right) \right.\right.\\&\left.\left.  \left( \cosh \left( y_{{16}}
 \right)  \right) ^{2}+3\,{y_{{1}}}^{2}y_{{9}} \left( 1+w \right) 
 \left( \cos \left( y_{{15}} \right)  \right) ^{2}-2\,\sin \left( y_{{
15}} \right) {y_{{1}}}^{2}y_{{6}} \left( 1+w \right) \cos \left( y_{{
15}} \right) - \left( y_{{9}}+1 \right)  \left( 1+w \right) {y_{{1}}}^
{2}-2\,wy_{{2}} \right) y_{{5}}
\right\}/\\&\left\{ -1/2\,y_{{1}} \left(  \left( w-1 \right)  \left( \cosh \left( y_{{16}}
 \right)  \right) ^{2}-w \right)
\right\}\,,\\
\dot{\Phi}_{3, 1}=&\left\{-18\,y_{{1}} \left(  \left( 2/3\,y_{{12}} \left( \beta+8/3\,\alpha
 \right) {y_{{6}}}^{3}+ \left(  \left( { {11}/{6}}\,{y_{{12}}}^{3}
+ \left( {{35}/{6}}\,{y_{{9}}}^{2}-1/2\,y_{{10}}-{ {11}/{6}}+
1/6\,y_{{4}} \right) y_{{12}}-1/2\,y_{{14}}-2\,y_{{13}} \right.\right.\right.\right.\right.\\&\left.\left.\left.\left.\left. -1/2\,y_{{9}}y_
{{13}} \right) \alpha+2\,y_{{12}}\beta\, \left( {y_{{9}}}^{2}+{y_{{12}
}}^{2}-y_{{4}}+2 \right)  \right) y_{{6}}+\alpha\, \left(  \left( -y_{
{7}}-1/6\,y_{{8}}+y_{{7}}y_{{9}} \right) y_{{12}}-1/2\,y_{{7}}y_{{13}}
 \right)  \right) {y_{{1}}}^{2}\right.\right.\\&\left.\left.+2\,y_{{12}} \left( \beta\,y_{{2}}-1/3
\,\alpha\,y_{{2}}+1/12 \right) y_{{6}} \right) \sqrt {3}-9\,y_{{1}}y_{
{5}}\sin \left( y_{{15}} \right)  \left( \cosh \left( y_{{16}}
 \right) -1 \right)  \left( \cosh \left( y_{{16}} \right) +1 \right) 
 \left( 1+w \right) \cos \left( y_{{15}} \right) \right.\\&\left.+ \left( 36\, \left( 
y_{{9}}+1 \right)  \left( \beta+8/3\,\alpha \right) {y_{{6}}}^{3}+36\,
y_{{7}} \left( \beta+8/3\,\alpha \right) {y_{{6}}}^{2}+ \left( 
 \left(  \left( 153\,y_{{9}}+72 \right) {y_{{12}}}^{2}+39\,y_{{12}}y_{
{13}}+153\,{y_{{9}}}^{3}+72\,{y_{{9}}}^{2}+ \left( 21\,y_{{10}}\right.\right.\right.\right.\right.\\&\left.\left.\left.\left.\left.+21\,y_
{{4}}-99 \right) y_{{9}}-108\,y_{{10}}-27\,y_{{11}}+3\,y_{{4}}-18
 \right) \alpha+108\, \left(  \left( y_{{9}}+1 \right) {y_{{12}}}^{2}+
2/3\,y_{{12}}y_{{13}}+{y_{{9}}}^{3}+{y_{{9}}}^{2}+ \left( -7/3\,y_{{4}
}\right.\right.\right.\right.\right.\\&\left.\left.\left.\left.\left.+2/3\,y_{{10}}+2 \right) y_{{9}}+2-1/3\,y_{{4}} \right) \beta
 \right) y_{{6}}+ \left( 33\,y_{{7}}{y_{{12}}}^{2}+51\,y_{{7}}{y_{{9}}
}^{2}+ \left( -54\,y_{{7}}-9\,y_{{8}} \right) y_{{9}}+ \left( -27\,y_{
{10}}+3\,y_{{4}}-33 \right) y_{{7}}-18\,y_{{8}} \right) \alpha\right.\right.\\&\left.\left.+36\,
\beta\,y_{{7}} \left( {y_{{9}}}^{2}+{y_{{12}}}^{2}-y_{{4}}+2 \right) 
 \right) {y_{{1}}}^{3}+ \left(  \left(  \left( -36\,y_{{2}}y_{{9}}-21
\,y_{{2}} \right) \alpha+ \left( 108\,y_{{2}}y_{{9}}+252\,y_{{2}}
 \right) \beta+9\,y_{{9}}+9 \right) y_{{6}}-12\,y_{{2}} \left( y_{{7}}
\right.\right.\right.\\&\left.\left.\left.+3/4\,y_{{8}} \right) \alpha+36\,\beta\,y_{{2}}y_{{7}}+3\,y_{{7}}
 \right) y_{{1}}+36\, \left( \beta-1/12\,\alpha \right) y_{{6}}y_{{3}}
\right\}/\left\{3\,\alpha\,{y_{{1}}}^{2}\right\}\,,\\
\dot{\Sigma}_{+2}=&\left\{-12\,y_{{6}}{y_{{1}}}^{3} \left(  \left( -3\,y_{{9}}y_{{12}}+3\,y_{{12
}}+2\,y_{{13}} \right) y_{{6}}+y_{{7}}y_{{12}} \right) \alpha\,\sqrt {
3}+9\, \left( 1+w \right) y_{{1}}y_{{5}} \left(  \left( \cos \left( y_
{{15}} \right)  \right) ^{2}-1/3 \right)  \left( \cosh \left( y_{{16}}
 \right)  \right) ^{2}\right.\\&\left.-9\,y_{{1}}y_{{5}} \left( 1+w \right)  \left( 
\cos \left( y_{{15}} \right)  \right) ^{2}+ \left(  \left( -64\,{y_{{6
}}}^{4}+ \left( -102\,{y_{{9}}}^{2}-66\,{y_{{12}}}^{2}-6\,y_{{4}}+156
\,y_{{9}}+88\,y_{{10}}+66 \right) {y_{{6}}}^{2}+ \left( 68\,y_{{7}}y_{
{9}}+108\,y_{{7}}\right.\right.\right.\right.\\&\left.\left.\left.\left.+24\,y_{{8}} \right) y_{{6}}+144\,{y_{{9}}}^{3}+144\,
{y_{{9}}}^{2}y_{{10}}+ \left( 144\,{y_{{12}}}^{2}+96\,y_{{12}}y_{{13}}
-36 \right) y_{{9}}+48\,y_{{10}}{y_{{12}}}^{2}+18\,{y_{{7}}}^{2}-66\,y
_{{10}}-36\,y_{{11}} \right) \alpha\right.\right.\\&\left.\left.+48\, \left( -1/2\,{y_{{6}}}^{4}+
 \left( -3+1/2\,y_{{10}}+3/2\,y_{{9}}-3/2\,{y_{{12}}}^{2}+3/2\,y_{{4}}
-3/2\,{y_{{9}}}^{2} \right) {y_{{6}}}^{2}+y_{{7}}y_{{9}}y_{{6}}+9/2\,{
y_{{9}}}^{3}+ \left( -6\,y_{{4}}+9/2\,y_{{10}} \right) {y_{{9}}}^{2}\right.\right.\right.\\&\left.\left.\left.+
 \left( -3/2\,y_{{4}}+9+3\,y_{{12}}y_{{13}}+9/2\,{y_{{12}}}^{2}
 \right) y_{{9}}+3/2\,y_{{10}} \left( {y_{{12}}}^{2}-y_{{4}}+2
 \right)  \right) \beta \right) {y_{{1}}}^{3}+ \left( -42\, \left( -4/
7\,{y_{{6}}}^{2}+y_{{9}}+4/7\,y_{{10}}\right.\right.\right.\\&\left.\left.\left.+3/7\,y_{{11}} \right) y_{{2}}
\alpha+ \left( -72\,\beta\,y_{{2}}-6 \right) {y_{{6}}}^{2}+ \left( 504
\,\beta\,y_{{2}}+18 \right) y_{{9}}+72\,y_{{10}}y_{{2}}\beta+6\,y_{{10
}}+ \left( 3\,w+3 \right) y_{{5}} \right) y_{{1}}+72\,y_{{3}} \left( 
\beta-1/12\,\alpha \right) y_{{9}}
\right\}\\&/\left\{6\,\alpha\,{y_{{1}}}^{2}\right\}\,,\\
\dot{\Sigma}_{-2}=& -\left\{9\,\sqrt {3} \left(  \left(  \left(  \left( -16\,{y_{{12}}}^{3}-16\,y_
{{13}}{y_{{12}}}^{2}+ \left(  \left( -{ {28}/{3}}+4\,y_{{9}}
 \right) {y_{{6}}}^{2}-{ {44}/{9}}\,y_{{6}}y_{{7}}-16\,{y_{{9}}}^{
2}+ \left( -{ {32}/{3}}\,y_{{10}}+16/3\,y_{{4}} \right) y_{{9}}-8/
3\,y_{{4}}\right.\right.\right.\right.\right.\right.\\&\left.\left.\left.\left.\left.\left.+4 \right) y_{{12}}-{ {40}/{9}}\,y_{{13}}{y_{{6}}}^{2}+
 \left( -16/3\,{y_{{9}}}^{2}+{ {22}/{3}}-8/3\,y_{{4}} \right) y_{{
13}}+4\,y_{{14}} \right) \alpha-16/3\, \left( 9/2\,{y_{{12}}}^{3}+9/2
\,y_{{13}}{y_{{12}}}^{2}+ \left( 3/2\,{y_{{6}}}^{2}\right.\right.\right.\right.\right.\right.\\&\left.\left.\left.\left.\left.\left.+y_{{6}}y_{{7}}+9/2
\,{y_{{9}}}^{2}+ \left( -6\,y_{{4}}+3\,y_{{10}} \right) y_{{9}}-3/2\,y
_{{4}}+9 \right) y_{{12}}+3/2\, \left( {y_{{9}}}^{2}+1/3\,{y_{{6}}}^{2
}-y_{{4}}+2 \right) y_{{13}} \right) \beta \right) {y_{{1}}}^{3}+
 \left( 14/3\,y_{{2}} \left( y_{{12}}\right.\right.\right.\right.\right.\\&\left.\left.\left.\left.\left.+4/7\,y_{{13}}+3/7\,y_{{14}}
 \right) \alpha+ \left( -56\,\beta\,y_{{2}}-2 \right) y_{{12}}-8\,y_{{
13}}y_{{2}}\beta-2/3\,y_{{13}} \right) y_{{1}}-8\, \left( \beta-1/12\,
\alpha \right) y_{{3}}y_{{12}} \right) \sqrt {3}+ \left( y_{{5}}
 \left( \left(\cos \left( y_{{15}} \right)\right)^2 \right.\right.\right.\right.\\&\left.\left.\left.\left.-1 \right)    \left( 1+w \right)  \left( \cosh \left( y_{
{16}} \right)  \right) ^{2}-y_{{5}} \left( 1+w \right)  \left( \cos
 \left( y_{{15}} \right)  \right) ^{2}+ \left(  \left( -{ {22}/{3}
}\,{y_{{6}}}^{2}{y_{{12}}}^{2}-{ {64}/{9}}\,{y_{{6}}}^{4}+ \left( 
8\,y_{{10}}+{ {22}/{3}}-14/3\,y_{{4}}\right.\right.\right.\right.\right.\right.\\&\left.\left.\left.\left.\left.\left.-{ {34}/{3}}\,{y_{{9}}}^{
2}+12\,y_{{9}} \right) {y_{{6}}}^{2}+ \left( 4\,y_{{7}}y_{{9}}+12\,y_{
{7}}+8/3\,y_{{8}} \right) y_{{6}}+2\,{y_{{7}}}^{2} \right) \alpha-8\,
 \left( {y_{{9}}}^{2}+{y_{{12}}}^{2}+1/3\,{y_{{6}}}^{2}-y_{{4}}+2
 \right) \beta\,{y_{{6}}}^{2} \right) {y_{{1}}}^{2}\right.\right.\right.\\&\left.\left.\left.+8/3\,\alpha\,y_{{2
}}{y_{{6}}}^{2}+ \left( -8\,\beta\,y_{{2}}-2/3 \right) {y_{{6}}}^{2}+
 \left( 1+w \right) y_{{5}} \right) y_{{1}} \right)
\right\}/\left\{18\,\alpha\,{y_{{1}}}^{2}\right\}\,,\end{align*}\begin{align*}
\dot{\eta}=& \left\{y_{{1}} \left( \cos \left( y_{{15}} \right) \cosh \left( y_{{16}}
 \right) \sinh \left( y_{{16}} \right) \sqrt {3}y_{{12}}-3\,\cos
 \left( y_{{15}} \right) \cosh \left( y_{{16}} \right) \sinh \left( y_
{{16}} \right) y_{{9}}+2\,\sin \left( y_{{15}} \right) \cosh \left( y_
{{16}} \right) \sinh \left( y_{{16}} \right) y_{{6}}\right.\right.\\&\left.\left.- \left( \cosh
 \left( y_{{16}} \right)  \right) ^{2}\sqrt {y_{{4}}}+\sqrt {y_{{4}}}
 \right) \sin \left( y_{{15}} \right) 
\right\}/\left\{\cosh \left( y_{{16}} \right) \sinh \left( y_{{16}} \right) 
\right\}\,,\\
\dot{r}=& -\left\{2\, \left( w\cos \left( y_{{15}} \right)  \left( \cosh \left( y_{{16}}
 \right) -1 \right)  \left( \cosh \left( y_{{16}} \right) +1 \right) 
\sqrt {y_{{4}}}+3/2\,\cosh \left( y_{{16}} \right) \sinh \left( y_{{16
}} \right)  \left(  \left( 1/3\,y_{{12}}-1/3\, \left( \cos \left( y_{{
15}} \right)  \right) ^{2}y_{{12}} \right) \sqrt {3}\right.\right.\right.\\&\left.\left.\left.+w+ \left( \cos
 \left( y_{{15}} \right)  \right) ^{2}y_{{9}}-2/3\,\cos \left( y_{{15}
} \right) \sin \left( y_{{15}} \right) y_{{6}}-1/3\,y_{{9}}-1/3
 \right)  \right) y_{{1}}
\right\}/\left\{\left( w-1 \right)  \left( \cosh \left( y_{{16}} \right)  \right) ^{2
}-w
\right\}\,.
  \end{align*}
Subject to the constraints:
      \begin{align*}
E_{00}=& \,6\,{y_{{1}}}^{4}y_{{6}} \left(  \left( 3\,y_{{9}}y_{{12}}-y_{{13}}
 \right) y_{{6}}+y_{{7}}y_{{12}} \right) \alpha\,\sqrt {3}-3\,{y_{{1}}
}^{2}y_{{5}} \left( 1+w \right)  \left( \cosh \left( y_{{16}} \right) 
 \right) ^{2}+ \left(  \left( -16\,{y_{{6}}}^{4}+ \left( -51\,{y_{{9}}
}^{2}-33\,{y_{{12}}}^{2}-19\,y_{{4}}\right.\right.\right.\\&\left.\left.\left.+18\,y_{{10}}+3 \right) {y_{{6}}}^
{2}+ \left( -18\,y_{{7}}y_{{9}}+4\,y_{{7}}+2\,y_{{8}} \right) y_{{6}}-
36\,{y_{{9}}}^{4}+ \left( -72\,{y_{{12}}}^{2}-24\,y_{{4}}+9 \right) {y
_{{9}}}^{2}+ \left( 12\,y_{{10}}+6\,y_{{11}}+12\,y_{{4}} \right) y_{{9
}}\right.\right.\\&\left.\left.-36\,{y_{{12}}}^{4}+ \left( -12\,y_{{4}}+9 \right) {y_{{12}}}^{2}+
 \left( 12\,y_{{13}}+6\,y_{{14}} \right) y_{{12}}-{y_{{7}}}^{2}-3\,{y_
{{10}}}^{2}+12\,y_{{10}}y_{{4}}-3\,{y_{{13}}}^{2} \right) \alpha+24\,
\beta\, \left( -1/4\,{y_{{6}}}^{4}\right.\right.\\&\left.\left.+ \left( 1/2\,y_{{4}}-3/2\,{y_{{12}}
}^{2}-3/2\,{y_{{9}}}^{2}-3/2 \right) {y_{{6}}}^{2}+y_{{6}}y_{{7}}-9/4
\,{y_{{9}}}^{4}+ \left( -9/2\,{y_{{12}}}^{2}-9/2+3/2\,y_{{4}} \right) 
{y_{{9}}}^{2}+ \left( -6\,y_{{4}}+3\,y_{{10}} \right) y_{{9}}\right.\right.\\&\left.\left.-9/4\,{y_
{{12}}}^{4}+ \left( -9/2+3/2\,y_{{4}} \right) {y_{{12}}}^{2}+3\,y_{{12
}}y_{{13}}+3/2\,y_{{4}}+3/4\,{y_{{4}}}^{2} \right)  \right) {y_{{1}}}^
{4}+ \left( 6\, \left( {y_{{9}}}^{2}+{y_{{12}}}^{2}+1/3\,{y_{{6}}}^{2}
 \right) y_{{2}}\alpha-72\, \left( {y_{{9}}}^{2}\right.\right.\\&\left.\left.+{y_{{12}}}^{2}+1/3\,{
y_{{6}}}^{2}-3/2 \right) y_{{2}}\beta+3\,y_{{5}}w-3\,{y_{{9}}}^{2}-3\,
{y_{{12}}}^{2}-{y_{{6}}}^{2}-3\,y_{{4}}+3 \right) {y_{{1}}}^{2}+36\,y_
{{3}}y_{{1}}\beta-18\,{y_{{2}}}^{2}\beta
\,,\\
E_{01}=&\,3\,{y_{{1}}}^{2} \left( 24\,{y_{{4}}}^{3/2}{y_{{1}}}^{2}y_{{9}}\beta+
 \left( 2\,{y_{{1}}}^{2}\sqrt {3}y_{{12}}{y_{{6}}}^{2}\alpha+ \left( 
 \left( -16\,\alpha-24\,\beta \right) {y_{{9}}}^{3}+ \left(  \left( -
16\,{y_{{12}}}^{2}-{ {34}/{3}}\,{y_{{6}}}^{2}+4 \right) \alpha-24
\,\beta\, \left( {y_{{12}}}^{2}\right.\right.\right.\right.\right.\\&\left.\left.\left.\left.\left.+1/3\,{y_{{6}}}^{2}+2 \right)  \right) 
y_{{9}}+6\,\alpha\, \left( -{y_{{6}}}^{2}-y_{{6}}y_{{7}}+y_{{10}}+1/3
\,y_{{11}} \right)  \right) {y_{{1}}}^{2}+ \left( 2\,\alpha\,y_{{2}}-
24\,\beta\,y_{{2}}-2 \right) y_{{9}} \right) \sqrt {y_{{4}}}\right.\\&\left.+\cosh
 \left( y_{{16}} \right) \sinh \left( y_{{16}} \right) \cos \left( y_{
{15}} \right) y_{{5}} \left( 1+w \right)  \right) 
\,,\\
E_{03}=&\,3\,{y_{{1}}}^{2} \left( -12\, \left( \beta-1/12\,\alpha \right) {y_{{1
}}}^{2}y_{{6}}{y_{{4}}}^{3/2}+ \left( -6\, \left(  \left( y_{{9}}-1/2
 \right) y_{{12}}-1/2\,y_{{13}} \right) {y_{{1}}}^{2}y_{{6}}\alpha\,
\sqrt {3}+ \left(  \left( 4\,\beta+{ {32}/{3}}\,\alpha \right) {y_
{{6}}}^{3}+ \left(  \left( 17\,{y_{{9}}}^{2}\right.\right.\right.\right.\right.\\&\left.\left.\left.\left.\left.+11\,{y_{{12}}}^{2}-9\,y_{
{9}}-9\,y_{{10}}-2 \right) \alpha+12\,\beta\, \left( {y_{{9}}}^{2}+{y_
{{12}}}^{2}+2 \right)  \right) y_{{6}}-\alpha\, \left( y_{{8}}+3\,y_{{
7}} \right)  \right) {y_{{1}}}^{2}+y_{{6}} \left( -\alpha\,y_{{2}}+12
\,\beta\,y_{{2}}+1 \right)  \right) \sqrt {y_{{4}}}\right.\\&\left.+\cosh \left( y_{{
16}} \right) \sinh \left( y_{{16}} \right) \sin \left( y_{{15}}
 \right) y_{{5}} \left( 1+w \right)  \right) \,.
  \end{align*}

\balance
\twocolumn

\section{Appendix}\label{appc}

The curvature invariants $R$, $R_{ab}R^{ab}$, and $R_{abcd}R^{abcd}$, expressed in terms of the ENV \eqref{ENV1} and \eqref{vnq1}, are given by:
\begin{align*}
 &R=\left\{-2\,H^2\, \left( -\left({\Phi_3}\right)^{2}-3\,\left({\Sigma_{+}}\right)^{2}-3\,\left({\Sigma_-}\right)^{2}+3\,\Omega_{{K}}\right.\right.\nonumber\\&\left.\left.-3\,\dot{H}/H-6 \right) \right\}/\left\{3\right\},\nonumber\\
&R_{ab}R^{ab}=\left\{ 2\,{H}^{4} \left( -2\,\Phi_3 \left(  \left( 3\,\Sigma_{+}\Sigma_--\Sigma_{-1}H \right)\Phi_3\right.\right.\right.\nonumber\\&\left.\left.\left.+3\,\Sigma_-\Phi_{3 ,0}H \right) \sqrt {3}+6\,\left({\Phi_3}\right)^{4}+ \left( 14\,\left({\Sigma_{+}}\right)^{2}+15\,\right.\right.\right.\nonumber\\&\left.\left.\left.-9\,\Omega_{{K}}+6\,\dot{H}/H-6\,\Sigma_{+1}H+15\,\left({\Sigma_-}\right)^{2} \right)\left( {\Phi_3}\right)^{2}\right.\right.\nonumber\\&\left.\left.+6\,\Phi_{3 ,0}H \Phi_3\left( \Sigma_{+}+1 \right) +18\,\left({\Sigma_{+}}\right)^{4}+ \left( 36\,\left({\Sigma_-}\right)^{2}-36\,\Omega_{{K}}\right.\right.\right.\nonumber\\&\left.\left. \left.+18\,\dot{H}/H+45 \right)\left( {\Sigma_{+}}\right)^{2}+18\,\Sigma_{+}\Sigma_{+1}H+18\,\left({\Sigma_-}\right)^{4}\right.\right.\nonumber\\&\left.\left.+\left( 45+18\,\dot{H}/H \right) \left({\Sigma_-}\right)^{2}+18\,\Sigma_-\Sigma_{-1}H+6\,\left({\dot{H}/H}\right)^{2}\right.\right.\nonumber\\&\left.\left.+6\, \dot{H}/H \left( 3-\Omega_{{K}} \right)+18+6\,\left({\Omega_{{K}}}\right)^{2}+\left({\Phi_{3 ,0}H}\right)^{2}\right.\right.\nonumber\\&\left.\left.+3\,\left({\Sigma_{-1}H}\right)^{2}-18\,\Omega_{{K}}+3\,\left({\Sigma_{+1}H}\right)^{2} \right) \right\}/\left\{9\right\},\nonumber\\
&R_{abcd}R^{abcd}=\left\{4\, \left[ -6\, \left( 2\,\Sigma_-\Phi_3\Sigma_{+}+ \left( 9\,\Sigma_--3\,\Sigma_{-1}H \right) \right.\right.\right.\nonumber\\&\left.\left.\left.\Phi_3+12\,\Sigma_-\Phi_{3 ,0}H \right) \Phi_3\sqrt {3}+27\,\left({\Sigma_{+}}\right)^{4}-12\,\left({\Sigma_{+}}\right)^{3}\right.\right.\nonumber\\&\left.\left.+\left( 36\,\left({\Phi_3}\right)^{2}+54\,\left({\Sigma_-}\right)^{2}-30\,\Omega_{{K}}+12\,\dot{H}/H-12\,\Sigma_{+1}H\right.\right.\right.\nonumber\\&\left.\left.\left.+36 \right) \left({\Sigma_{+}}\right)^{2}+ \left( -6\,\left({\Phi_3}\right)^{2}+8\,\Phi_3\Phi_{3 ,0}H+36\,\left({\Sigma_-}\right)^{2}\right.\right.\right.\nonumber\\&\left.\left.\left.+24\,\Sigma_-\Sigma_{-1}H+24\,\Sigma_{+1}H \right) \Sigma_{+}+11\,\left({\Phi_{3}}\right)^{4}+\left( {\Phi_3}\right)^{2}\right.\right.\nonumber\\&\left.\left. \left( 24\,\left({\Sigma_-}\right)^{2}-8\,\Omega_{{K}}+4\,\dot{H}/H-14\,\Sigma_{+1}H+12 \right)\right.\right.\nonumber\\&\left.\left.+8\,\Phi_3\Phi_{{3,0}}H+27\,\left({\Sigma_-}\right)^{4}+ \left( 36+12\,\dot{H}/H-6\,\Omega_{{K}}\right.\right.\right.\nonumber\\&\left.\left.\left.+12\,\Sigma_{+1}H \right) \left({\Sigma_-}\right)^{2}+24\,\Sigma_-\Sigma_{-1}H+6\,{\Sigma_{-1}H}^{2}-6\,\Omega_{{K}}\right.\right.\nonumber\\&\left.\left.+3\,{\Omega_{{K}}}^{2}+3\,\left({\dot{H}/H}\right)^{2}+6\,\dot{H}/H+6\,\left({\Sigma_{+1}H}\right)^{2}+6\right.\right.\nonumber\\&\left.\left.+2\,\left({\Phi_{{3,0}}}\right)^{2} \right] {H}^{4}\right\}/\left\{9\right\}.
\end{align*}

\balance

\bibliographystyle{utcaps}

\end{document}